\documentclass[twocolumn,times]{aastex62}  

\usepackage{graphicx}
\usepackage{float}
\usepackage{color}
\usepackage{soul}
\usepackage{xspace}
\usepackage{amsmath}
\usepackage{subcaption}
\usepackage[T1]{fontenc}

\begin{document}

\title{Characterizing the Directionality of Gravitational Wave Emission from Matter Motions within Core-collapse Supernovae}

\author[0000-0002-4983-4589]{Michael A. Pajkos}
\affiliation{TAPIR, Mailcode 350-17, California Institute of Technology, Pasadena, CA 91125}

\author[0009-0007-8505-6087]{Steven J. VanCamp}
\affiliation{Department of Physics and Astronomy, Michigan State University, East Lansing, MI 48824, USA}

\author[0000-0002-1473-9880]{Kuo-Chuan Pan}
\affiliation{Department of Physics, National Tsing Hua University, Hsinchu 30013, Taiwan}
\affiliation{Institute of Astronomy, National Tsing Hua University, Hsinchu 30013, Taiwan}
\affiliation{Center for Informatics and Computation, National Tsing Hua University, Hsinchu 30013, Taiwan}
\affiliation{National Center for Theoretical Sciences, National Tsing Hua University, Hsinchu 30013, Taiwan}

\author[0000-0003-1938-9282]{David Vartanyan}
\affiliation{Carnegie Observatories, 813 Santa Barbara St, Pasadena, CA 91101, USA; NASA Hubble Fellow}

\author[0000-0003-4557-4115]{Nils Deppe}
\affiliation{Department of Physics, Cornell University, Ithaca,
  NY, 14853, USA}
\affiliation{Cornell Center for Astrophysics and Planetary
    Science, Cornell University, Ithaca, New York 14853, USA}

\author[0000-0002-5080-5996]{Sean M.~Couch}
\affiliation{Department of Physics and Astronomy, Michigan State University, East Lansing, MI 48824, USA}
\affiliation{Department of Computational Mathematics, Science, and Engineering, Michigan State University, East Lansing, MI 48824, USA}
\affiliation{Facility for Rare Isotope Beams, Michigan State University, East Lansing, MI 48824, USA}




\shorttitle{Directionality of GWs from CCSNe}
\shortauthors{Pajkos et al.}

 \begin{abstract}

We analyze the directional dependence of the gravitational wave (GW) emission from 15 3D neutrino radiation hydrodynamic simulations of core-collapse supernovae.  Using spin weighted spherical harmonics, we develop a new analytic technique to quantify the evolution of the distribution of GW emission over all angles.  We construct a physics-informed toy model that can be used to approximate GW distributions for general ellipsoid-like systems, and use it to provide closed form expressions for the distribution of GWs for different CCSN phases.  Using these toy models, we approximate the PNS dynamics during multiple CCSN stages and obtain similar GW distributions to simulation outputs.  When considering all viewing angles, we apply this new technique to quantify the evolution of preferred directions of GW emission.  For nonrotating cases, this dominant viewing angle drifts isotropically throughout the supernova, set by the dynamical timescale of the protoneutron star.  For rotating cases, during core bounce and the following tens of ms, the strongest GW signal is observed along the equator.  During the accretion phase, comparable---if not stronger---GW amplitudes are generated along the axis of rotation, which can be enhanced by the low T/|W| instability.  We show two dominant factors influencing the directionality of GW emission are the degree of initial rotation and explosion morphology.  Lastly, looking forward, we note the sensitive interplay between GW detector site and supernova orientation, along with its effect on detecting individual polarization modes.
 
\end{abstract}
%

\section{Introduction}

The end of massive stellar evolution is marked by a core-collapse supernova (CCSN).  These stellar explosions, and in some cases implosions, are dynamic events influenced by a variety of physical processes, from neutrino processes, to hydrodynamic turbulence, to magnetic field interactions.  Moments after the supernova is launched, the birth of a compact object called the protoneutron star (PNS) occurs, which will ultimately cool to a neutron star or---with sufficient mass accretion---will collapse to a black hole.  The PNS contains the neutron rich remnants of the once-stellar iron core and is characterized by nuclear matter at densities $\gtrsim 2 \times 10^{14}$ g cm$^{-3}$ near the center.  As turbulent downflows interact with the PNS surface and convection within the PNS develops, the PNS can oscillate, generating gravitational waves (GWs) \citep{sotani:2016, kotake:2017}.  The principal source driving PNS oscillations---PNS convection \citep{andresen:2019,mezzacappa:2023} or accretion downflows \citep{vartanyan:2023}---remains an ongoing discussion.  In the case of accreting matter, the luminosity of the GWs is proportional to the square of the rate at which turbulent kinetic energy is accreted by the PNS \citep{thorne:1996} (Also see \citet{muller:2017} for order of magnitude estimates of GW features).  

These oscillations can exhibit different `modes' which correspond to unique frequencies of emission, based on the restoring force driving the PNS.  Some examples include g-modes driven by gravity, p-modes driven by pressure, r-modes driven by rotation, and w-modes driven by oscillations in spacetime.  As each mode has a unique restoring force, GW observables can encode different characteristics describing the supernova system, such as the average density of the PNS (g-modes) or degree of rotation (r-modes) \citep{unno:1989}.  

Beyond PNS oscillations, other sources of GWs within CCSNe include the `bounce signal' from a deformed rotating PNS \citep{dimm:2008}, hydrodynamic turbulence \citep{pajkos:2019}, asymmetric emission of neutrinos \citep{vartanyan:2020}, compact object ejection \citep{burrows:1996}, and GWs from fluid instabilities like the standing accretion shock instability (SASI) \citep{kuroda:2016, andresen:2019}, or the low T/|W| instability \citep{shibagaki:2020}.  While generating predicted waveforms for CCSNe can be valuable for improving detectability of the next Galactic event, connecting characteristics of the signals to internal source physics is vital.  
Each of these sources also encodes physical characteristics of the supernova center, a region previously unobservable with electromagnetic (EM) radiation.  GWs from fluid instabilities provide information regarding the strength of convection and the timescale on which it occurs \citep{andresen:2017,takiwaki:2016,radice:2019}.  Neutrino sources provide the degree of asymmetric neutrino production (and to some extent mass accretion) \citep{vartanyan:2020}.  The bounce signal is directly related to the rotational content of the PNS just after core bounce \citep{dimm:2008}.  Furthermore, GWs from CCSNe depend on the hot nuclear equation of state (EOS) \citep{pan:2018}.  Several works have also proposed a joint analysis of GWs and neutrinos, to better constrain physical conditions of the PNS \citep{nagakura:2022}, SASI activity \citep{kuroda:2017}, EOS insights \citep{vartanyan:2019}, and stellar compactness along with explosion properties \citep{warren:2020}.  

With the recent Supernova 2023ixf discovery from Itagaki \citep{teja:2023} and the fourth observing run for LIGO \citep{abbott:2015} underway--- along with observations by Virgo \citep{acernese:2015} and KAGRA \citep{akutsu:2019}---supporting GW observers with theoretical predictions is paramount.  With various physical insights from CCSN gravitational radiation, considering these signals in the context of observability is important to distinguish between signal features that are interesting and signal features that are detectable, therefore more valuable.  One often considered factor is the GW amplitude, or the GW strain, commonly denoted $h$.  Depending on the specifications of the GW observatory, detectors have certain sensitivities, or a minimum threshold below which GW amplitudes fall below the noise floor.  Supernova models often produce time domain waveforms (TDWFs) to observe the predicted GW amplitudes over time for a given event.  

The GW frequency content is equally valuable.  GW detectors have limiting factors that constrain the frequency range of observable signals.  For example, for the Laser Interferometer Gravitational-wave Observatory (LIGO), shot noise imparted by the individual laser photons on the test mass limit the upper limit of the frequency range $\sim 8$ kHz.  By contrast, seismic noise establishes the lower limit of the frequency range $\sim 10$ Hz \citep{aasi:2013}.  One common tool for determining the frequency content of the CCSN GW signal includes spectrograms, which display the frequency content of the signal as time evolves.  Characteristic strain plots are another method that displays the frequency distribution of the signal summed over a given time interval.  Characteristic strain plots are often used to compare cumulative GW frequency content to GW detector sensitivity curves \citep{moore:2014}.  

Directly measuring the polarization of the GW signal is an ongoing area of research.  According to Einstein's theory of general relativity (GR), GWs exhibit two polarization states: plus ($h_+$) and cross ($h_\times$) modes.  These states can serve as a set of bases, a linear combination of which can describe any GW signal; as a parallel with EM radiation, GWs too can be circularly or elliptically polarized \citep[e.g.,][]{shibagaki:2021}.  In principle, arrays of GW detectors at different orientations can be used to detect GW polarization. 
 However, current detector sensitivities have difficulty discerning GW polarization even from sources with stronger amplitudes than CCSNe, such as black hole mergers \citep{gair:2013}. 
 
 One tool that has emerged to predict the polarization from numerical models includes `polograms', which describe the relative strength between the $h_+$ and $h_\times$ modes.  \citet{hayama:2016} connect the circular polarization of GWs with rotation near the center of the supernova.  \citet{hayama:2018} conclude the circular polarization from nonrotating CCSNe can encode information regarding the SASI and PNS g-mode oscillations. \citet{chan:2021} quantify the degree of circular polarization through quantities such as Stoke's parameters, within existing GW detection pipelines, such as Coherent WaveBurst (cWB) \citep{klimenko:2021}.  

Before the advent of modern high performance computing platforms, simplified analytic models were used to estimate GW production from compact objects.  One example is treating rotating compact objects as rotating, triaxial ellipsoids \citep{chin:1965, chau:1967, saenz:1978, zimmerman:1979, galtsov:1984, thorne:1996, pacheco:2010}, with deformations from magnetic fields \citep{bonazzola:1996, konno:2000, palomba:2001} and even neutron star (NS) crust interaction \citep{ushomirsky:2002}.  This simplified description has also been used to describe so called `bar-mode' instabilities proposed to form in CCSN centers with sufficient rotation \citep{new_2000}.  Beyond simple physical expressions, with increased access to modern computing resources, more recent multidimensional simulations have noted the impact of viewing angle on expected GW amplitudes.

In the context of the low T/|W| instability, \citet{ott:2005} note a stronger GW strain along the axis of rotation for short term purely hydrodynamic models without radiation transport.  \citet{scheidegger:2008} and \citet{scheidegger:2010} find similar results for rotating, magnetized models.  Similarly, \citet{kuroda:2014, takiwaki:2018, powell:2020, shibagaki:2021} corroborate these results when comparing polar and equatorial GW strains. 
 \citet{kotake:2011} consider the role of rotation influencing neutrino contributions to GW signals.  While providing valuable first steps towards understanding the directionality of GWs from CCSNe, these works only consider viewing angles along the equator and axis of rotation.  Furthermore, certain observationally-focused works consider different CCSN source orientations \citep{abbott:2020,szczepanczyk:2021,szczepanczyk:2023}.  Even with these considerations, there stands open questions: are there viewing angles---not necessarily aligned with the cardinal axes---along which GW amplitudes are strongest, and, if so, do they evolve in time?  Furthermore, what are the physical processes that dictate these preferred viewing angles?

Due to cheaper computational cost, the use of axisymmetric simulations has the advantage of pushing to later times for a wider variety of initial conditions, but can only compute $h_+$, due to the geometry of the domain.  Axisymmetric models have also been shown to develop large asymmetries along the symmetry axis, leading to GW amplitudes larger by an order of magnitude, compared to 3D works 
 \citep{oconnor:2018b}.  Nevertheless, extensive work has been completed leveraging these 2D works in the context of exploding CCSN scenarios \citep{murphy:2009,yakunin:2010,muller:2013,yakunin:2015,moro:2018}.  
\citet{marek:2009b}, \citet{pan:2018}, and \citet{eggenberger:2021} explore EOS-dependent features of CCSN evolution and the corresponding imprint on the GW signal.  \citet{Pajkos:2021} use multiple components of the GW signal to constrain the progenitor compactness.

 Recently, \citet{mezzacappa:2020} conducted a 3D study outlining the different sources of GW emission from matter motion within a 15 $M_\odot$ star. 
 \citet{raynaud:2022} use GWs to constrain convective motion within magnetized PNSs.    \citet{mezzacappa:2023} conduct a numerical experiment with three 3D models, constraining the physical engine generating GWs at different frequencies.  \citet{afle:2023} use multidimensional models to analyze the detectability of f-mode GW frequencies with third generation GW detectors.  More exotically, \citet{urrutia:2023} explore the impact of viewing angle of GW emission from long GRB jets.

At later times in the supernova evolution, \citet{muller:1997} note the large imprint of neutrino asymmetries on the expected CCSN GW signal; \citet{vartanyan:2020} and \citet{vartanyan:2023} investigate the directional dependence of GW emission generated directly from asymmetric neutrino emission.  \citet{richardson:2022} investigate the directional dependence of GW amplitudes from neutrinos and matter, for a single model.  \citet{mukhopadhyay:2022} investigate the observability of these low frequency signals for the DECi-hertz Interferometer Gravitational Wave Observatory (DECIGO) \citep{kawamura:2011}.  Likewise, work continues to detect CCSN GWs with next generation GW detectors \citep{srivastava:2019}, with some 
 focusing on decihertz frequencies \citep{arcasedda:2020}.  Various works have also investigated the expected contributions to the stochastic GW background from CCSNe \citep{buonanno:2005,finkel:2022}.

The main goal of this work is to characterize the time evolution of the directional dependence of GW emission for both rotating and nonrotating nonmagnetized CCSNe.  To fill the need to understand how the preferred direction of GW observations evolves in time for a variety of initial conditions, we present a new method to understanding \textit{the directionality} of GW emission from CCSNe for arbitrary viewing angles.  We determine an evolving preferred direction of GW emission during CCSNe, for both rotating and nonrotating cases.  We observe possible correlations with the presence of the low T/|W| instability.  We use simplified analytic estimates to show the distribution of GW amplitudes over viewing angle, depending on the supernova phase.  Identifying preferred viewing angles of dominant GW emission would help inform future population studies of GW sources to the GW background.  Likewise, understanding the emergence of a preferred direction of GW emission---if any---during specific phases of the supernova evolution (e.g., bounce, prompt convection, accretion phase, explosion, etc.) would connect CCSN source physics to observability results seen in GW CCSN observation papers \citep[e.g.,][]{szczepanczyk:2023}.  For example, two relevant questions we answer: what are the CCSN physical conditions that leave detectability relatively agnostic to source orientation?  For each phase of CCSNe, what are progenitor characteristics that allow for a preferred direction of GW emission, which may in turn affect detectability?

This paper is organized as follows:  in Section \ref{sec:methods} we review the methods used to set up our simulations and perform our analysis.  In Section \ref{sec:tdwf} we review current techniques used to analyze GW signals.  Section \ref{sec:strainSurface} introduces a new visualization method to investigate the directional dependence of the GW emission at a given instance in time.  Section \ref{sec:toy_models} uses simplified analytic arguments to obtain closed form solutions for the GW direction for different phases of the supernova. 
 Section \ref{sec:scatter3D} visually uses this method to track the time evolution of the directionality of GWs. Section \ref{sec:quantify_scatter3D} quantifies these observations and connects their source physics.  Section \ref{sec:surface_decomp} quantifies the evolution of the directionality when accounting for both polarizations. 
  Section \ref{sec:physicality} discusses the physical implications generating the directional dependence.  Section \ref{sec:observability} discusses future benefits for observability.  Section \ref{sec:open_angle} provides a worked example discussing the influence of directionality on a GW bounce signal detection.  Section \ref{sec:injection} discusses directionality considerations in the context of signal injection into data analysis pipelines.  Finally, in Section \ref{sec:summary}, we summarize. 


\section{Methods}
\label{sec:methods}

\subsection{Numerical Models}

\begin{table*}[t]
\centering
\begin{tabular}{c|c|c|c|c|c|c|c}
 Label & M($M_\odot$)  & $\Omega_0$(rad s$^{-1}$) & \textit{A}($10^3$ km) & EOS & $\nu$ Treatment & $t^\mathrm{pb}_\mathrm{end}$ (s) & Reference \\
\hline

\texttt{m20} & 20 & 0.0 & - & SFHo & M1 & 0.501 & \citet{oconnor:2018b}\\
\texttt{m20LR} & 20 & 0.0 & - & SFHo & M1 & 0.637 & \citet{oconnor:2018b}\\
\texttt{m20p} & 20 & 0.0 & - & SFHo & M1 & 0.528 & \citet{oconnor:2018b}\\
\texttt{m20pLR} & 20 & 0.0 & - & SFHo & M1 & 0.518 & \citet{oconnor:2018b}\\
\texttt{m20vLR} & 20 & 0.0 & - & SFHo & M1 & 0.483 & \citet{oconnor:2018b}\\
\texttt{s40o0} & 40  & 0.0  & -  & LS220 & Ye($\rho$)+IDSA & 0.776 & \citet{pan:2021} \\
\texttt{s40o0.5} & 40  & 0.5  & 1  & LS220 & Ye($\rho$)+IDSA & 0.935 & \citet{pan:2021} \\
\texttt{s40o1} & 40  & 1  & 1  & LS220 & Ye($\rho$)+IDSA & 0.458 & \citet{pan:2021} \\
\texttt{s20o2} & 20  & 2  & 1.034  & LS220 & Ye($\rho$)+IDSA & 0.100 & \citet{hsieh:2023} \\
\texttt{s20o3} & 20  & 3  & 1.034  & LS220 & Ye($\rho$)+IDSA & 0.094 & \citet{hsieh:2023} \\
\texttt{v9BW} & 9  & 0.0  & - & SFHo & M1 & 1.100 & \citet{burrows:2020} \\
\texttt{v11} & 11  & 0.0  & - & SFHo & M1 & 4.500 & \citet{vartanyan:2023}\\
\texttt{v12} & 12  & 0.0  & - & SFHo & M1 & 0.902 & \citet{radice:2019}\\
\texttt{v12.25} & 12.25  & 0.0  & - & SFHo & M1 & 2.000 & \citet{vartanyan:2023} \\
\texttt{v23} & 23  & 0.0  & - & SFHo & M1 & 6.200 & \citet{vartanyan:2023} \\
\end{tabular}
\caption{Setup information for all 15 models in this study.  Column labels represent the following: M--zero age main sequence mass, $\Omega_0$--central rotation rate at collapse, $A$ differential rotation parameter, EOS, neutrino ($\nu$) treatment, and $t^\mathrm{pb}_\mathrm{end}$--simulation end time (post-bounce), and corresponding reference.  }

\label{table:everyone}
\end{table*}

 This work examines 15 3D neutrino radiation hydrodynamic simulations of CCSNe.  The first set of five models are referred to as the \texttt{mesa} set, which are nonrotating $20\, M_\odot$ models used in \citet{oconnor:2018b}.  They make use of the SFHo EOS \citep{hempel:2012,steiner:2013}, M1 neutrino transport \citep{oconnor:2018}, and are evolved for $\sim 0.5$ sec post-bounce (pb).  Five additional simulations, one nonrotating and four rotating, are analyzed from $40\, M_\odot$ models from \citet{pan:2021} and 20 $M_\odot$ models.  These use the LS220 EOS \citep{lattimer:1991}.  To account for neutrino interactions, they use parameterized deleptonization \citep{lieb:2005} through collapse, with the isotropic diffusion source approximation (IDSA) for neutrino transport \citep{liebendorfer:2009} after bounce; we abbreviate this combination as Ye($\rho$)+IDSA.  These 10 models are completed with the FLASH multiphysics code \citep{dubey:2009,fryxell:2010}.  Likewise, they use the Newtonian multipole solver from \citet{couch:2013a}, supplemented by the general relativistic effective potential (GREP) proposed by \citet{marek:2006}.  We also make use of five models from \citet{burrows:2020}, \citet{radice:2019}\footnote{For access to these data see \url{ https://dvartany.github.io/data/}}, and \citet{vartanyan:2023} \footnote{For access to these data see \url{https://vartanyandavid7.wixsite.com/dvyan/data}}.  These simulations make use of the Fornax code \citep{skinner:2019} and have comparable microphysics to FLASH.  For neutrino interactions, Fornax accounts for inelastic scattering off electrons and nucleons.  The neutrino physics used in the version of FLASH from the \texttt{mesa} suite does not account for inelastic processes \citep{oconnor:2018b,oconnor:2018}, which may contribute to why the models from the \texttt{mesa} suite fail to explode, whereas many models from \citet{vartanyan:2023} successfully explode.  The specific initial conditions are outlined in the aforementioned works.  One advantage of using models from FLASH and Fornax is they employ drastically different grids: FLASH uses the Paramesh \citep{macneice:2000} adaptive mesh refinement library on a Cartesian grid, and Fornax uses a static, spherical, dendritic grid.  This diversity of the dataset also allows us to consider multiple zero-age main-sequence masses, rotation rates, EOSs, grid perturbations at collapse, neutrino treatments, and GWs from multiple supernova phases through explosion.  For a concise review of the relevant input parameters, see Table \ref{table:everyone}.  As a point of emphasis, the GW analyses in this work examine GW sources that arise from matter contributions only.  For insightful works regarding GW emission from neutrino asymmetries, see \citet{vartanyan:2020}, \citet{richardson:2022}, and \citet{vartanyan:2023}.

\subsection{Gravitational Wave Analysis}

As the gravitational treatments used in these models do not formally evolve a spacetime metric, the GW generation must be calculated during a post processing step.  We make use of the generic quadrupole formulae presented in \citet{oohara:1997}
\begin{equation}
h_+ = \frac{G}{c^4 D}\Big(\ddot{\mathcal{Q}}_{\theta\theta}-\ddot{\mathcal{Q}}_{\phi\phi}\Big),
\label{eq:h+}
\end{equation}
and
\begin{equation}
h_\times = \frac{2G}{c^4 D}\ddot{\mathcal{Q}}_{\theta\phi},
\label{eq:hx}
\end{equation}
where
$\ddot{\mathcal{Q}}_{ij}$ represents the second time derivative of the reduced mass quadrupole moment in an orthonormal basis, $D$ is the distance to the source, $G$ is Newton's gravitational constant, and $c$ is the speed of light.  For all calculations of $h_+$ and $h_\times$, we assume a fiducial distance of $D = 10$ kpc.  Expanding the angular $\mathcal{Q}$ values in terms of Cartesian components yields
\begin{align}
\mathcal{Q}_{\theta\theta} =& \Big(\mathcal{Q}_{xx}\cos^2\phi + \mathcal{Q}_{yy}\sin^2\phi + \mathcal{Q}_{xy}\sin2\phi\Big)\cos^2\theta\nonumber\\ 
& + \mathcal{Q}_{zz}\sin^2\theta \nonumber \\ &- \Big(\mathcal{Q}_{xz}\cos\phi + \mathcal{Q}_{yz}\sin\phi\Big)\sin2\theta, \label{eq:Qtt}\\
\mathcal{Q}_{\phi\phi} =& \mathcal{Q}_{xx}\sin^2\phi + \mathcal{Q}_{yy}\cos^2\phi - \mathcal{Q}_{xy}\sin2\phi,\label{eq:Qpp}\\
\mathcal{Q}_{\theta\phi} =& \Big(\mathcal{Q}_{yy} - \mathcal{Q}_{xx}\Big)\cos\theta\sin\phi\cos\phi + \mathcal{Q}_{xy}\cos\theta\cos2\phi \nonumber\\
&+ \mathcal{Q}_{xz}\sin\theta\sin\phi - \mathcal{Q}_{yz}\sin\theta\cos\phi.\label{eq:Qtp}
\end{align}
In practice, simulation data tracks each of the $\mathcal{\dot{Q}}_{ij}$ components through numerical integration \citep{finn:1990}, and a finite difference in time is performed to construct $\ddot{\mathcal{Q}}_{ij}$.  These values are then applied to all altitudinal angles of $ 0 < \theta \leq \pi$ and azimuthal angles $0 < \phi \leq 2\pi$ to construct the surface plots outlined in Section \ref{sec:strainSurface}.  As a note, $\theta$ is measured with respect to the positive z axis, and $\phi$ is measured with respect to the positive x axis.  The rotating models in this work align the rotation axis with the positive z axis.

\subsection{Spin Weighted Spherical Harmonic Decomposition}
\label{ssec:spher_harm}

A key part of our analysis lies in quantifying the distribution of GW strain at each point in time.  To perform this decomposition, similar to \citet{gossan:2016} and \citet{ajith:2007}, we calculate coefficients for spin -2 weighted spherical harmonics characterized by coefficients 
\begin{equation}
    _{-2} a_{lm} = D\int_0^{2\pi} d\phi \int_0^\pi d\theta \sin\theta (h_+ - i h_\times) _{-2}Y^*_{lm}(\theta, \phi),
\end{equation}
where $_{-2}Y^*_{lm}(\theta, \phi)$ is the complex conjugate of the spin -2 weighted spherical harmonics, $D$ is the source distance (to cancel out the factor of $1/D$ in Equation (\ref{eq:h+}) and Equation (\ref{eq:hx})), and $i$ is the imaginary number.  In the plots below, we select the real component of $_{-2} a_{lm}$.  The use of spin weighted spherical harmonics is advantageous because it provides an orthogonal set of basis functions to describe multidimensional surfaces and has been used to great success describing GW distributions in the numerical relativity community \citep{boyle:2014}.  Likewise, for varying values of $m$, they conveniently describe differing altitudinal angles; for example, $m = 0$ peaks at $\theta = \pi / 2$ (the equator of a sphere), whereas $m = \pm 2 $ peaks at $\theta = 0, \pi$ (the pole of the sphere), a characteristic that will prove useful for this work. The expressions for $_{-2}Y_{lm}(\theta, \phi)$ can be found in Table \ref{table:spin_spher_harms} \citep{ajith:2007}.

\section{Visualizing Gravitational Wave Emission}

In this section, we begin with existing methods quantifying GW data, and sequentially generalize these techniques to eventually consider GW directionality.  All visualizations in this section make use of models \texttt{s40o[0, 0.5, 1]}.

\subsection{Time Domain Waveforms}
\label{sec:tdwf}

\begin{figure*}
    \centering
    \includegraphics[]{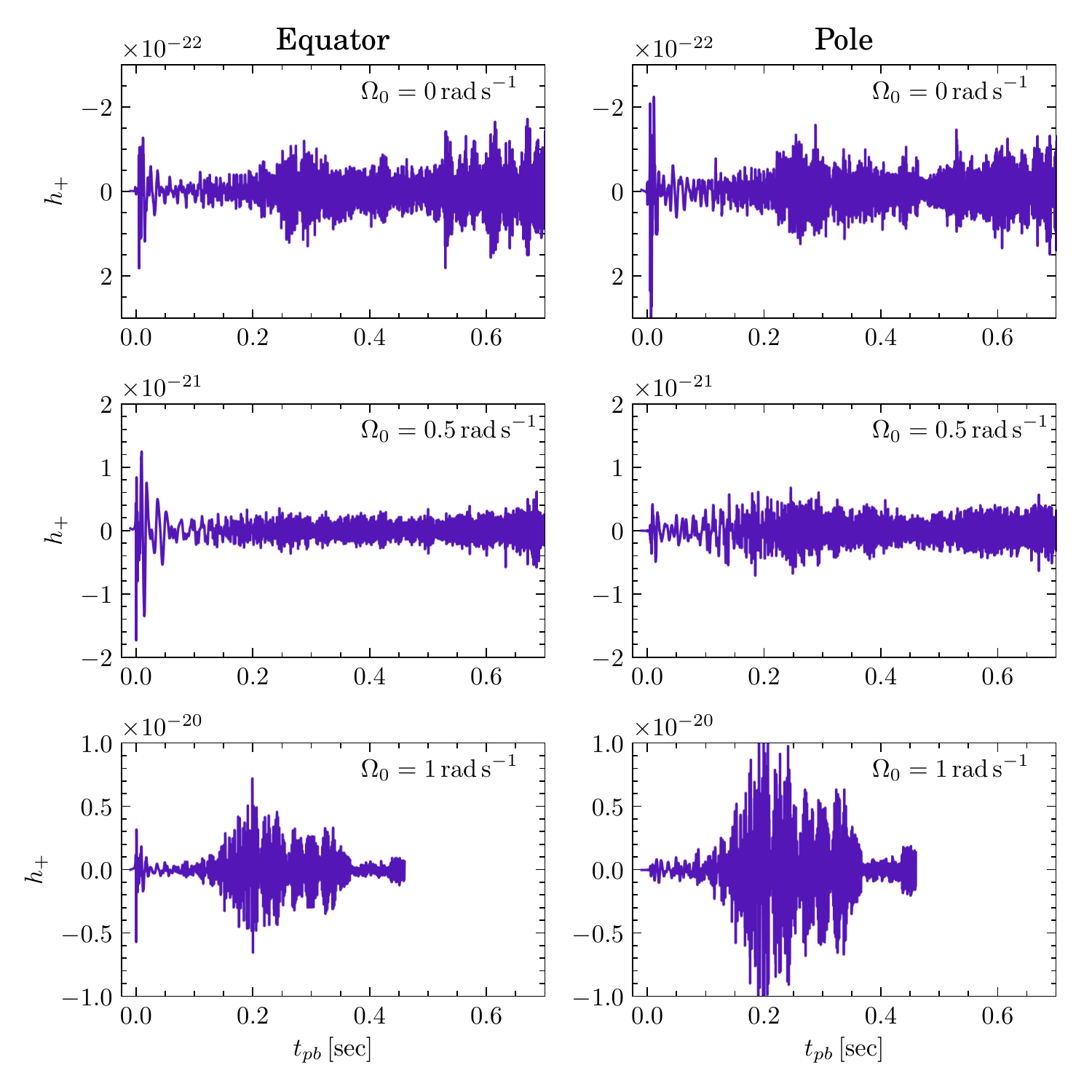}
    \caption{Time domain waveforms (plus polarization) for \texttt{s40o0} (top), \texttt{s40o0.5} (middle), and \texttt{s40o1} (bottom) \citep{pan:2021}.  The left (right) column corresponds to an observed $h_+$ when viewed along the equator (pole).  While TDWFs are useful for analyzing GW emission throughout the supernova simulation, they are inherently limited to GW signals at a single, fixed viewing angle.  This motivates the need for additional analysis methods that can identify a dominant viewing angle beyond the equator or pole and determine if this direction evolves in time.  Note the different scales between each y axis.  Figure from \citet{pajkos:2022}.}
    \label{fig:TDWF}
\end{figure*}

We begin with the TDWF from \texttt{s40o0}, \texttt{s40o0.5}, and \texttt{s40o1}.  Figure \ref{fig:TDWF} displays $h_+$ throughout the simulation duration---note the different y scales between each case.   The left column corresponds to the expected GW signal for an observer along the line of site of the CCSN equator ($\theta = \pi /2$) and along the positive x axis ($\phi = 0$).  The right column corresponds to an observer along the supernova north pole ($\theta = 0$), which represents the axis of rotation for rotating models.  The top row represents the nonrotating model $\texttt{s40o0}$, the central row represents the slow rotating model $\texttt{s40o0.5}$, and the bottom row represents the fast rotating model $\texttt{s40o1}$.  As rotation increases, coherent motion of the fluid generates stronger oscillations of the PNS.  Likewise, in the fast rotating model, beginning $\sim 150$ ms pb, the presence of the low T/|W| instability excites the PNS to generate GWs with larger amplitudes as well \citep{pan:2018}.  

When comparing signals of a given rotation rate, but differing viewing angle, there are only moderate differences for $\texttt{s40o0}$.  This result is expected because there is no centrifugal support to deform the PNS along a particular plane (e.g., a more oblate PNS in the xy plane for an axis of rotation in the z direction).  However, as rotation increases, the GW signal detected by observers at different viewing angles becomes more pronounced.  For the rotating cases, the GW signal just after bounce, $t_{pb} = 0$ sec, commonly called the bounce signal, has a large GW amplitude when viewed along the equator, but has an amplitude of 0 when viewed along the axis of rotation.  This effect is explained through Equation (\ref{eq:h+}) and Equation (\ref{eq:hx}).  When viewed along the equator, the oblate PNS will have a cross section similar to an ellipse.  As the infalling material deforms the PNS at the time of bounce, the cross section changes to a more spherical shape, and $\ddot{\mathcal{Q}}_{xz}$, $\ddot{\mathcal{Q}}_{yz}$, and $\ddot{\mathcal{Q}}_{zz}$ will take on correspondingly larger values---a higher amplitude bounce signal is generated.  By contrast, when observed along the axis of rotation, the PNS cross section is nearly circular.  As the infalling material does not have an azimuthal dependence just after bounce, the cross section of the PNS retains its circular shape.  The dominant $\ddot{\mathcal{Q}}_{xx}$, $\ddot{\mathcal{Q}}_{xy}$, and $\ddot{\mathcal{Q}}_{yy}$ retain smaller values, thereby preventing an observed GW signal.  Hundreds of milliseconds after bounce, the GW amplitude related to the PNS oscillations also displays variations in amplitude, depending on viewing angle.

Examining TDWFs can be valuable in investigations such as this one.  However, they are inherently limited to observing one viewing angle at a time.  In general, there is no guarantee that the maximum GW signal must be along the equator or pole.  Furthermore, as the mechanism of GW generation can change throughout the supernova evolution, in principle, a viewing angle of preferred GW emission may evolve through time.

\subsection{Visualizing GWs in Multiple Dimensions}
\label{sec:strainSurface}

\begin{figure*}
\centering
    \begin{subfigure}[t]{0.45\linewidth}
\includegraphics[width=\linewidth]{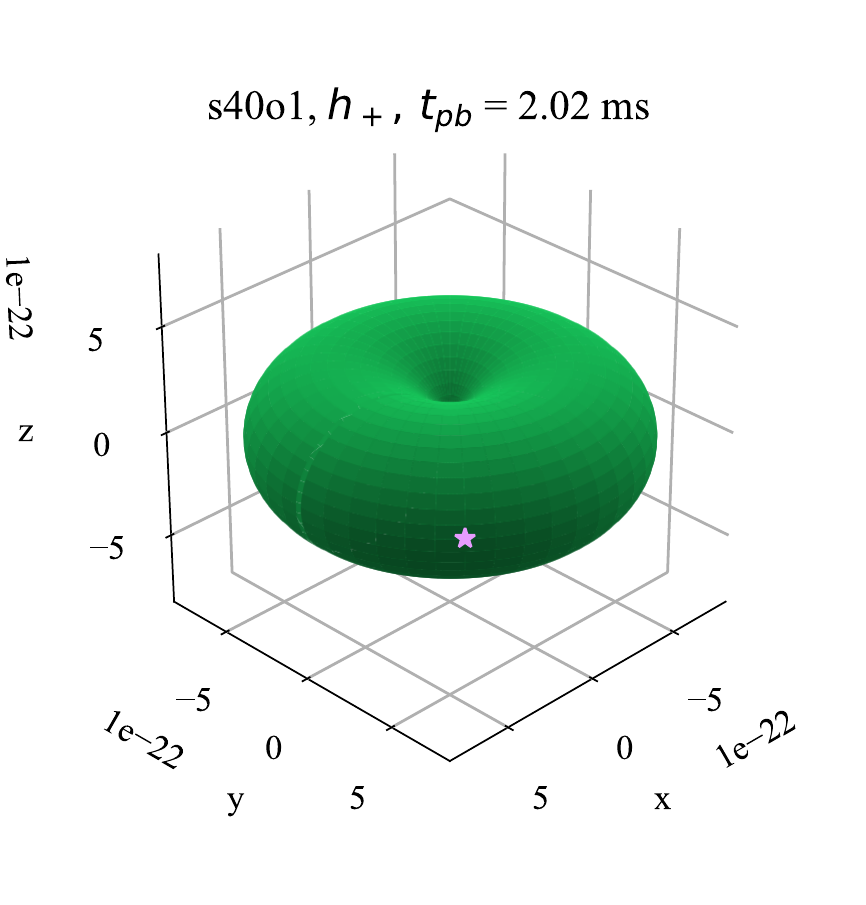}
    \caption{Strain surface plot for $h_+$ just after bounce.  The star denotes the direction along which the largest $h_+$ would be received, what we denote as the `preferred' viewing angle.}
\label{fig:surface_sim_bounce}
    \end{subfigure}\hfill
    \begin{subfigure}[t]{0.45\linewidth}
\includegraphics[width=\linewidth]{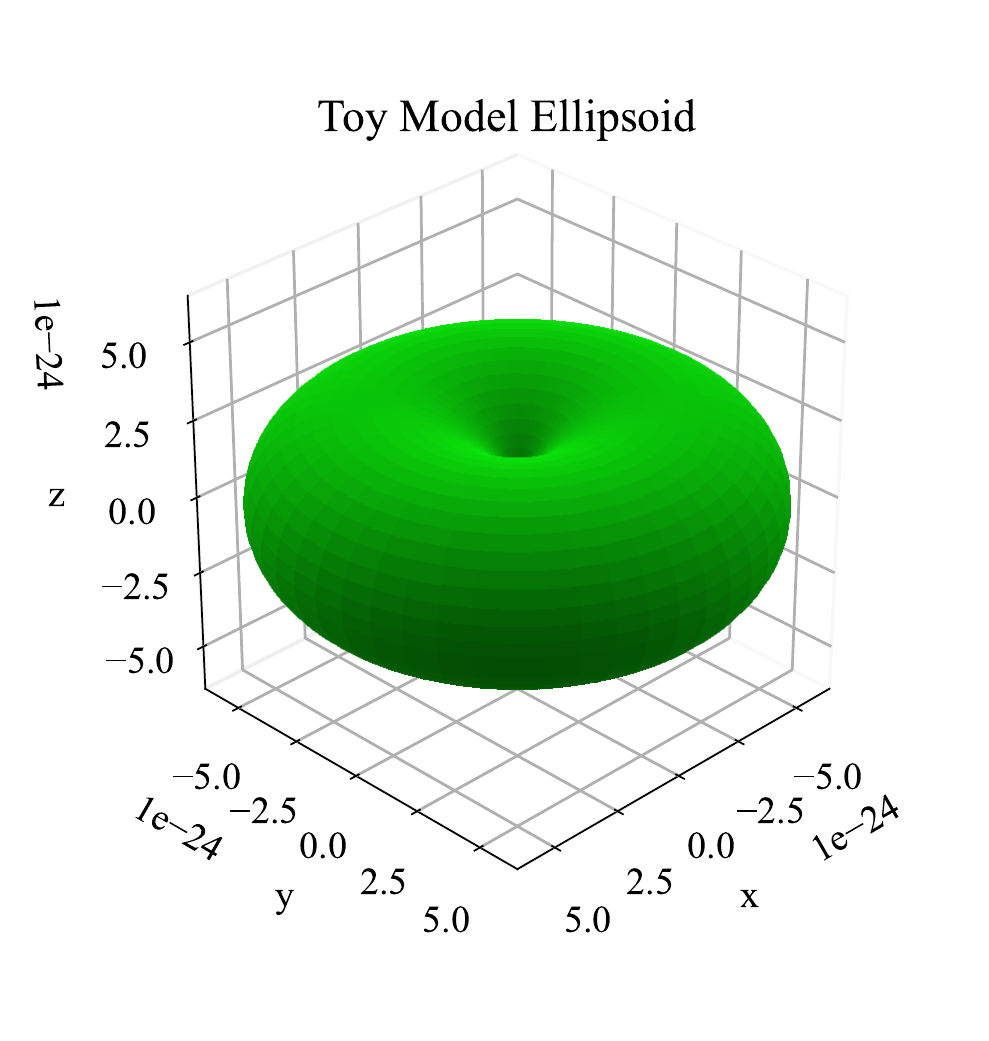}
    \caption{Strain surface plot for a toy model of an axisymmetric prolate ellipsoid of mass 1 M$_\odot$ with polar radius $c = 62$ km and equatorial radii $a = b = 60$ km, with velocity at the pole $\dot{c} = 600$ km s$^{-1}$, $\dot{a} = \dot{b} = 0$ km s$^{-1}$, replicating the strain surface distribution for bounce seen in Figure \ref{fig:surface_sim_bounce}.  We note the fluid-like nature of the PNS (deviating from a solid-body ellipsoid) makes assigning a single polar velocity difficult---thus the discrepancies in the magnitude of the estimate.  Nevertheless, our goal is to describe the geometry of the GW distribution, which agrees quite well.}
\label{fig:surface_toy_bounce}
    \end{subfigure}

    \begin{subfigure}[t]{0.49\linewidth}
\includegraphics[width=\linewidth]{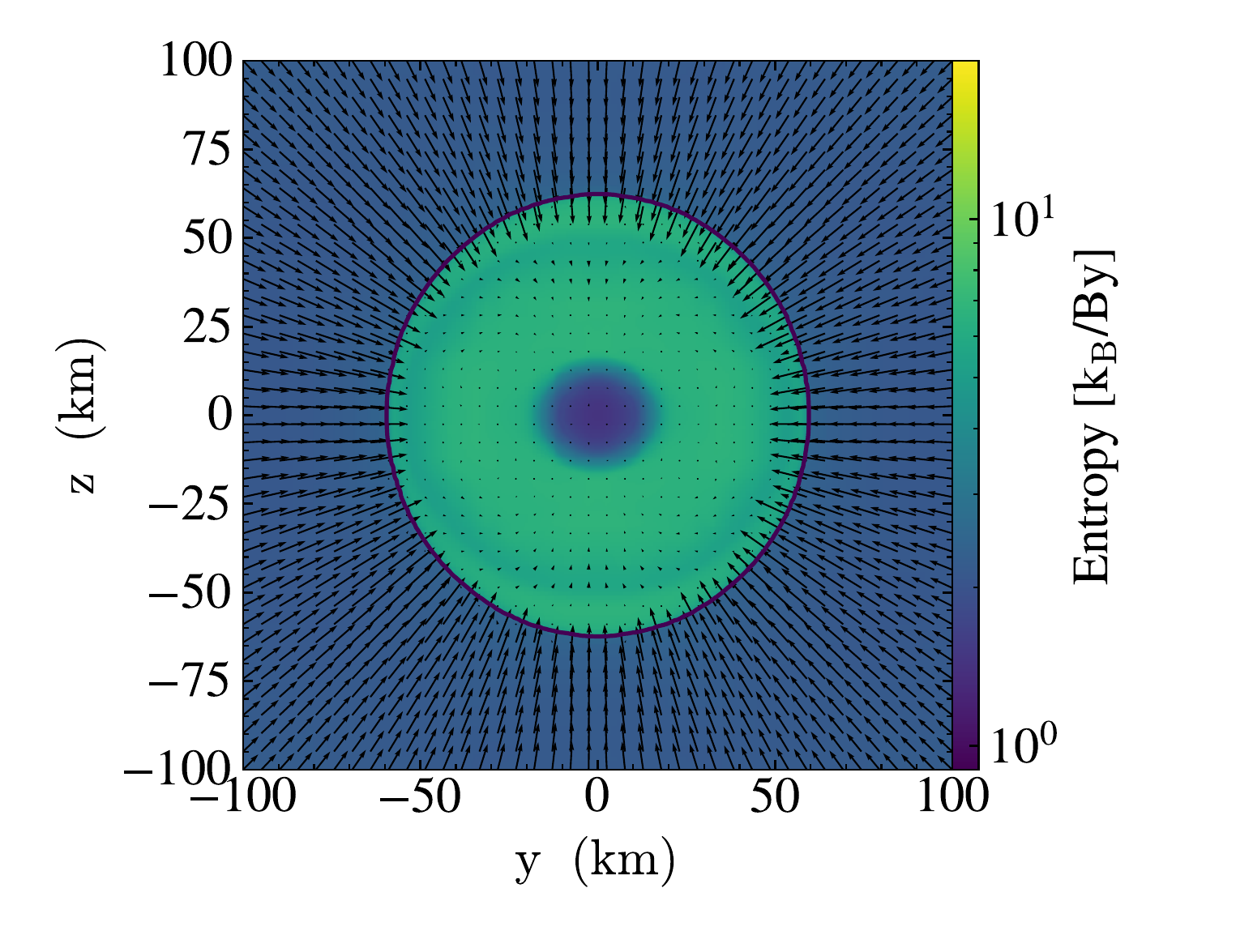}
    \caption{Entropy slice through the pole (z axis) from model \texttt{s40o1} 2 ms after bounce. Notice the PNS surface (solid black line: $10^{11}$ g cm$^{-3}$) slightly prolate, after being deformed from the infalling matter.  The PNS will ringdown, finally achieving an equilibrium as an oblate spheroid, due to centrifugal effects.  For scale, every cm on the page corresponds to $5.6\times10^{9}$ cm s$^{-1}$.}
\label{fig:entr_po_slice_bounce}
    \end{subfigure}\hfill
    \begin{subfigure}[t]{0.49\linewidth}
\includegraphics[width=\linewidth]{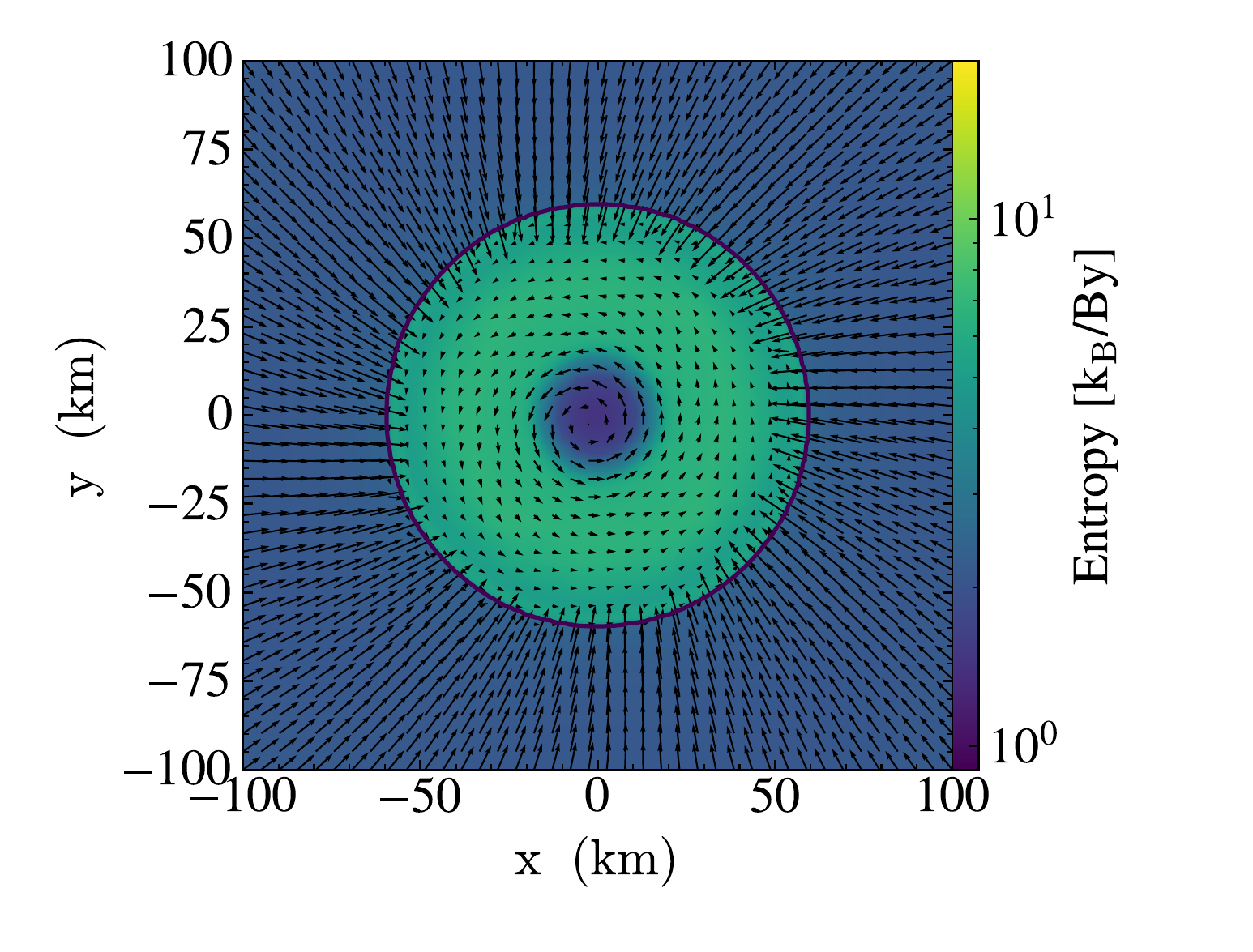}
    \caption{Entropy slice through the equator from model \texttt{s40o1} 2 ms after bounce display the axisymmetry of the PNS before convective instabilities create asymmetries.}
\label{fig:entr_eq_slice_bounce}
    \end{subfigure}
\caption{Strain surface plots and entropy slices for model \texttt{s40o1} during the bounce phase.  The x and y axes form the equator of the supernova; the z axis indicates the axis of rotation.  For the top two panels, the distance from the origin to a point on the green surface represents the detected $h_+$ along that direction.  The bottom two panels are entropy slices, with overlaid velocity profiles (arrows).}
    \label{fig:GWsurfaces_bounce}
    \end{figure*}

To address the need to understand GW emission along every viewing angle, we introduce the \textit{strain surface plot}.  This figure is inspired from those in \citet{vartanyan:2020} that display the GW emission by color and concavity on a 3D surface.  It can be thought of as an extension to 3D of Figure 3.7 of \citet{maggiore:2007} and applied to strain.  In practice, it can be difficult to compare color or concavity by eye, so we offer another form of visualization, with examples provided in Figure \ref{fig:surface_sim_bounce}, Figure \ref{fig:surface_sim_ringdown}, Figure \ref{fig:surface_sim_accretion}, and Figure \ref{fig:surface_sim_tilt}.  

These surfaces are taken from different points in time for \texttt{s40o1}.  To interpret these plots, consider Figure \ref{fig:surface_sim_bounce} at $t_{pb} \sim 2$ ms.  The distance from the origin to any point on the green surface corresponds to $h_+$ along that viewing angle.  The purple star indicates the direction along which there is a maximum $h_+$.  The x and y axes form the plane of the supernova equator, and the z axis is the axis of rotation.  Figure \ref{fig:surface_sim_bounce} displays similar $h_+$ emission when viewed at any angle in line with the equator.  By contrast, when viewed along the axis of rotation, the surface does not extend far from the origin, indicating a weak GW amplitude.  Physically, this behavior is justified.  As explained in Section \ref{sec:tdwf} and shown in the top row of TDWFs in Figure \ref{fig:TDWF}, the bounce signal is detected when observed along the equator due to the geometry of the deformed PNS; a more detailed explanation of the PNS dynamics is reserved for Section \ref{sec:toy_models}.  Figure \ref{fig:surface_sim_bounce} offers a compact way to conclude this directional dependence at a given point in time without relying on multiple TDWFs.

In Figure \ref{fig:surface_sim_ringdown}, we notice a deviation from azimuthal symmetry, around 13 ms post-bounce.  At this point, a combination of the ringdown from the bounce and emergence of prompt convection begins to skew the symmetry of the emission.  The direction of maximum $h_+$ amplitude is roughly aligned with the x axis at this time.  Furthermore, a moderate GW amplitude can be observed along the z axis, in contrast to the bounce phase.  

Figure \ref{fig:surface_sim_accretion} displays GW amplitudes nearly 270 ms later.  The configuration of the $h_+$ surface preferentially lies along the axis of rotation, with a distinct four lobed structure.  There are also non-negligible deformations of the PNS along all three spatial degrees of freedom, likely due to the turbulent nature of the accreting matter.  During the accretion phase, for models \texttt{s40o.5} and \texttt{s40o1}, this four-lobed structure varies in amplitude, but retains similar morphology.  
Figure \ref{fig:surface_sim_tilt} displays the strain surface plot for $\texttt{s40o0}$ roughly 230 ms pb.  We see the emergence of nonaxisymmetric behavior, paired with GW maxima misaligned from all coordinate axes.  This is due to accretion downflows striking the PNS along a direction misaligned from the principal axes.

For convenience, we offer methodology to translate from strain surface plots to TDWFs:

\begin{enumerate}
    \item Construct a strain surface plot at a time of interest $t$,
    \item identify a desired line of site $(\theta,\phi)$. In this example we pick a viewing angle in the equator: the x axis  $(\pi/2, 0)$,
    \item the distance from the origin to the green surface along this line of site represents the injected GW strain $h(\theta,\phi,t)$ for a detector,
    \item the values $(t,h)$ correspond to the ordered pair on the TDWF.
\end{enumerate}

As a point of emphasis, these surface plots represent GW strain detected by an \textit{observer far from the GW source}.  Equation (\ref{eq:h+}) and Equation (\ref{eq:hx}) are applicable only for distant observers from slow moving ($v \ll c$) sources.  These do not represent the GW amplitudes generated just inside the supernova.  To calculate these quantities, a more robust treatment of relativity is needed, with numerical models that track quantities of the spacetime metric.

\section{Analytic Estimates with Toy Models}
\label{sec:toy_models}

Armed with established GW strain surfaces from Section \ref{sec:strainSurface}, we now move to describing their origin by means of a simplified example.  In this section, we step through the different phases of GWs from CCSNe: bounce, ringdown, and accretion phase. We explain the directionality of GWs through the use of an intuitive example of a wobbling, oscillating ellipsoid toy model to reproduce $h_+$ along a given direction.  Considering both polarizations will be addressed in Section \ref{sec:surface_decomp}.

\subsection{Introducing the Wobbling, Oscillating Ellipsoid}

To quantify GW amplitudes with closed form solutions, we approximate the PNS as a uniform density, solid-body ellipsoid with three principal axes $a$, $b$, and $c$ for the $x$, $y$, and $z$ axes, respectively.  We acknowledge that PNSs retain complex convective structures and undergo differential rotation that can cause them to deviate from solid-body rotation; for more detail, see Appendix \ref{app:rotational_velocity}.  To more realistically account for the complex dynamics---while retaining closed form solutions---we allow the ellipsoid to evolve in three ways: rotation, tilt, and size changes.  To describe how each physical component of the ellipsoid contributes to GW generation, consider a simplified mathematical example.  As $h \propto \ddot{\mathcal{Q}}$, a simplified expression of the reduced mass quadrupole moment of an ellipsoid is $\mathcal{Q}\sim MR^2$, for a given mass $M$, and radius $R$.  Applying two time derivatives yields a useful reference estimate
\begin{equation}    
\Ddot{\mathcal{Q}}\sim \Ddot{M}R^2 + 4\dot{M}\dot{R}R + 2M\dot{R}^2 + 2M\Ddot{R}R.
\label{eq:h_estimate}
\end{equation}
Here the time derivatives represent different physical mechanisms that contribute to the GW generation: $\dot{M}$ the mass accretion rate, $\ddot{M}$ the rate of change of the mass accretion rate, $\dot{R}$ the velocity of the PNS surface, and $\Ddot{R}$ the acceleration of the PNS surface.  When considering these order of magnitude estimates, it is important to remember contributions of these terms---for example, velocities $\dot{R}$---must contribute to nonspherical mass motions. 
 As this work is primarily concerned with the dynamics of the PNS, we pay special attention to the fixed mass case, when $\ddot{M} = \dot{M} = 0$, and constant speed case $\ddot{R} = 0$.  The remaining quantities---$M$, $R$, and $\dot{R}$---are free parameters.  In the upcoming sections, we will assign values to these parameters to mimic conditions in our numerical simulations.  The values of $M$ are taken from our simulation outputs of the PNS baryonic mass at a given time.  The values of $R$ along each direction correspond to $a$, $b$, and $c$ values; they correspond to the simulation values of the PNS radius along the x, y, and z axes, respectively.  Values of velocities near the PNS surface, $\dot{R}$, along each direction correspond to $\dot{a}$, $\dot{b}$, and $\dot{c}$. For intuitive explanations, we also refer to a mass quadrupole moment matrix of the form
\begin{equation}
    I_\mathrm{ellipsoid} = \begin{pmatrix}
I_{xx} & 0 & 0\\
0 & I_{yy} & 0\\
0 & 0 & I_{zz}
\end{pmatrix}.
\end{equation}
In Equation \ref{eqn:ellipsoid_quad_mmnt} we explicitly define this quadrupole moment matrix for our toy model case, which will pick up off-diagonal components when considering a tilted, rotating ellipsoid. To connect the reduced mass quadrupole moment $\mathcal{Q}$ used in Equation (\ref{eq:h+}) and Equation (\ref{eq:hx}), with the mass quadrupole moment $I$ used in Appendix \ref{appendix:rotation_matrices}, one can perform 
\begin{equation}
    \mathcal{Q}_{ij} = I_{ij} - \frac{1}{3} \mathrm{Trace}(I_{ij}).
\end{equation}
This toy model will be used to draw connections between PNS dynamics during each phase of the supernova and the associated distribution of GWs.  Furthermore, it can be used to build an intuition to identify under which conditions different dynamics may dominate the GW signal, for example, surface oscillations versus nonaxisymmetric rotation.  For future astrophysics studies that explore GW sources approximated as rotating, dynamic, tilted, ellipsoids, this model can be used to estimate the directionality of GW emission, beyond simple first order estimates that rely on tilt angles $\theta_\mathrm{tilt} \ll 1$.

\subsection{The Bounce Phase}
\label{ssec:toy_bounce}
  Occurring at bounce, before significant hydrodynamic instabilities manifest, the centrifugally deformed PNS is nearly axisymmetric.  With a nearly spherical infall of matter, the ram pressure imparted on the PNS deforms it, making it less oblate---more spherical.  As the PNS rings down, the PNS will oscillate between a prolate and oblate configuration over the timescale of milliseconds, losing energy by exerting pressure on the overlying accreting fluid (or \textit{PdV} work), and eventually settle in an oblate configuration, depending on its degree of rotation.  When viewed at a viewing angle perpendicular to the axis of rotation (along the equatorial plane), the PNS cross section appears as an ellipse with varying principal axes.  At this viewing angle, the mass quadrupole moment is accelerating.  When viewed perpendicular to the equator (along the axis of rotation), the PNS cross section resembles a circle with varying radius.  However, the ratio between the principal axes remains nearly constant; no GWs are generated along the axis of rotation.  Although the PNS remains nearly axisymmetric during this bounce and subsequent ringdown, \textit{the dynamics} of the ringdown are what generate the GW burst mostly along the equator---the $\dot{R}$ and $\ddot{R}$ terms in Equation (\ref{eq:h_estimate}).  Hence, a clearer picture emerges: with axisymmetric bounce dynamics, a resulting axisymmetric GW distribution is generated for this set of coordinate axes \citep{kotake:2017}.  For clarity, refer to Figure \ref{fig:entr_po_slice_bounce}.  This image is a slice along the axis of rotation of model \texttt{s40o1} 2 ms after bounce.  Color corresponds to entropy, and arrows denote the velocity of the fluid.  The solid black line is the PNS surface chosen where the density $\rho = 10^{11}$ g cm$^{-3}$.  This visual shows a prolate spheroid with equatorial radius $\sim 60$ km and polar radius $\sim 62$ km.  Figure \ref{fig:entr_eq_slice_bounce} is taken at the same time but shows a slice through the equator, displaying the axisymmetry of the matter profile.  It is this axisymmetric matter configuration, with evolving principal axes, that motivates our toy ellipsoid model of $M_\mathrm{ellipsoid} = 1 M_\odot$, with polar radius $c = 62$ km and equatorial radii $a = b = 60$ km.  We select a velocity at the poles of $\dot{c} = 600$ km s$^{-1}$, taken from the simulation output.  Figure \ref{fig:surface_toy_bounce} displays the GW output with these toy parameters.  The correspondence between the distribution of GW generation agrees quite well with Figure \ref{fig:surface_sim_bounce}, displaying an axisymmetric GW bounce signal, with maximum amplitude along the equator and no GW amplitude along the pole.  Referring to Equation (\ref{eq:h_estimate}) it is the $2M\dot{R}^2$ term generating this GW distribution.  However, we note that magnitude of $h_+$ differs by nearly 2 orders of magnitude between the simulation data and toy model.  While the $2M\ddot{R}R$ term may partially contribute and additional model tuning could achieve closer numbers, our goal is not to exactly reproduce GW amplitudes with these toy models.  Our goal of this section is to give instructive examples connecting the \textit{directions} of the GW strain to the internal matter dynamics.

\subsection{The Ringdown Phase}
\label{ssec:toy_ringdown}

\begin{figure*}
\centering
    \begin{subfigure}[t]{0.45\linewidth}
\includegraphics[width=\linewidth]{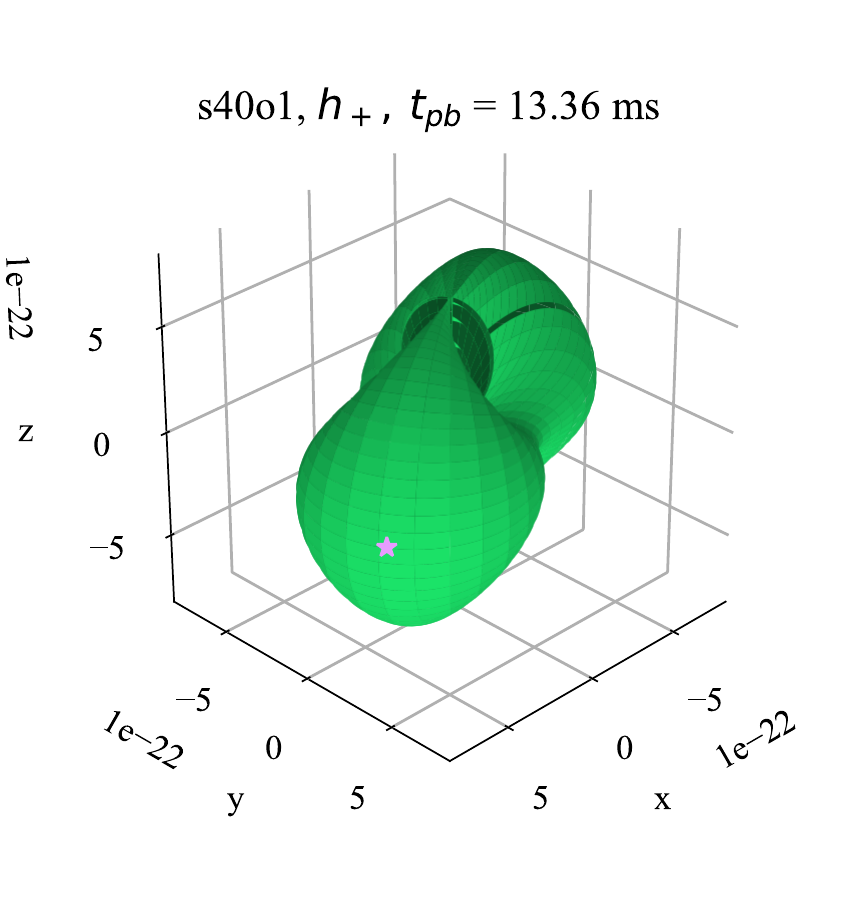}%
    \caption{Strain surface plot for $h_+$ during ringdown.  This rotating structure, which will modulate in scale for tens of ms during ringdown, will give the largest GW amplitudes along the equator of the supernova.}
\label{fig:surface_sim_ringdown}
    \end{subfigure}\hfill
    \begin{subfigure}[t]{0.45\linewidth}
\includegraphics[width=\linewidth]{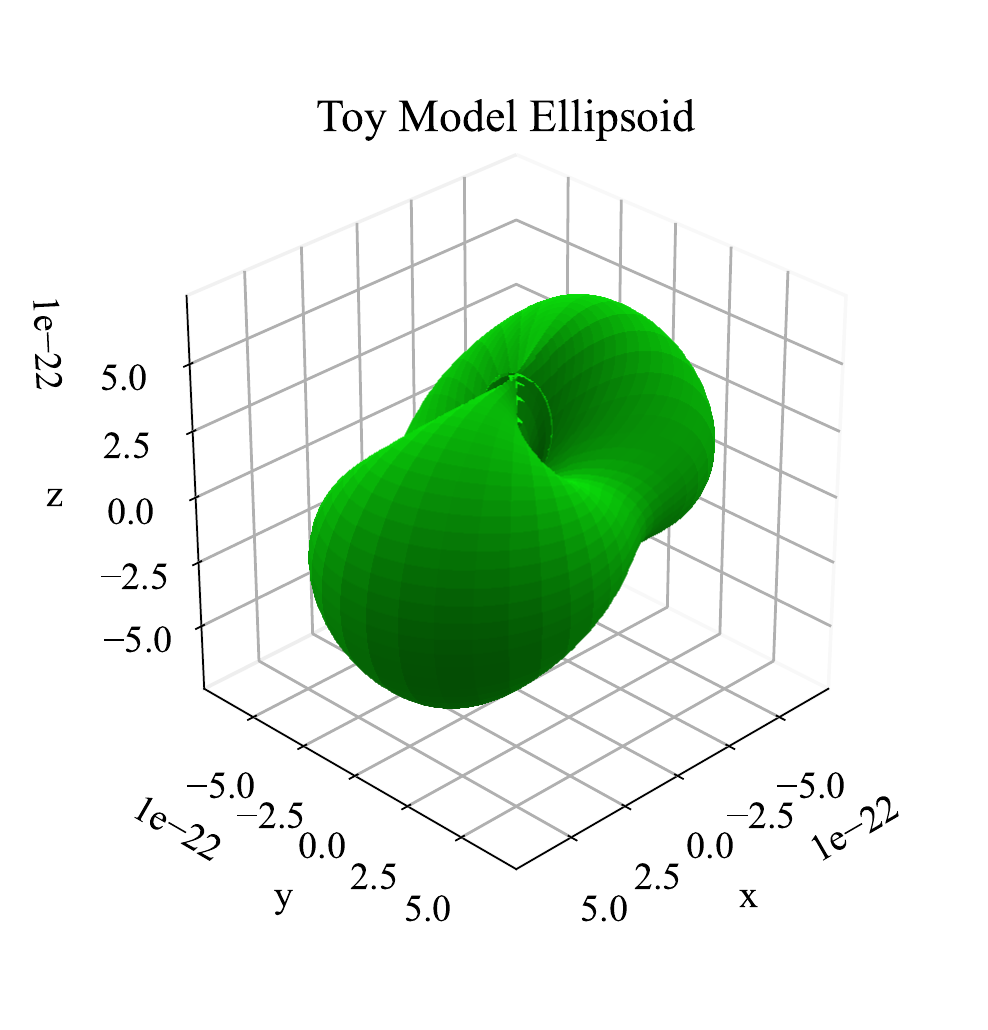}
    \caption{Strain surface plot for a toy model of an `elastic ellipsoid'.  Numbers taken from model \texttt{s40o1} inform toy ellipsoid parameters of mass 1.3 M$_\odot$, with principal axes $c = 77$ km, $a = 83$ km, $b = 81$, rotating with $\Omega = 100$ rad s$^{-1}$, $\dot{a} = \dot{b} = 5000$ km s$^{-1}$, $\dot{c} = 0$ km s$^{-1}$, which replicates the strain surface for rotating ringdown seen in Figure \ref{fig:surface_sim_ringdown}.  In particular, the rotating nonaxisymmetry (i.e., $a\neq b$) of the ellipsoid contributes to the `four lobed' structure.  The changing principal axes ($\dot{a} \neq 0$, $\dot{b} \neq 0$) contribute to the protruding GW surface along the equator.}
\label{fig:surface_toy_ringdown}
    \end{subfigure}

    \begin{subfigure}[t]{0.49\linewidth}
\includegraphics[width=\linewidth]{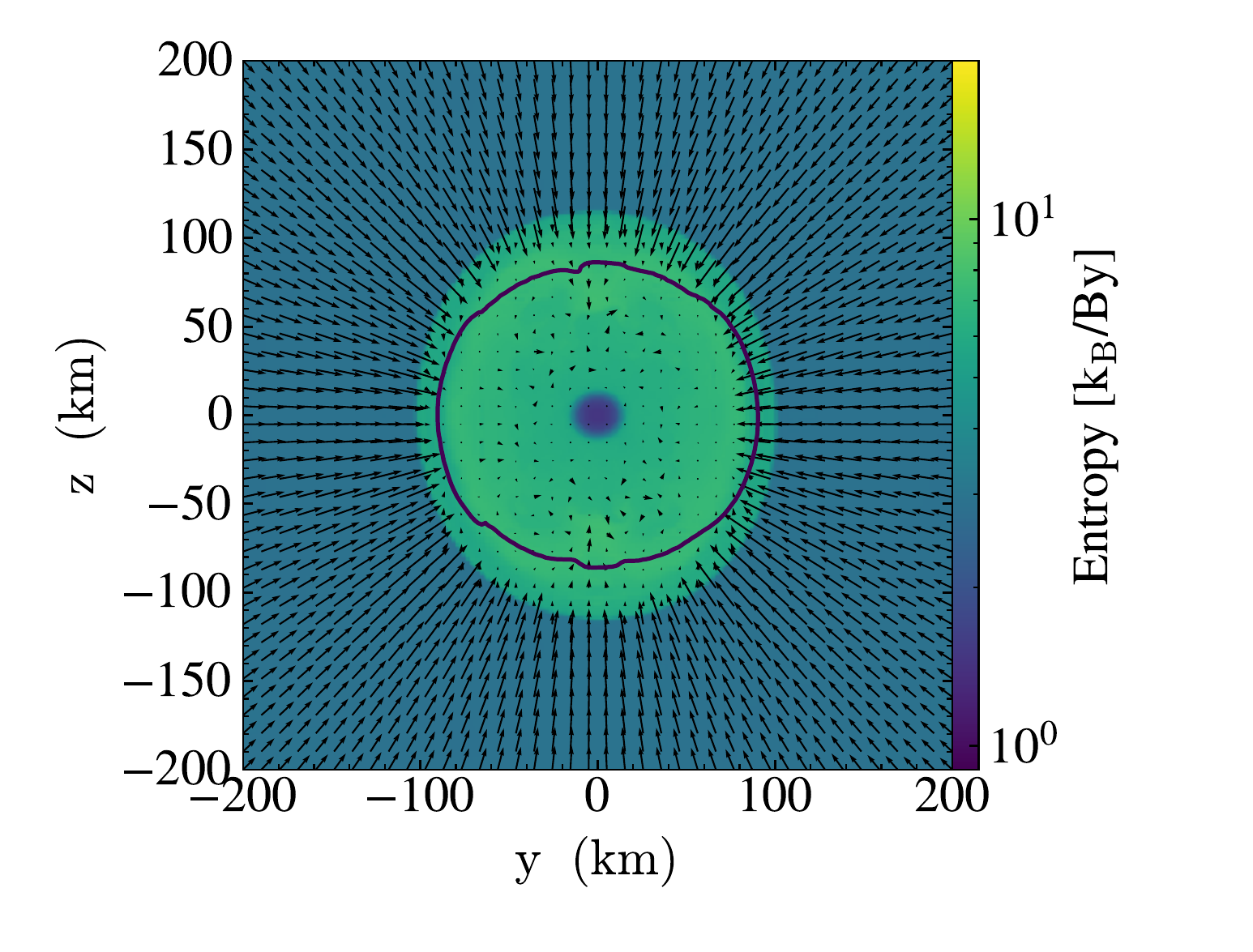}
    \caption{Entropy slice through the pole from model \texttt{s40o1} $\sim 13$ ms after bounce shows a slightly oblate PNS surface.  For scale, every cm on the page corresponds to $5.6\times10^{9}$ cm s$^{-1}$.}
\label{fig:entr_po_slice_ringdown}
    \end{subfigure}\hfill
    \begin{subfigure}[t]{0.49\linewidth}
\includegraphics[width=\linewidth]{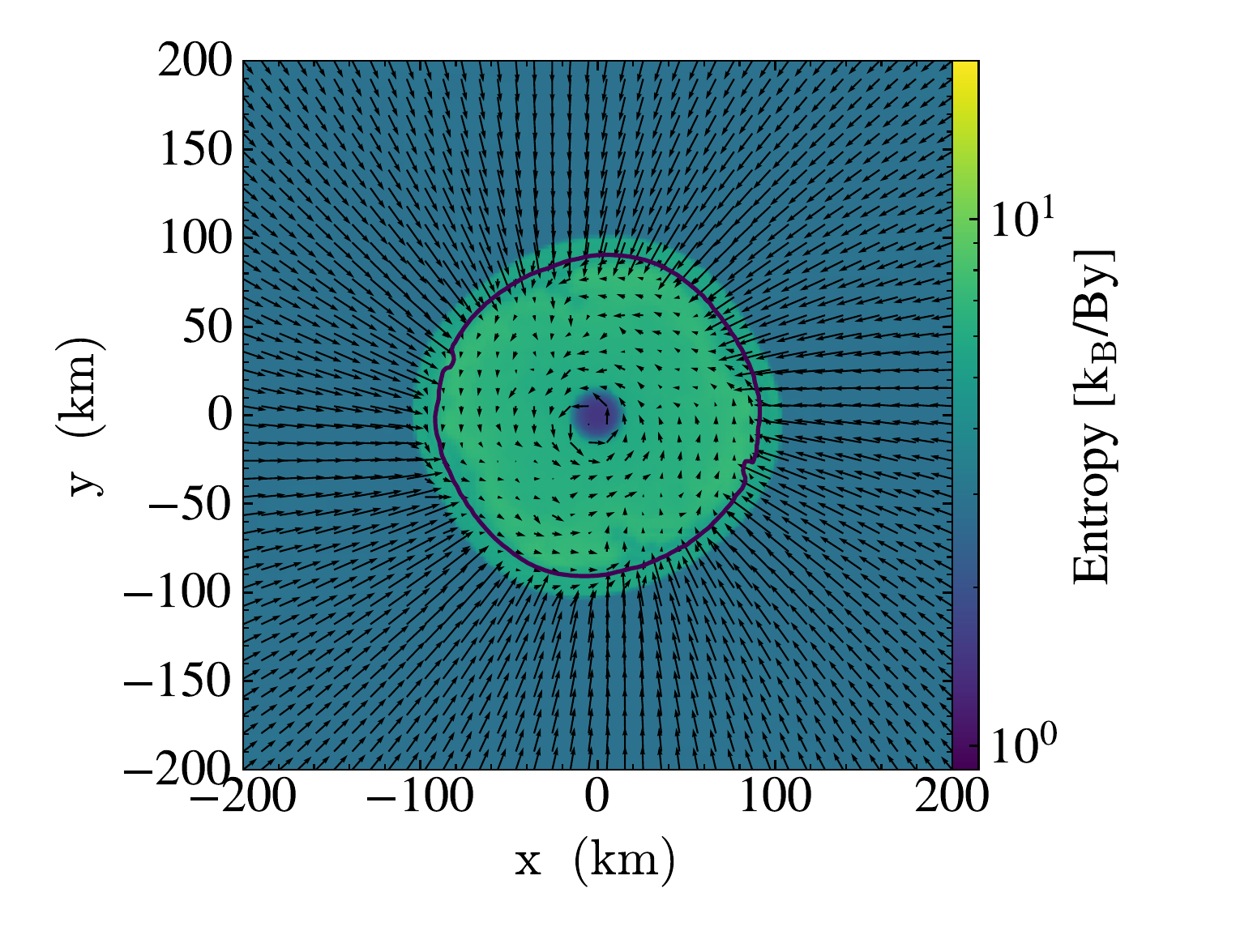}
    \caption{Entropy slice through the equator from model \texttt{s40o1} $\sim 13$ ms after bounce display the PNS surface deviating from axisymmetry (a circle in this cross section) as convective instabilities within the PNS create asymmetries.}
\label{fig:entr_eq_slice_ringdown}
    \end{subfigure}
\caption{Strain surface plots and entropy slices for model \texttt{s40o1} during the ringdown phase.  The x and y axes form the equator of the supernova; the z axis indicates the axis of rotation.  For the top two panels, the distance from the origin to a point on the green surface represents the detected $h_+$ along that direction.  The bottom two panels are entropy slices, with overlaid velocity profiles (arrows).}
    \label{fig:GWsurfaces_ringdown}
    \end{figure*}

We define the ringdown as the tens of ms following bounce, in which the rotating PNS finds a new equilibrium after being perturbed from redirecting initially infalling material to an outgoing shock.  For tens of ms pb, as the PNS achieves a new equilibrium, nonaxisymmetric structure may arise as the deleptonization due to the neutrino burst leaves a negative lepton gradient.  In the case when rotation is not dominant enough to stabilize the PNS system, according to the Solberg-Hoiland stability criterion \citep{endal:1978}, prompt convection can occur \citep{mazurek:1982}.
 Once again, we refer to simulation outputs from \texttt{s40o1} to understand the physical picture.  Figure \ref{fig:entr_po_slice_ringdown} displays a nearly circular cross section along the axis of rotation.  Compared to the bounce phase, the PNS is transitioning towards a more stable configuration that is less prolate compared to Figure \ref{fig:entr_po_slice_bounce}, as expected.  Small perturbations due to prompt convection allow for slight perturbations in the y-z plane as well.  In Figure \ref{fig:entr_eq_slice_ringdown}, the nonaxisymmetry becomes more clear.  In particular, prompt convection creates 4 maxima, causing deviations from the circular surface during the bounce phase.  While not a true ellipsoid, the degree of nonaxisymmetry still motivates our numerical choices for our toy model with $M = 1.3 M_\odot$, with equatorial radii $a = 83$ km and $b = 81$ km.  The polar radius is $c = 77$ km. By asserting the majority of the motion will occur along the equator, we choose $\dot{a} = \dot{b} = 5000$ km s$^{-1}$ and $\dot{c} = 0$. For the angular velocity, we observe a value of $\Omega = 100$ rad s$^{-1}$ from the simulation output. As a caveat, we acknowledge 5000 km s$^{-1}$ is a large value for an equatorial velocity tens of ms after bounce.  In our simulation output we only note typical values $\sim 100$ km s$^{-1}$.  However, because this toy model is solid-body, the $\ddot{Q}$ term will be more strongly influenced by the rotational motion about its axis of rotation, compared to a less solid-body, hydrodynamically evolving PNS.  Hence, a higher $\dot{a}$ and $\dot{b}$ are required to more strongly contribute to the $\dot{R}$ component of $\ddot{Q}\sim 2M\dot{R}^2$ of the toy model.  The exact angular velocity profile during the ringdown phase is shown in Appendix \ref{app:rotational_velocity}.

Figure \ref{fig:surface_toy_ringdown} shows the GW distribution from the toy model.  The combination of rotation,  paired with a nonaxisymmetric object, is what drives the second characteristic distribution of GWs during the ringdown phase.  Figure \ref{fig:surface_sim_ringdown} displays this distribution characterized by a maximum along the equator, similar to the bounce phase.  However, within the equator, a key difference arises $\pi/2$ radians from either GW maximum, showing $h_+$ amplitudes that are smaller by a factor of 2.  Additionally, nonnegligible $h_+$ along the pole arises.  Thus there is a superposition of two dynamics.  

The first form of dynamics is the remaining ringdown that will complete over the timescale of tens of ms, causing the prominent GW amplitudes along the CCSN equator, as described in the previous section.  The second is the effect of a rotating, nonaxisymmetric body.  Consider the work of \citet{creighton:2011} (see their Section 3.4), who outline the GW emission for a solid-body, fixed-size, rotating ellipsoid,
\begin{equation}
h_+ = \frac{4G(I_{xx} - I_{yy})\Omega^2}{c^4 D}\frac{1 + \cos^2\theta}{2}\cos2\phi, 
\label{eq:h+_rot}
\end{equation}
and
\begin{equation}
h_\times = \frac{4G(I_{xx} - I_{yy})\Omega^2}{c^4 D}\cos\theta\sin2\phi,
\label{eq:hx_rot}
\end{equation}
for an angular velocity $\Omega$, distance $D$, altitudinal angle $\theta$, and azimuthal angle $\phi$.  In this example, the terms generating the GWs are due to the rotation of the PNS.  At a given point in time, note $h_+$ varies with altitudinal angle $\theta$ as $1 + \cos^2\theta$.  This implies expected maxima closest to the axis of rotation, for a solid-body rotator.  Secondly, varying with azimuthal angle $\phi$, the $\cos 2\phi$ term indicates 4 GW maxima equally spaced by $\pi / 2$ radians.  In Figure \ref{fig:surface_toy_ringdown}, the $1 + \cos^2\theta$ term contributes to the nonzero GW amplitudes along the axis of rotation, arising from the nonzero difference  $I_{xx} - I_{yy}$, mathematically quantifying the deviation from axisymmetry.  The modulated GW amplitude when viewed along the CCSN equator is characterized by the $\cos 2\phi$ term in Equation (\ref{eq:h+_rot}).  The reason $h_+$ is no longer axisymmetric, is due to the mismatch between CCSN equatorial radii $a \neq b$ generating a nonzero $I_{xx} - I_{yy}$.  For the closed form accounting for both ringdown and nonaxisymmetric rotation, we appeal to our closed form solution from Appendix \ref{appendix:rotation_matrices}, in particular Equation (\ref{eqn:h+_general}).  When selecting the noted toy model parameters, the strain surface plot for this rotating, oscillating ellipsoid---Figure \ref{fig:surface_toy_ringdown}---shows good agreement with the distribution of $h_+$ compared to the simulation output: Figure \ref{fig:surface_sim_ringdown}.

\begin{figure*}[h!]
\centering
    \begin{subfigure}[t]{0.45\linewidth}
\includegraphics[width=\linewidth]{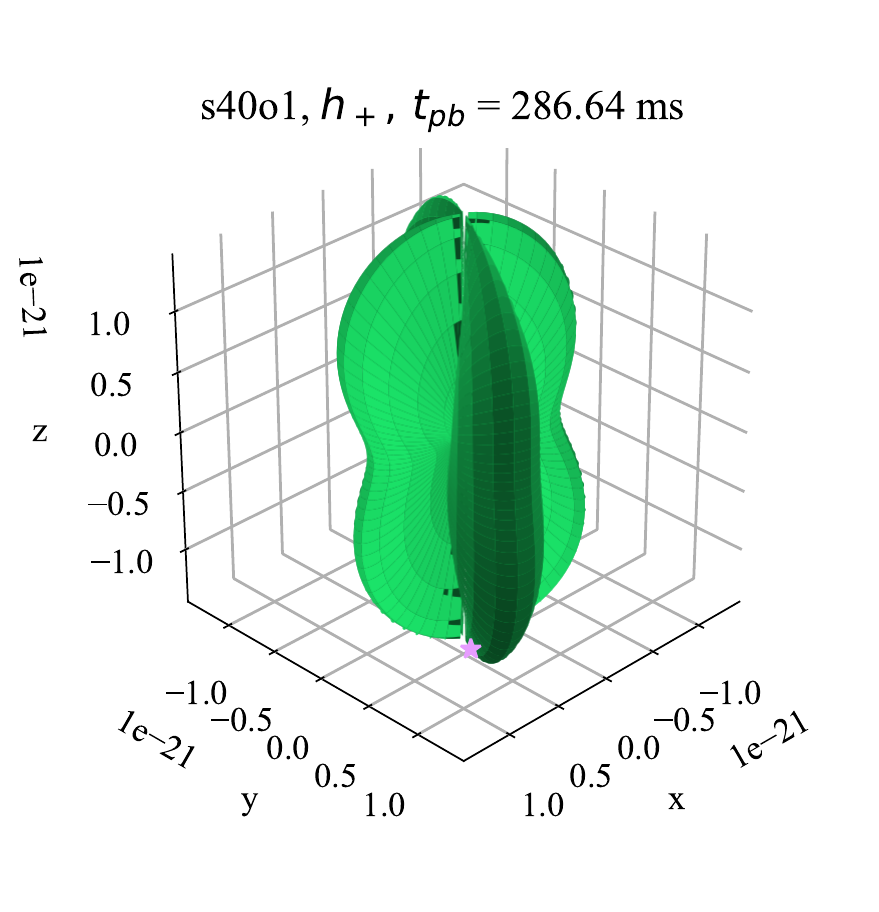} 
    \caption{
    Strain surface plot for $h_+$ during the accretion phase ($\sim 285$ ms pb).  This rotating, four-lobed structure persists for the majority of the accretion phase, producing the largest GW amplitudes along the axis of rotation.}
\label{fig:surface_sim_accretion}
    \end{subfigure}\hfill
    \begin{subfigure}[t]{0.45\linewidth}
\includegraphics[width=\linewidth]{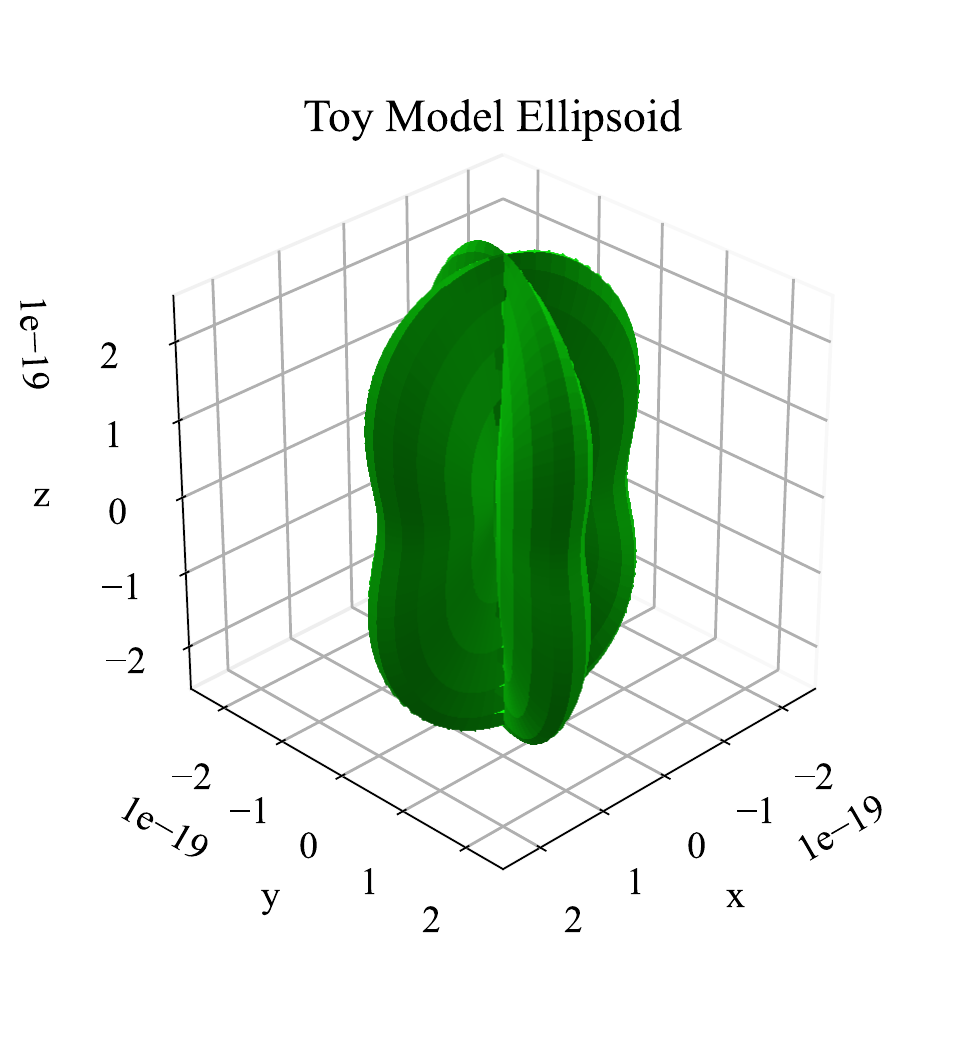}
    \caption{Strain surface plot for a `rigid, solid-body ellipsoid'.  Data taken from simulation outputs motivate toy model parameters of mass 2.1 M$_\odot$, with principal axes $c = 40$ km, $a = 62$ km, $b = 55$ km, rotating with $\Omega = $1000 rad s$^{-1}$, $\dot{a} = \dot{b} = \dot{c} = 0$ km s$^{-1}$, replicating the strain surface during the accretion phase, seen in Figure \ref{fig:surface_sim_accretion}.  Once again, our choice of the `solid-body' approximation captures the morphology of the GW strain, while overestimating the magnitude.}
\label{fig:surface_toy_accretion}
    \end{subfigure}

    \begin{subfigure}[t]{0.49\linewidth}
\includegraphics[width=\linewidth]{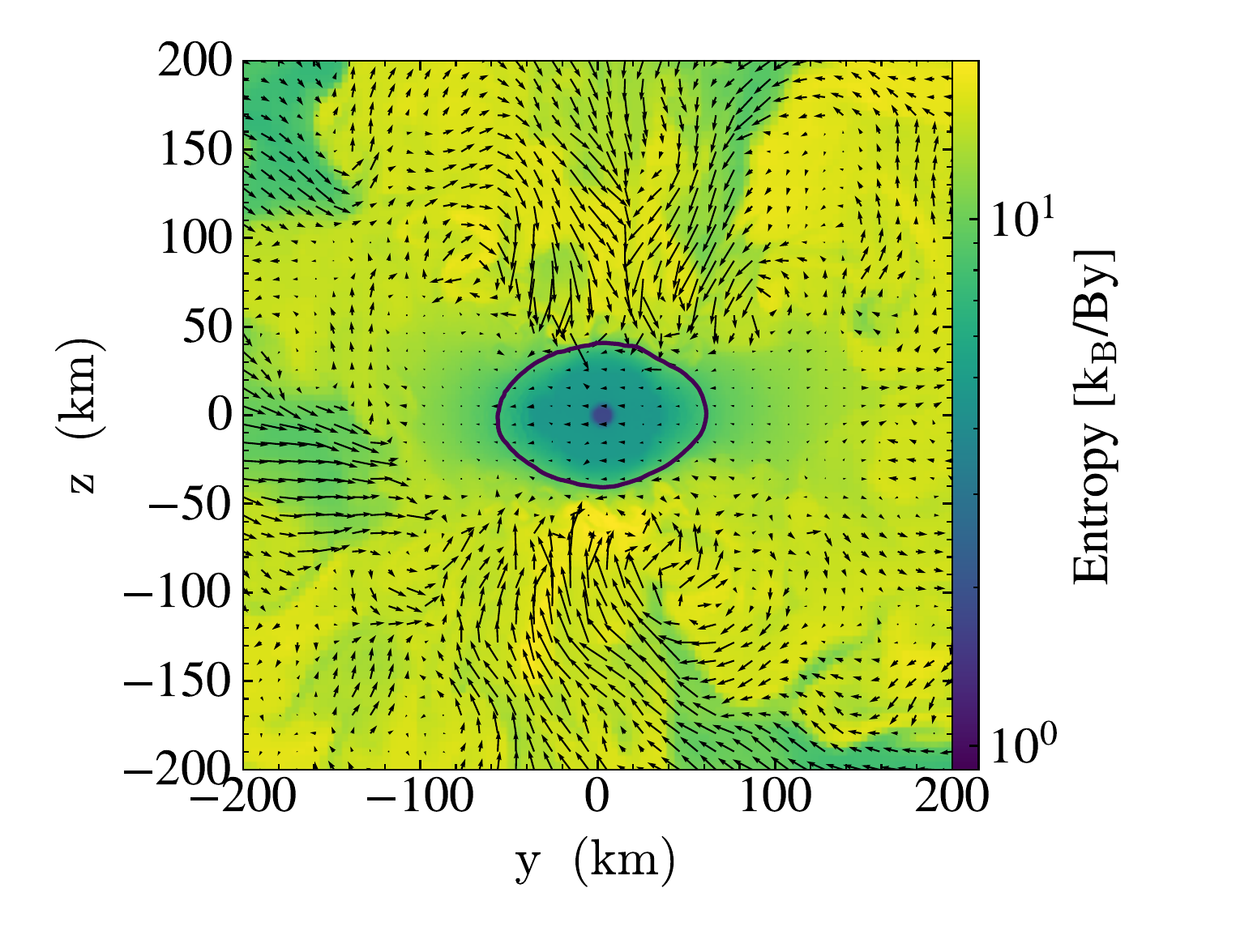}
    \caption{Entropy slice through the CCSN pole for model \texttt{s40o1} $\sim 285$ ms pb.  Arrows show the magnitude of the velocity in the yz plane, and the solid black line is the density contour for the PNS surface of $\rho = 10^{11}$ g cm$^{-3}$.  After the accretion of sufficient angular momentum from overlying material, the oblateness of the PNS increases.  For scale, every cm on the page corresponds to $5.6\times10^{9}$ cm s$^{-1}$.}
\label{fig:entr_po_slice_accretion}
    \end{subfigure}\hfill
    \begin{subfigure}[t]{0.49\linewidth}
\includegraphics[width=\linewidth]{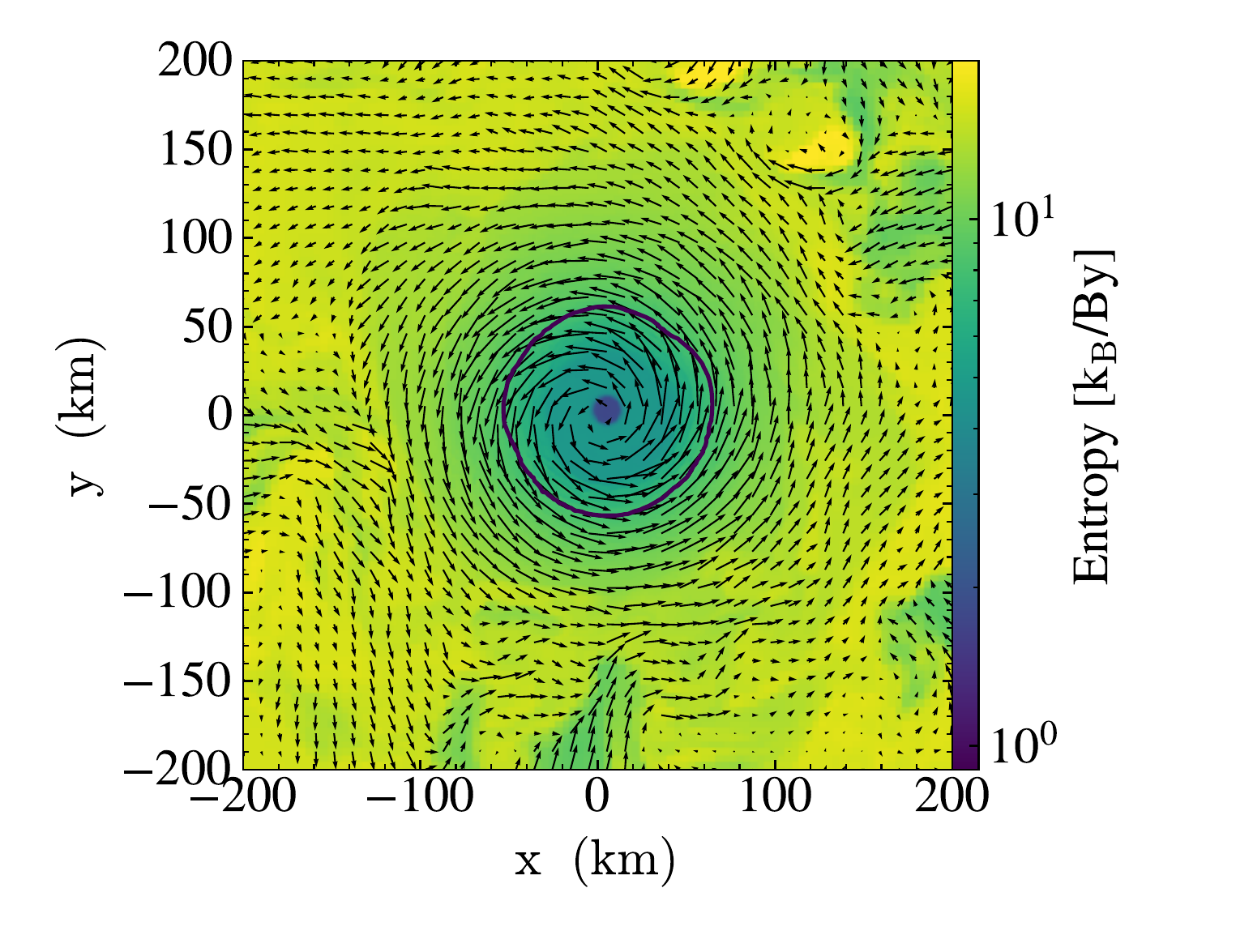}
    \caption{Entropy slice through the equator from model \texttt{s40o1} $\sim 285$ ms after bounce.  Accretion downflows striking the PNS surface cause slight deviations from axisymmetry (a circle in this slice).  The oblateness of the PNS, deviation of axisymmetry in the azimuthal direction, and rotation motivate our choice for modeling the source of the GW emission as a rotating ellipsoid.}
\label{fig:entr_eq_slice_accretion}
    \end{subfigure}
\caption{Strain surface plots and entropy slices for model \texttt{s40o1} during the accretion phase.  The x and y axes form the equator of the supernova; the z axis indicates the axis of rotation.  For the top two panels, the distance from the origin to a point on the green surface represents the detected $h_+$ along that direction.  The bottom two panels are entropy slices, with overlaid velocity profiles (arrows).  A clearer 3D picture of the dynamics can be seen in Figure \ref{fig:3D_vmag}.  }
    \label{fig:GWsurface_solid_rot}
    \end{figure*}

\begin{figure*}
\centering
\includegraphics[width=0.75\linewidth]{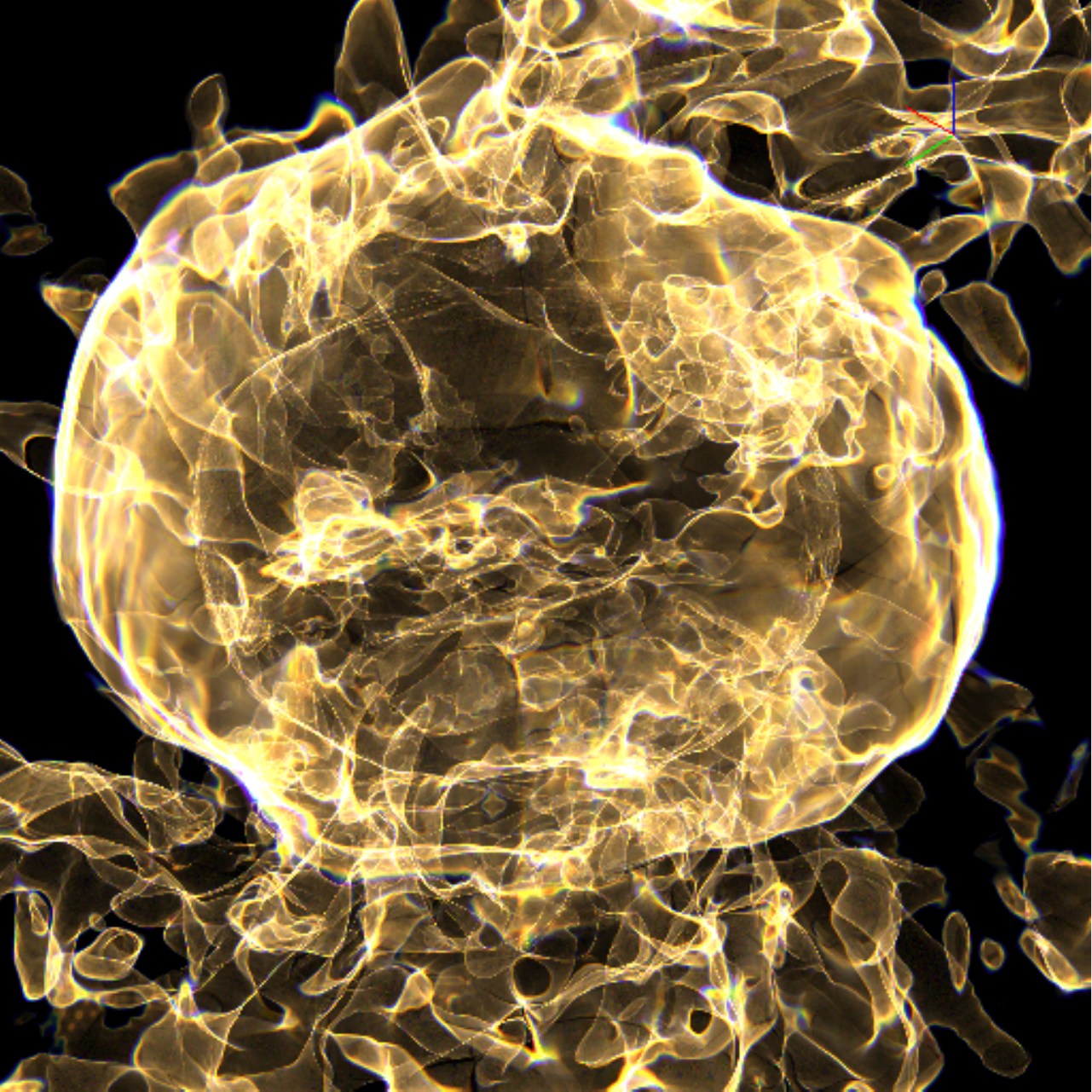} 
    \caption{Magnitude of the spatial velocity---color values centered $\sim 3.9\times 10^{9}$ cm s$^{-1}$---near the PNS for model \texttt{s40o1}, taken $\sim 286$ ms pb.  The panel spans 200 km.  The axis of rotation is aligned vertically with the page.  The 3D structure more clearly displays the coherent flows around the PNS, whose oblateness can be seen by the inner contour, complementing the entropy slices seen in Figure \ref{fig:GWsurface_solid_rot}.}
\label{fig:3D_vmag}
\end{figure*}

\begin{figure*}
\centering
    \begin{subfigure}[t]{0.45\linewidth}
\includegraphics[width=\linewidth]{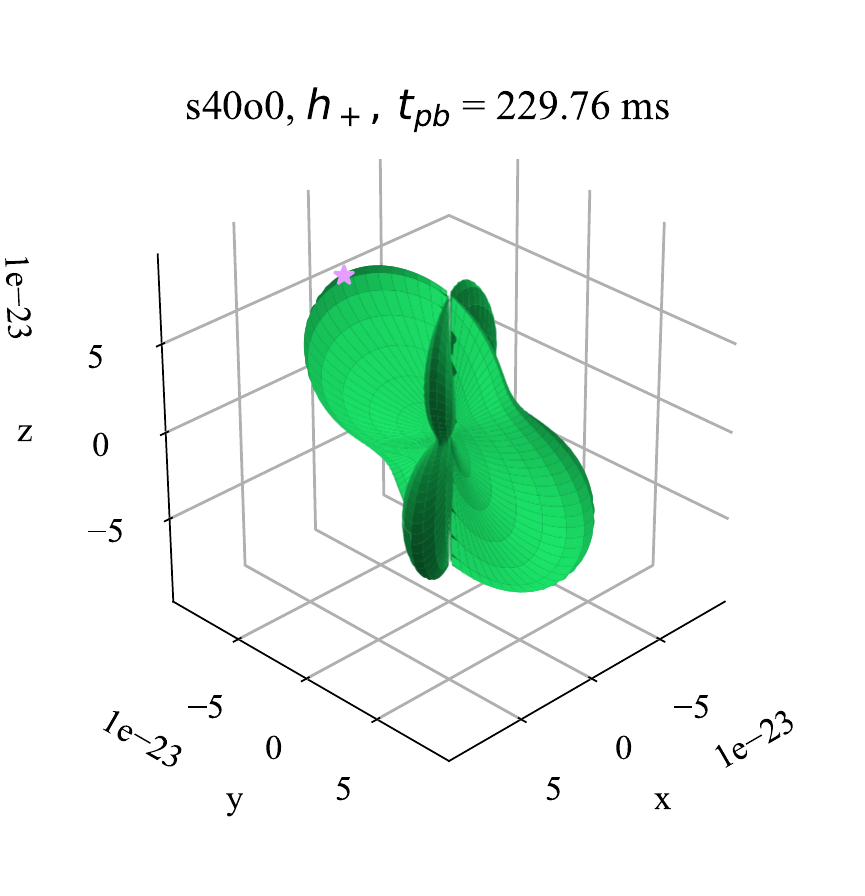} 
    \caption{
    Strain surface plot of $h_+$ for nonrotating model \texttt{s40o0} during the accretion phase ($\sim 230$ ms pb).  We attribute the skewed maximum GW amplitude (not aligned with the coordinate axes) with downflows perturbing the PNS at a misaligned angle.}
\label{fig:surface_sim_tilt}
    \end{subfigure}\hfill
    \begin{subfigure}[t]{0.45\linewidth}
\includegraphics[width=\linewidth]{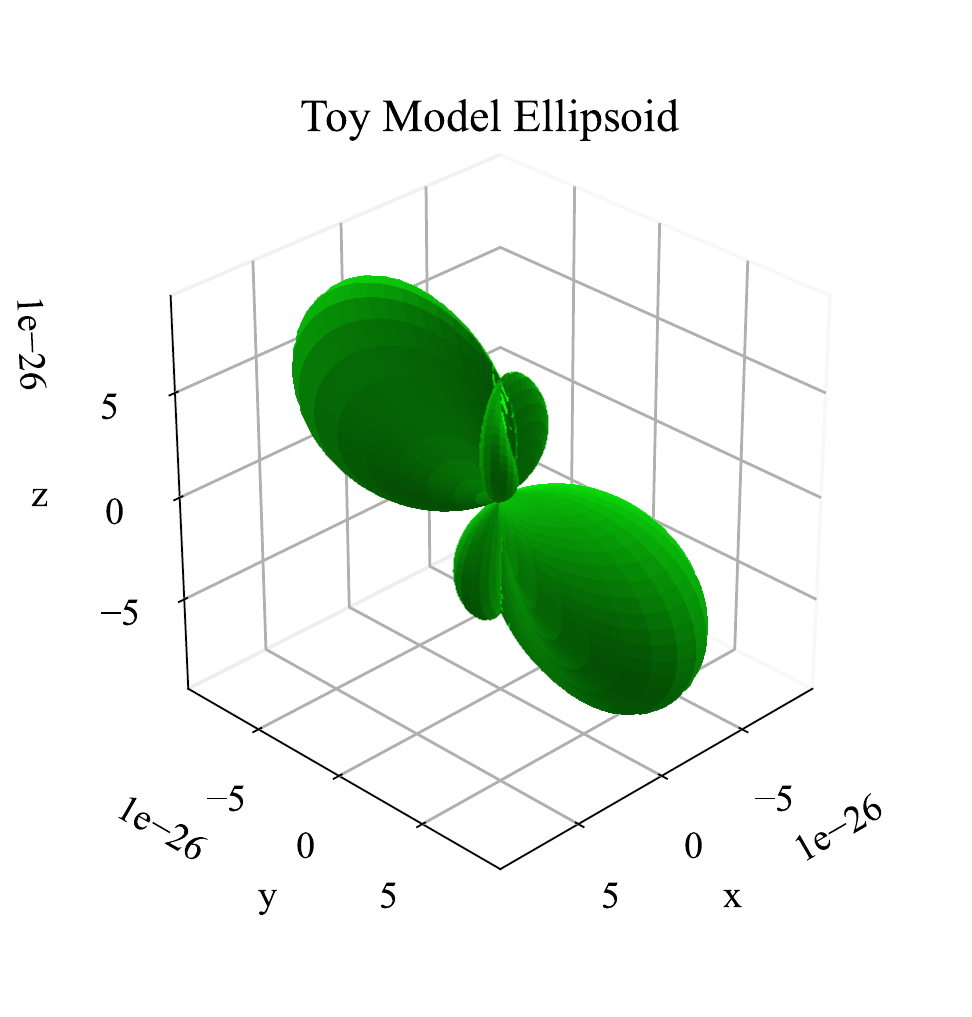}
    \caption{Strain surface plot for nonrotating, tilted ellipsoid.  Data taken from simulation outputs motivate toy model parameters of mass 2.1 M$_\odot$, with principal axes $a = b = c = 45$ km, nonrotating, $\dot{a} = \dot{b} = 0$ km s$^{-1}$, and $\dot{c} = 50$ km s$^{-1}$, tilted by $\pi / 4$ radians, replicating the strain surface during the accretion phase, seen in Figure \ref{fig:surface_sim_tilt}.  Once again, our choice of the toy model approximation captures the morphology of the GW strain, while differing in the magnitude.}
\label{fig:surface_toy_tilt}
    \end{subfigure}

    \begin{subfigure}[t]{0.49\linewidth}
\includegraphics[width=\linewidth]{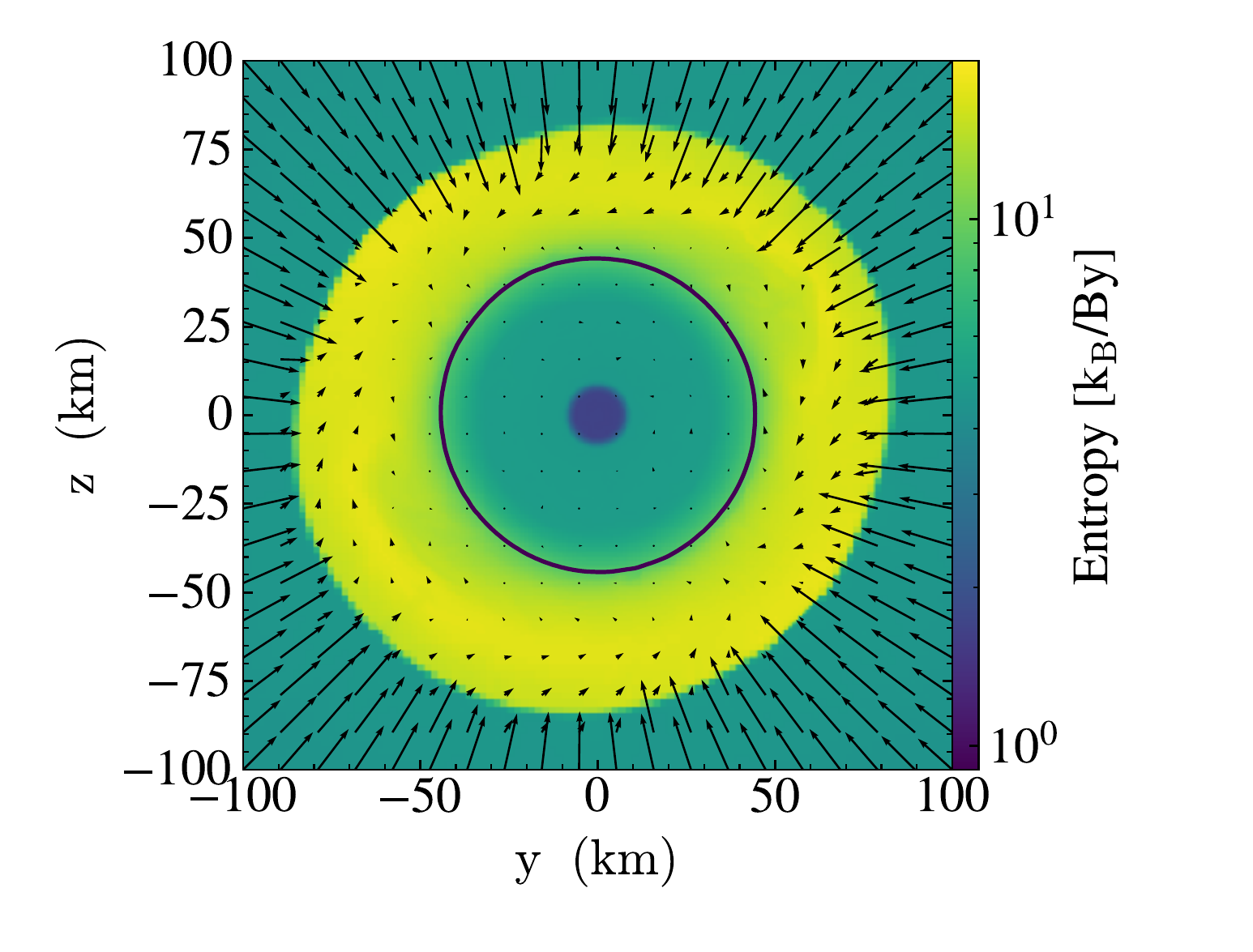}
    \caption{Entropy slice through the CCSN pole for model \texttt{s40o0} $\sim 230$ ms pb.  Arrows show the magnitude of the velocity in the yz plane, and the solid black line is the density contour for the PNS surface of $\rho = 10^{11}$ g cm$^{-3}$.  The skewed angle of the post shock region is what motivates our choice for the tilted ellipsoid.  For scale, every cm on the page corresponds to $5.6\times10^{9}$ cm s$^{-1}$.}
\label{fig:entr_po_slice_tilt}
    \end{subfigure}\hfill
    \begin{subfigure}[t]{0.49\linewidth}
\includegraphics[width=\linewidth]{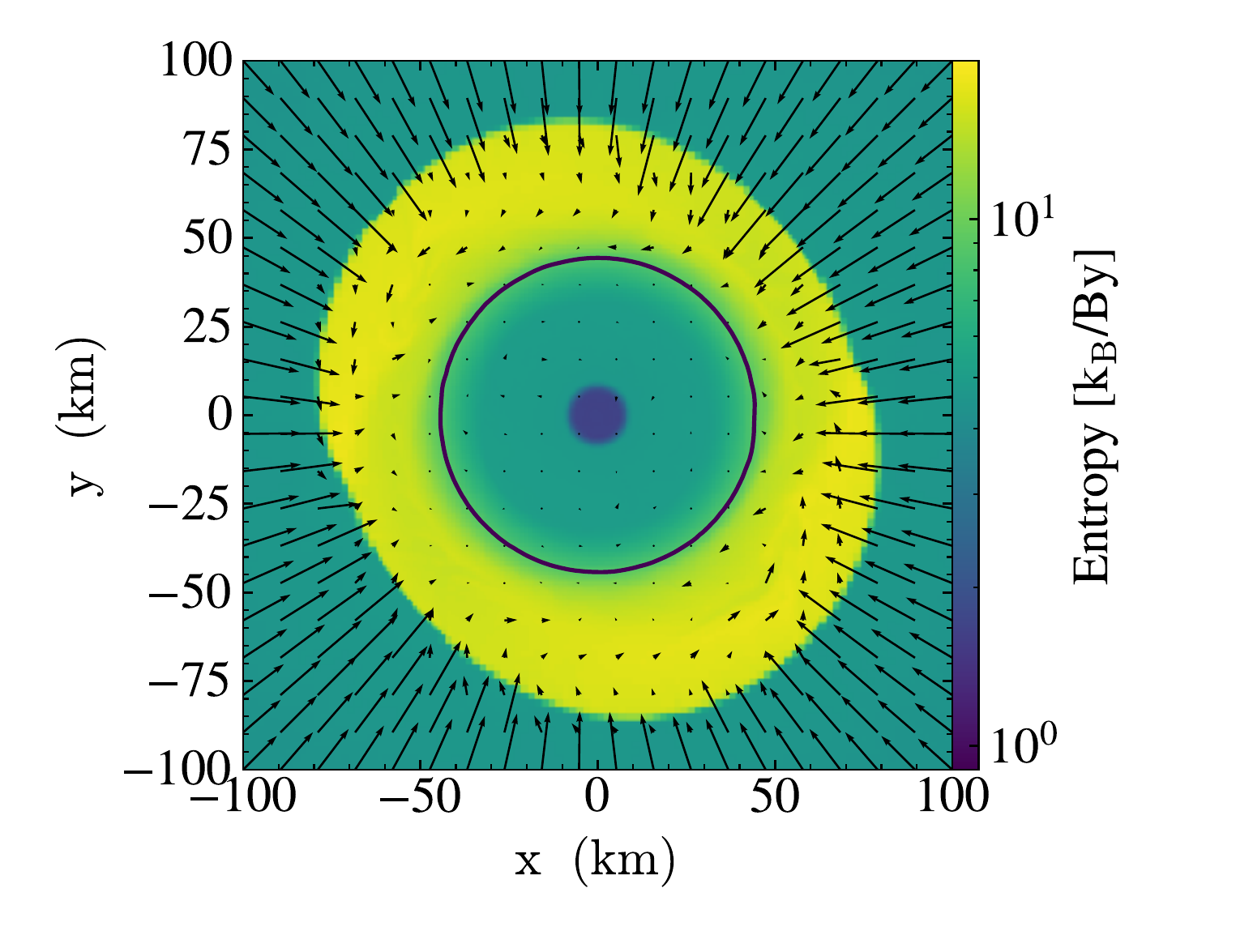}
    \caption{Entropy slice through the equator from model \texttt{s40o0} $\sim 230$ ms after bounce.  Similar asymmetries to Figure \ref{fig:entr_po_slice_tilt} of the shock are visible when viewing the supernova equator.}
\label{fig:entr_eq_slice_tilt}
    \end{subfigure}
\caption{  Strain surface plots and entropy slices for model \texttt{s40o0} during the accretion phase.  The x and y axes form the equator of the supernova; the z axis indicates the pole.  For the top two panels, the distance from the origin to a point on the green surface represents the detected $h_+$ along that direction.  The bottom two panels are entropy slices, with overlaid velocity profiles (arrows).}
    \label{fig:GWsurfaces}
    \end{figure*}

\subsection{The Accretion Phase with Strong Rotation}
\label{ssec:toy_rotation}

As ringdown subsides and continued accretion onto PNS ensues, the PNS will be deformed from matter downflows onto its surface and---depending on the degree of rotation---possibly from PNS convection.  Figure \ref{fig:entr_po_slice_accretion} shows the entropy profile $\sim 285$ ms pb for model \texttt{s40o1}.  With continued angular momentum accretion, the PNS has become extremely oblate with polar radius  $\sim 40$ km and equatorial radii $\sim60$ km.  The velocity profile of the overlying material also shows the deformation of the PNS in the upper right quadrant due to downflows along the axis of rotation.  The equatorial slice in Figure \ref{fig:entr_eq_slice_accretion} shows slight deviations from axisymmetry as the PNS attempts to attain equilibrium after deformations from turbulent downflows.  To gain a clearer understanding regarding the multidimensionality of the flow, we refer to Figure \ref{fig:3D_vmag}.  This view into the inner 100s of km of the CCSN denotes higher values of the magnitude of the total spatial velocity in brighter hues.  The most inner contour appears near the PNS surface, illustrating how oblate the compact object has become from ample angular momentum accretion.  Just above the PNS surface, even higher velocity material is present.  As the PNS surface is nearly concentric with the brighter, higher velocity material just above, one is provided with another depiction of the angular momentum accretion generating deviations of the PNS from spherical symmetry.  This oblateness paired rotating nonaxisymmetry motivates our choice of toy model parameters $M = 2.1 M_\odot$, $c = 40$ km, $a = 62$ km, $b = 55$ km, $\dot{a} = \dot{b} = \dot{c} = 0$, and $\Omega = 1000$ rad s$^{-1}$.  Observing the velocity field in Figure \ref{fig:entr_po_slice_accretion} and Figure \ref{fig:entr_eq_slice_accretion} motivates our approximation of $\dot{a} = \dot{b} = \dot{c} = 0$.  In reality, the PNS will be dynamic; however, this assumption implies that rotation is dominant, or that the rotation of the nonaxisymmetric features (rather than their radial motion) contribute more importantly to the $2M\dot{R}^2$ term in Equation (\ref{eq:h_estimate}).

Figure \ref{fig:surface_toy_accretion} shows the GW distribution for the solid rotating ellipsoid, displaying a roughly four lobed `clover-like'  structure, with relative GW maxima along the axis of rotation.  To explain the behavior in the strain surface plots, we appeal to Equation (\ref{eq:h+_rot}).  Entering as $1 + \cos^2 \theta$, for $\theta = 0,\pi$ this term exhibits a maximum, consistent with the axis of rotation.  To explain the four lobed structure, observe the $\cos 2\phi$ term.  Similar to the explanation during the ringdown phase, as an observer circumscribes the CCSN in the azimuthal direction, the $\cos 2\phi$ term will exhibit two sets of maxima and minima: one occurring at $\phi = 0,\pi$ and $\pi/2,3\pi/2$.  These two sets are reflected by the orthogonal lobes depicted in Figure \ref{fig:surface_toy_accretion}.  However, the key difference between the ringdown and accretion phase is that rotation dominates the dynamics.  Thus, the GW amplitude at $\phi = 0,\, \pi/2,\, \pi,$ and $3\pi/2$ are of similar amplitude.  The behavior of $h_\times$ displays similar maxima along the axis of rotation and 4 GW maxima in the azimuthal direction.  However, $h_\times$ is out of phase by $\pi / 4$ radians, compared to $h_+$.  When comparing the toy model of Figure \ref{fig:surface_toy_accretion} with the simulation output of Figure \ref{fig:surface_sim_accretion}, we notice good agreement in the overall GW distributions.  Once again, the amplitudes differ by two orders of magnitude, indicating non-solid-body effects becoming more dominant.  Similar to before, the solid-body rotation in the toy model contributes more matter rotating, leaving a larger $\ddot{Q} \sim 2 M\dot{R}^2$, compared to a more differentially rotating PNS.  As a point of emphasis, we acknowledge rotating PNSs do not exactly mimic a solid, rotating ellipsoid.  We only use this simplified model to describe nonaxisymmetries, which revolve around a single axis, in a closed expression.  Detailed illustrations of the angular velocity profile can be found in Appendix \ref{app:rotational_velocity}.

\subsection{The Accretion Phase Without (or with Weak) Rotation}
\label{ssec:toy_accretion}

The last degree of freedom we explore is due to the tilt of an ellipsoid.  In the nonrotating or slowly rotating case, during the accretion phase, the PNS will still be perturbed by turbulent downflows and possibly more influenced by PNS convection.  However, without centrifugal support to maintain a mass quadrupole moment, the $\dot{a}$, $\dot{b}$, and $\dot{c}$ terms will directly contribute to the $2M\dot{R}^2$ term from Equation (\ref{eq:h_estimate}) and there will be no (less) rotational contribution from $\Omega$ for the non (slowly) rotating case.  Observing Figure \ref{fig:entr_po_slice_tilt}, in this cross section, we note a deviation of the shock from circular symmetry and a nearly circular PNS cross section.  Similarly, when viewing perpendicular to the equatorial plane, Figure \ref{fig:entr_eq_slice_tilt} shows deviations along the azimuthal direction as well.  It is this nonspherical postshock region that motivates our choice of an impulse to the PNS that is not aligned with the coordinate axes of the grid.

The physical picture is modeled with a nonrotating, spherical PNS.  We assume a strong downflow perturbs the PNS at an angle $\theta = \pi / 4$ from the z axis.  The ellipsoid principal axis is now aligned with $\theta = \pi / 4$ and receives an impulse causing $c$ to evolve with $\dot{c} = 50$ km s$^{-1}$.  Figure \ref{fig:surface_toy_tilt} shows a modified picture, compared to before.  The influence of tilt allows GW maxima to occur along directions misaligned from the x, y, and z axes.  This behavior is contained in the analysis in Appendix \ref{appendix:rotation_matrices}.  Originating from rotating the initial matrix $I_\mathrm{ellipsoid}$, there will be nondiagonal components introduced for an observer outside the corotating ellipsoid frame.  By introducing these off diagonal components, the expressions for $h_+$ and $h_\times$ become less concise, and introduce a quadrupole matrix of the form Equation (\ref{eq:tilted_quad_matrix}). 
 For stochastic accretion during the accretion phase, we expect this preferred angle to vary in time, depending on the direction from which the accretion flow originates.  In Figure \ref{fig:surface_sim_tilt}, we notice similar behavior, with a direction of maximum GW amplitude that does not occur along the x, y, or z axes. 

\subsection{A Note on Similar Dynamics with Different Orientations}

\begin{figure*}
\centering
\includegraphics[width=0.95
\linewidth]{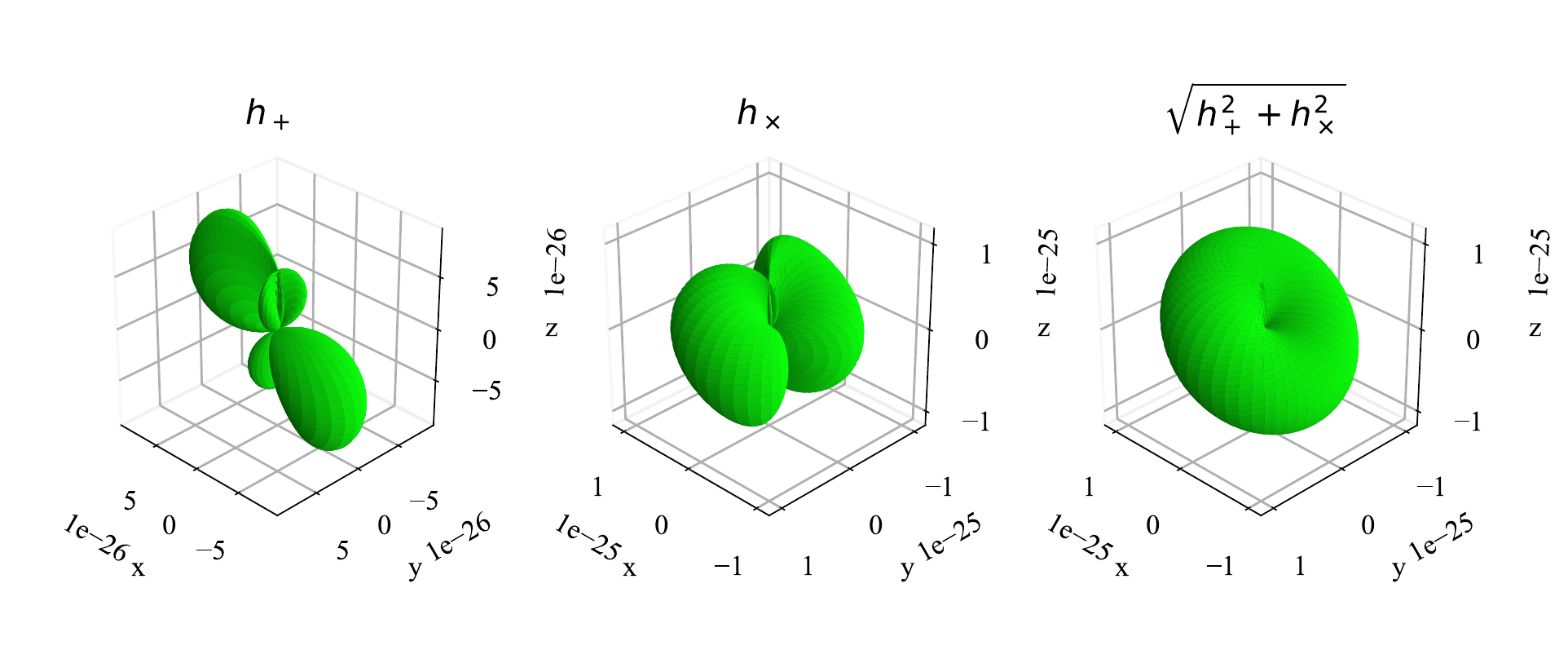} 
    \caption{Strain surface plots of the tilted toy model ellipsoid from Section \ref{ssec:toy_accretion}.  $h_+$, $h_\times$, and $\sqrt{h_+^2 + h_\times^2}$ are shown from left to right.  While similar to the dynamics of the bounce phase, we note different contributions to each polarization.  Whereas a coordinate system aligned with the principal ellipsoid axis along $c$ will only produce $h_+$ emission (Figure \ref{fig:surface_toy_bounce}), a differently oriented GW source will produce GW amplitudes with differing contributions in each polarization.  When considering $\sqrt{h_+^2 + h_\times^2}$ on the right, the axisymmetric GW amplitudes are recovered along the axis tilted by $\pi / 4$ from the z axis.  In short, two systems with similar dynamics will produce similar angular distributions of total amplitudes $\sqrt{h_+^2 + h_\times^2}$, but differing amounts in each polarization.}
\label{fig:hp_hc_tilt}
\end{figure*}

Notice the dynamics of the bounce toy model from Section \ref{ssec:toy_bounce} and those from the accretion phase in Section \ref{ssec:toy_accretion} are similar: they both involve a velocity $\dot{c}$ along the principal axis.  The main difference is the nonrotating accretion case has the principal axis along $c$ tilted by $\pi / 4$ radians with respect to the z axis, whereas the principal axis along $c$ is aligned with the z axis in the bounce case.  Intuitively, one would expect a similar strain surface plot, simply tilted by $\pi / 4$ radians.  However, it is clear the strain surface plots in Figure \ref{fig:surface_toy_bounce} and Figure \ref{fig:surface_toy_tilt} are different.  This difference arises because the relative contributions to the quadrupole moments have changed.  These differences in $\ddot{I}_{ij}^{\prime\mathrm{TT}}$ propagate through to modify the angular dependence of $h_+$ and $h_\times$ in Equations (\ref{eqn:h+_general}-\ref{eqn:hx_general}).  To illustrate this point, see Figure \ref{fig:hp_hc_tilt}.  The three panels show the distributions of GW amplitudes for the tilted ellipsoid case from Section \ref{ssec:toy_accretion}.  From left to right, the panels show $h_+$, $h_\times$, and $\sqrt{h_+^2 + h_\times^2}$.  The right panel displays the expected behavior of the GW distribution.  In particular, since the motion of the ellipsoid is occurring along the axis $\pi / 4$ radians from the z axis, the GW axisymmetry appears about this axis, as expected for transverse waves.  As this tilt amounts to a change in coordinates, the contributions to the distribution of each polarization change (observer dependent), yet the distribution of the total physical signal $\sqrt{h_+^2 + h_\times^2}$ must remain unchanged (physics dependent).  Thus, even for well-templated bounce signals, which arise from nearly axisymmetric dynamics, depending on the orientation, GW detectors may pick up plus and cross polarization modes.

 A simple yet important subtlety arises for the future of GW astronomy with multiple GW detectors at different orientations.  Consider uniquely oriented sets of observatories, who could detect both polarization modes---we acknowledge with current age detectors this is observationally challenging.  For a given event, two sets of detectors with two coordinates systems will detect differing amounts of signal \textit{in each polarization}.  In particular, the key feature influencing the difference in observed GW polarizations for two given observatories is their relative orientation, rather than their relative position on Earth. Thus, when discussing the polarization content of a given event, the relevant physical quantity each will report, to make an accurate comparison, should be the total content of the signal $\sqrt{h_+^2 + h_\times^2}$.  As a point of emphasis for future CCSN GW theory works, in an attempt to further support a GW era where both polarizations can be detected, we recommend reporting information regarding both polarization modes; only reporting a single polarization mode would not give an orientation invariant quantity.



\section{Finding Preferred Directions Over Time}
\label{sec:scatter3D}

\begin{figure*}
\centering
    \begin{subfigure}[t]{0.49\linewidth}
\includegraphics[width=\linewidth]{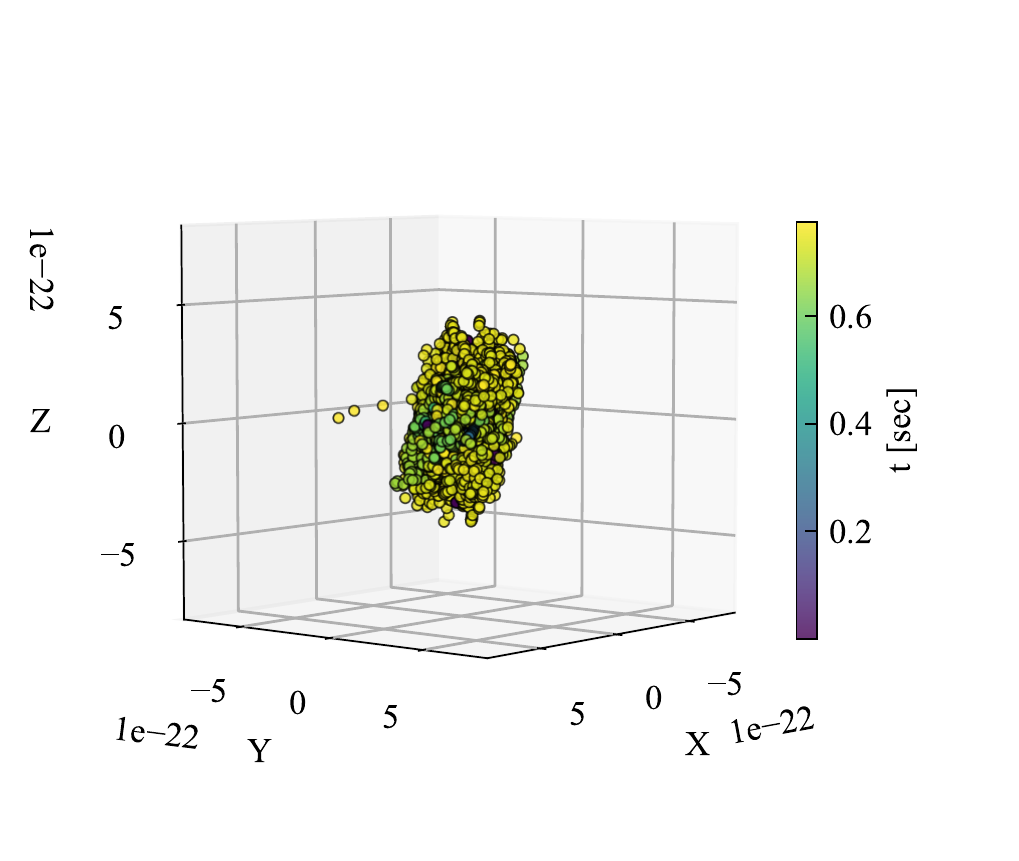} 
    \caption{Dominant viewing angles of $h_+$ for \texttt{s40o0}.  The distribution of dominant GW viewing angles remains nearly isotropic for the duration of the supernova evolution.}
\label{fig:scatterA}
    \end{subfigure}\hfill
    \begin{subfigure}[t]{0.49\linewidth}
\includegraphics[width=\linewidth]{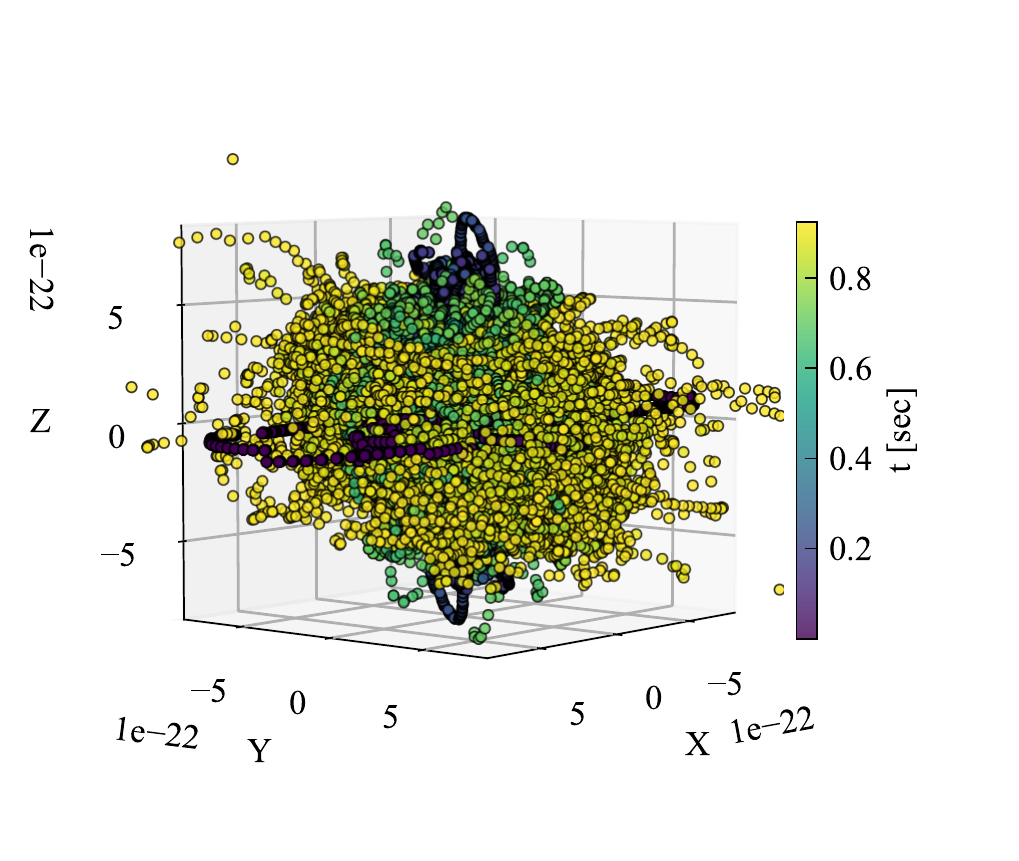}
    \caption{Dominant viewing angles of $h_+$ for \texttt{s40o0.5}.  The dark (early time) points distributed along the equatorial plane are generated from the bounce signal and ringdown.  The dominant signal direction then aligns with axis of rotation and finally relaxes to an isotropic distribution.}
\label{fig:scatterB}
    \end{subfigure}

    \begin{subfigure}[t]{0.49\linewidth}
\includegraphics[width=\linewidth]{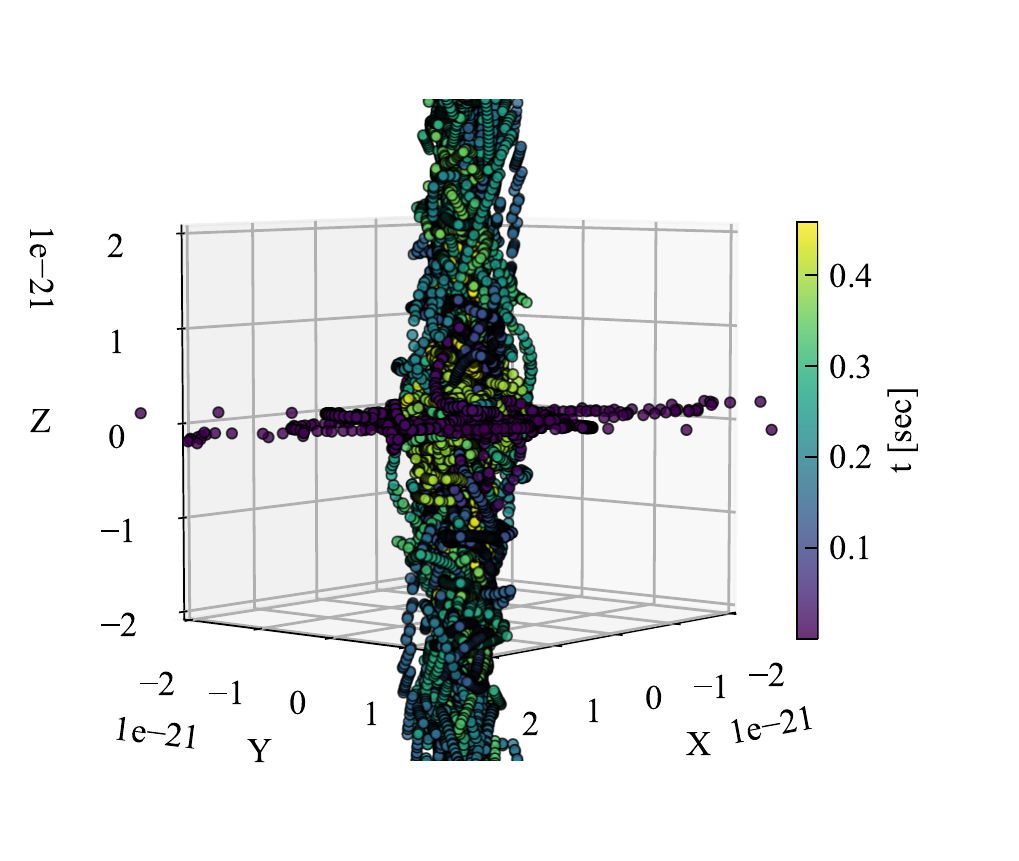}
    \caption{Dominant viewing angles of $h_+$ for the fast rotating model \texttt{s40o1}.  The preferred GW direction is noticeably columnated and follows coherent paths that precess around the axis of rotation.}
\label{fig:scatterC}
    \end{subfigure}\hfill
    \begin{subfigure}[t]{0.49\linewidth}
\includegraphics[width=\linewidth]{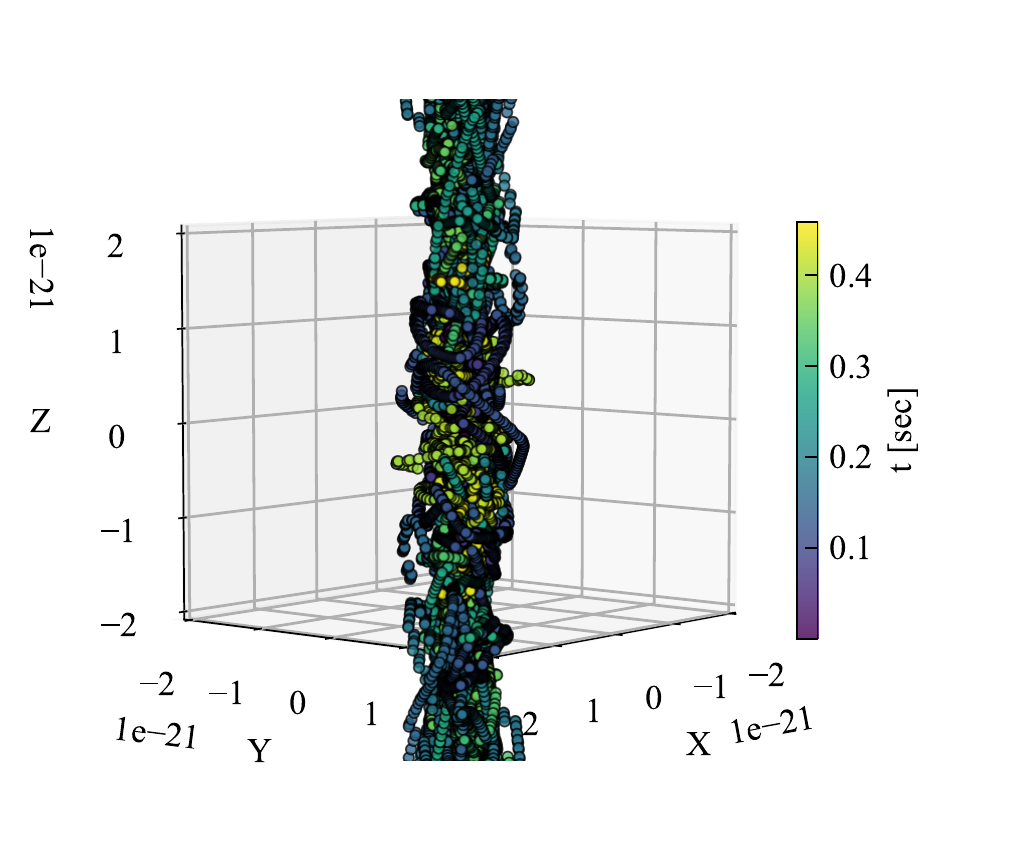}
    \caption{Same fast rotating model as in Figure \ref{fig:scatterB}, however, for the cross polarization $h_\times$.}
\label{fig:scatterD}
    \end{subfigure}
\caption{These strain surface scatter plots, or $S_3$ plots, compile the direction of maximum GW emission (e.g., purple stars in Figure \ref{fig:surface_sim_bounce}, Figure \ref{fig:surface_sim_ringdown}, Figure \ref{fig:surface_sim_accretion}, and Figure \ref{fig:surface_sim_tilt}) throughout the supernova evolution from simulation outputs.  Brighter points correspond to later times.  The xy plane forms the equator of the supernova.  The z axis identifies the axis of rotation for rotating cases.}
    \label{fig:tri3Dscatter}
    \end{figure*}

\begin{figure*}
    \centering    \includegraphics[width=\linewidth]{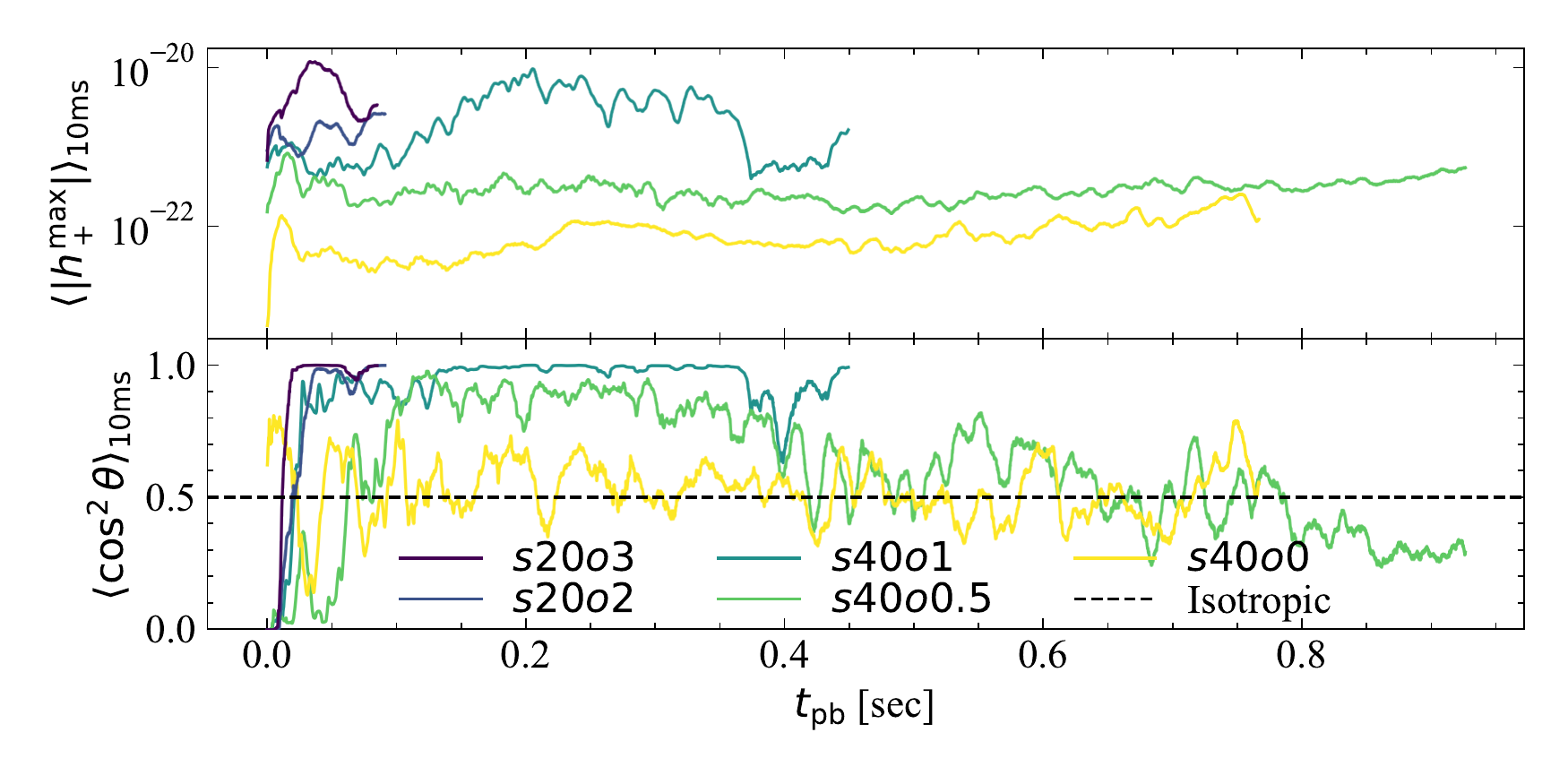}
    \caption{(Top): Average $h^\mathrm{max}_+$ amplitude for a 10 ms window from $t_\mathrm{pb}$ to $t_\mathrm{pb}$ + 10 ms.  Larger amplitudes correspond to greater time averages of GW amplitudes.  (Bottom): Average $|h^\mathrm{max}_+| $ weighted by $\cos^2\theta$ and normalized by $\langle |h^\mathrm{max}_+| \rangle_{10 \mathrm{ms}}$.  Denoted in the black dashed line, for a spherical distribution on the unit sphere, $\langle \cos^2\theta\rangle_{10\mathrm{ms}} = 1/2$.  Values above this line indicate an $\langle |h^\mathrm{max}_+| \rangle_{10 \mathrm{ms}}$ distribution that is more prolate; values below this line indicate a distribution that is more oblate.  Seen in the rotating models (darker hues), as rotation rate increases, the preferred GW directions favor the axis of rotation, or $\cos^2\theta = 1$. }
    \label{fig:GW_costheta_evolution}
\end{figure*}

\begin{figure*}
    \centering
    \includegraphics[width=\linewidth]{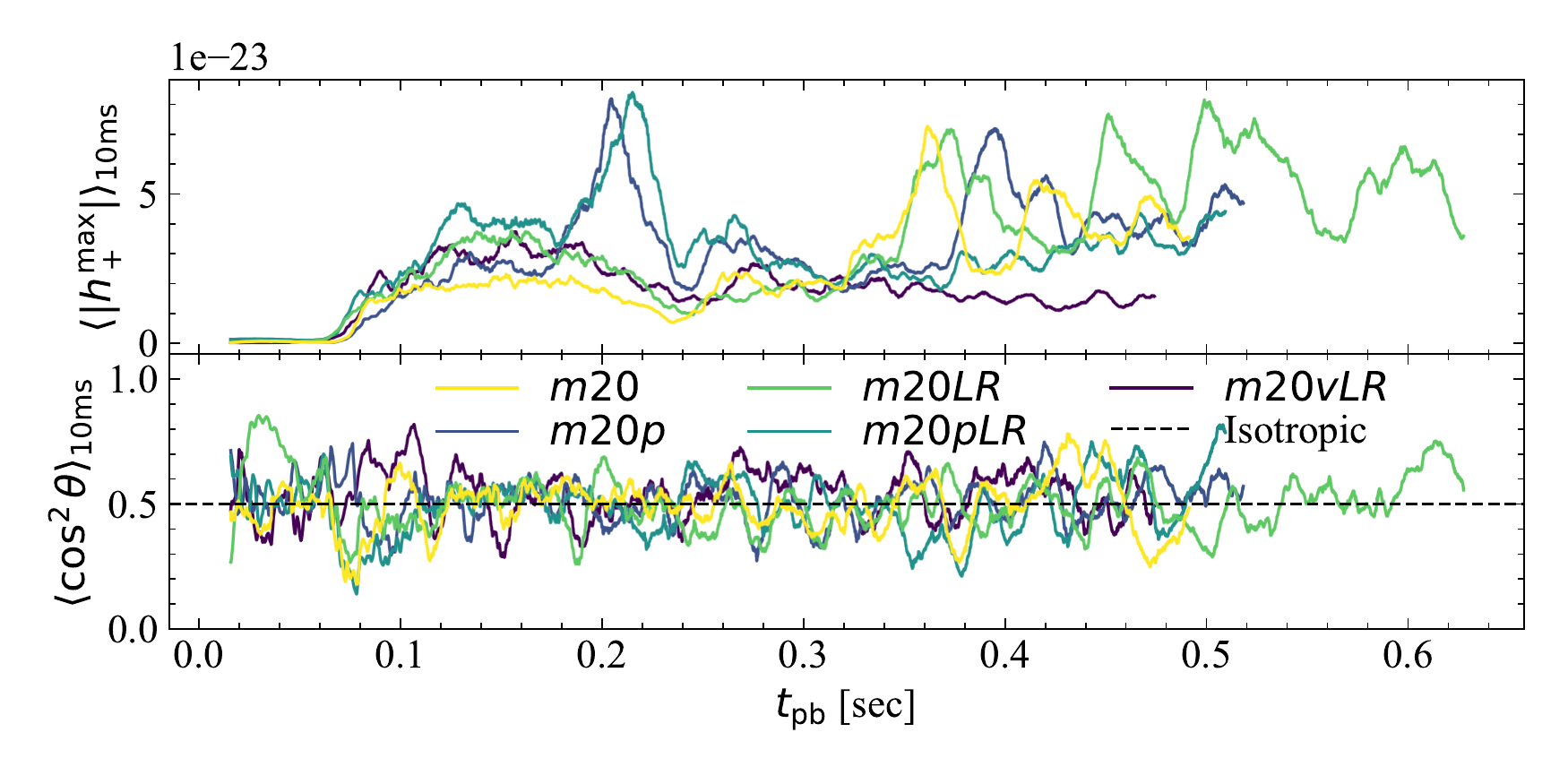}
    \caption{Similar to Figure \ref{fig:GW_costheta_evolution}, but for the \texttt{mesa} models from \citet{oconnor:2018b}.  (Top) Compared to the rotating 40 $M_\odot$ models, the \texttt{mesa} suite has GW amplitudes nearly an order of magnitude lower.  (Bottom) all GW distributions remain nearly isotropic, with a slight preference towards the poles. }
    \label{fig:GW_costheta_evolution_mesa}
\end{figure*}

\begin{figure*}[t!]
    \centering
    \includegraphics[width=\linewidth]{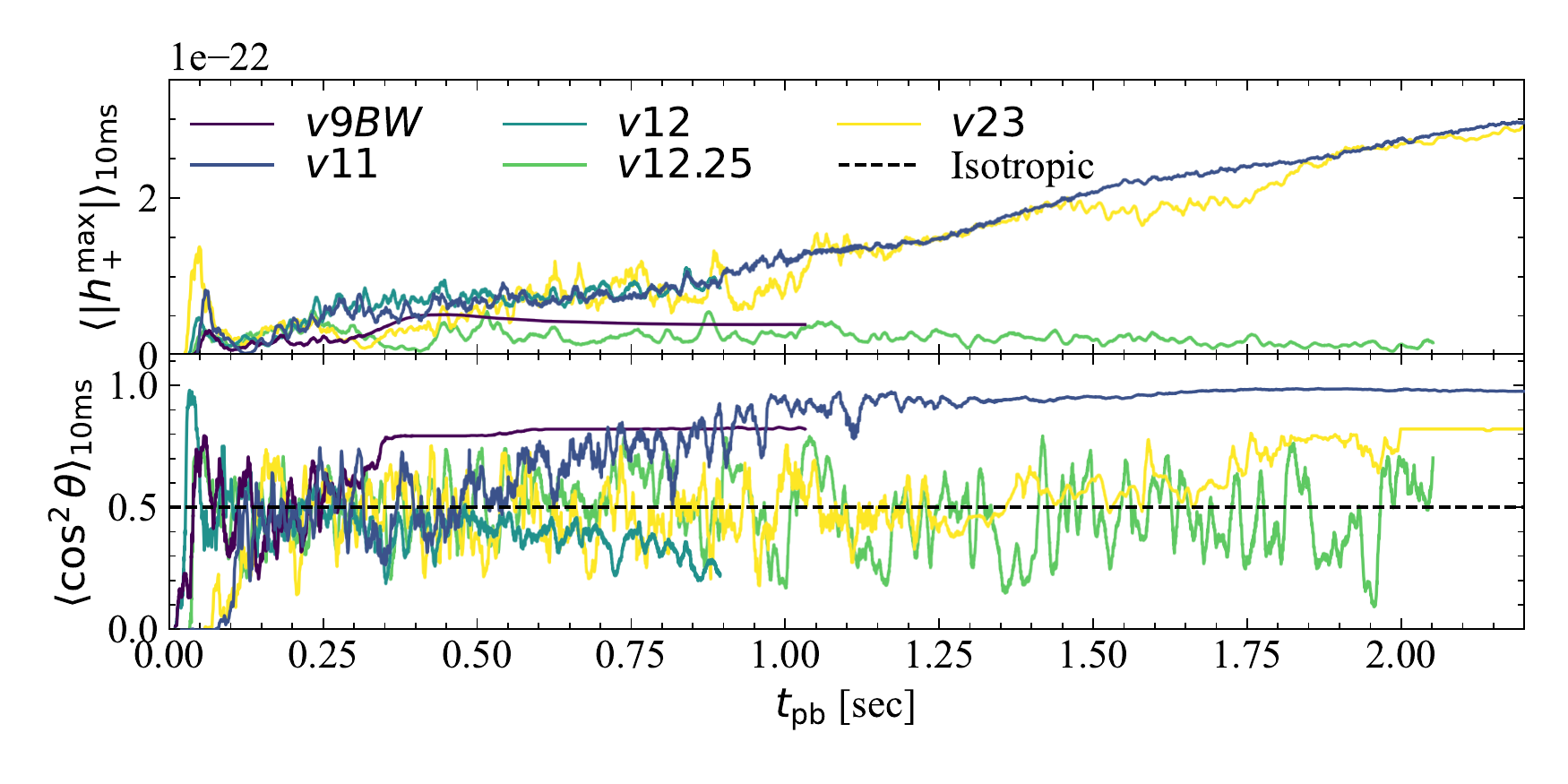}
    \caption{Similar to Figure \ref{fig:GW_costheta_evolution} and Figure \ref{fig:GW_costheta_evolution_mesa}, but for the nonrotating models \texttt{v[9BW, 11, 12, 12.25, 23]}.  (Top) Compared to the \texttt{mesa} models, this suite has similar GW amplitudes during the accretion phase, $t_\mathrm{pb} \lesssim 300$ ms.  After, the extreme matter asymmetries dominate the GW signal, by contributing to direct offset `GW memory' effects.  (Bottom) During the accretion phase, the preferred direction oscillates about $\langle \cos^2\theta\rangle_{10\mathrm{ms}} \sim 0.5$.  As the matter asymmetries become more drastic, different morphologies and orientations affect directionality to settle at different values of $\langle \cos^2\theta\rangle_{10\mathrm{ms}}$.  We note the only nonexploding model of this set, \texttt{v12.25}, retains its stochastic, nonsecular  behavior.}
    \label{fig:GW_costheta_evolution_vartanyan}
\end{figure*}

Strain surface plots can be powerful diagnostic tools for analyzing the directionality of GW emission at a specific instance in time.  However, CCSNe are dynamic systems whose GW generation evolves over seconds.  To track the evolution of the preferred direction emitting gravitational radiation, refer to Figure \ref{fig:tri3Dscatter}.  Each point in this 3D `strain surface scatter plot'---henceforth referred to as an $S_3$ plot---refers to the direction along which a maximum $h_+$ would be measured; that is, the purple stars (like those seen in Figure \ref{fig:surface_sim_bounce}, Figure \ref{fig:surface_sim_ringdown}, Figure \ref{fig:surface_sim_accretion}, and Figure \ref{fig:surface_sim_tilt}) from all time steps are recorded from the entire simulation duration and plotted.  Points that have brighter colors occur later in the supernova evolution.

The nonrotating model, \texttt{s40o0}, is represented in Figure \ref{fig:scatterA}.  As dark points (early times) are close to the center of the $h_+$ distribution, this indicates a relative period of quiescence for the early time GW emission.  As the supernova develops, GW amplitudes become more pronounced, shown through the brighter points (later times) near the outer parts of the distribution.  We do not observe any particular structure or morphology from this distribution for the \texttt{s40o0} model for either the plus or cross polarization.

The slowly rotating model, \texttt{s40o0.5}, is represented in Figure \ref{fig:scatterB}.  The slow rotating case exhibits noticeable $h_+$ amplitudes from the early evolution.  These points originate from the bounce signal, generating a dark purple distribution along the equator.  During the ringdown, one can imagine a strain surface, similar to Figure \ref{fig:surface_sim_ringdown} rotating and tracing a path along the equator that modulates in distance from the axis of rotation.  In the medium brightness, blue points display columnated behavior along the axis of rotation, which we attribute to the coherent matter motion circulating around the PNS due to rotation.  The final few hundred ms of evolution displays a relatively stochastic preferred GW direction.  In total, the entire distribution of preferred GW directions is slightly more prolate than the nonrotating case.   

The plus polarization for the fast rotating model, \texttt{s40o1}, is represented in Figure \ref{fig:scatterC}.  Similar to the slow rotating case, the darkest points display preferred directions along the supernova equator, consistent with the dominant bounce signal.  As time evolves, however, the dominant direction for $h_+$ becomes noticeably columnated along the axis of rotation.  This conclusion is important because it breaks the narrative set by many axisymmetric CCSN GW works that investigate the detectability of GWs from CCSNe, assuming the `optimal' source orientation for CCSNe lies along the equator.  This is indeed true for detecting the bounce signal.  However, as illustrated in Figure \ref{fig:scatterC}, the dominant viewing angle changes $\sim90^\circ$ to the axis of rotation.  Furthermore, GW amplitudes along this direction are larger by the nearly a factor of two.  \textit{These results draw two main conclusions: (1) GW amplitudes along the axis of rotation can be comparable, if not greater, than GW measurements viewed along the equator for rotating CCSNe.  (2) There is no single optimal orientation to observe GWs over the entire duration of a CCSN.}  Point (1) is in accordance with \citet{powell:2020} and \citet{shibagaki:2021}, which is present in model \texttt{s40o1}.  Similar to the slow rotating case, the dominant viewing angle also traces a coherent path around the axis of rotation.  We notice this behavior for both $h_+$ and $h_\times$, Figure \ref{fig:scatterD}.

As noted in Section \ref{sec:methods}, we also investigate the dominant viewing angle evolution for the $20 \, M_\odot$ models from \citep{oconnor:2018b}.  Similar to the nonrotating case for the $40\, M_\odot$ model, they display an isotropic distribution of dominant viewing angles in $h_+$ as well.  Models \texttt{v[9BW, 11, 12, 23]} display linear structures protruding from isotropic distributions, which are caused by large explosion asymmetries. 


\section{Quantifying the Directionality of Max GW Strain}
\label{sec:quantify_scatter3D}

We quantify the directionality of the maximum GW amplitude by decomposing the $S_3$ plots into useful metrics, shown in Figure \ref{fig:GW_costheta_evolution}, Figure \ref{fig:GW_costheta_evolution_mesa}, and Figure \ref{fig:GW_costheta_evolution_vartanyan}.  We begin by selecting a subset of the points in a 10 ms window.  This window was chosen, the light travel time between both LIGO detectors, to provide an appropriate balance between capturing the stochasticity of CCSN GWs, while being able to follow the directional evolution of the GW signal.  This window is then progressed forward in time, representing a moving average of the data from the $S_3$ plots.  Figure \ref{fig:GW_costheta_evolution} is for the \texttt{s20} and \texttt{s40} models. Figure \ref{fig:GW_costheta_evolution_mesa} is for the nonrotating \texttt{mesa} models, and Figure \ref{fig:GW_costheta_evolution_vartanyan} is for the \texttt{v[9BW, 11, 12, 12.25, 23]} models.

\subsection{Quantifying GW Maxima}

The top panel of all three figures refers to the mean value of the maximum GW amplitudes during this window, denoted $\langle |h^\mathrm{max}_+|\rangle_\mathrm{10\,ms}$.  Hence, it is a proxy for GW activity.  Beginning with the simple case, we examine the \texttt{s20} and \texttt{s40} models.  The top panel notes increasing average GW maxima for increased $\Omega_0$ values assigned at collapse.  While rotation can provide a stabilizing effect to convection and resulting GW signal in 2D \citep{endal:1978,pajkos:2019}, nonaxisymmetries may emerge in 3D, and can produce stronger GW amplitudes.  As shown, this coherent matter motion circumscribing the PNS surface bolsters GW emission by roughly an order of magnitude between nonrotating and rapidly rotating cases.  

In the top panel of Figure \ref{fig:GW_costheta_evolution_mesa}, the \texttt{mesa} models do not exhibit much GW activity, noted by the y axis scale smaller by nearly two orders of magnitude compared to Figure \ref{fig:GW_costheta_evolution}.  
Of the five models, \texttt{m20p} and \texttt{m20pLR} show early bursts of GW emission $\sim 210$ ms post-bounce due to the accretion of velocity perturbations placed in their overlying shells.  As noted in \citet{oconnor:2018} the larger values of lateral (i.e., nonradial) components of kinetic energy excite stronger PNS oscillations.  

The data set from \citet{vartanyan:2023} begins with relatively low GW amplitudes.  These models are not initialized with any artificial perturbations, relying on turbulent instabilities to seed PNS oscillations later in the CCSN evolution.  After an initial quiescent phase, typical GW amplitudes of order $10^{-23}$ are driven during the supernova accretion phase.  For models \texttt{v[9BW, 11, 12]}, beyond $\sim 350$ ms pb, the GW amplitude becomes dominated by the `memory effect', driven by large scale matter asymmetries in the explosion ejecta.  This onset occurs around 1.75 sec pb for \texttt{v23}. Once again, we reiterate, the GW amplitudes calculated for the \citet{vartanyan:2023} models are only from matter contributions, not neutrinos; though, neutrino asymmetries can produce similar linear evolution in the GW amplitudes. 



\subsection{Quantifying GW Maxima Along the Poles}

The bottom panel of Figure \ref{fig:GW_costheta_evolution}, Figure \ref{fig:GW_costheta_evolution_mesa}, and Figure \ref{fig:GW_costheta_evolution_vartanyan} selects GW maxima points from $S_3$ plots and weights them by $\cos^2\theta$; lastly, it is normalized by $\langle |h^\mathrm{max}_+|\rangle_\mathrm{10\,ms}$.  This quantity tracks the tendency of the $h_+$ to cluster near the poles of the CCSN, or the axis of rotation.  In the limit of infinite points isotropically distributed on the unit sphere, $\langle \cos^2 \theta \rangle$ = 1/2, marked by the black dashed line.  For points tightly clustered around the poles, $\langle \cos^2 \theta \rangle \sim 1$.  For GW amplitudes emitted near the plane of the supernova equator, $\langle \cos^2 \theta \rangle \sim 0$.  In Figure \ref{fig:GW_costheta_evolution}, model \texttt{s40o0} displays small oscillations about the isotropic distribution, displaying no major directionality preference.  For \texttt{s40o0.5}, $\langle \cos^2 \theta \rangle_\mathrm{10\, ms}$ begins near a value of zero, consistent with the bounce and ringdown producing maximum strains preferentially toward the supernova equator.  Over the timescale of 100 ms, this GW emission transitions to preferential strains along the pole.  From $\sim 100$ ms pb to $\sim 400$ ms pb, the preferred direction remains aligned with the axis of rotation.  As noted in \citet{pan:2021}, model \texttt{s40o0.5} exhibits strong mass accretion ($\dot{M}$) up until $\sim 400$ ms pb.  Paralleled with the $S_3$ plot in Figure \ref{fig:scatterB}, this $\sim 300$ ms interim corresponds to the dark blue, columnated collection of points.  This is due to dynamics of PNS rotation, paired with oscillations from the high $\dot{M}$, allowing PNS nonaxisymmetries to mimic rotating ellipsoid behavior---see Section \ref{ssec:toy_rotation}.  After this point, the mass accretion rate plateaus to a value of $\dot{M} \sim 500 \, M_\odot$/sec.  As mass accretion stagnates, the rate at which the PNS receives angular momentum plateaus as well.  This reduction in the rate of angular momentum transfer allows for GW emission to deviate from preferential emission along the axis of rotation, allowing stochastic convective processes and downflows to excite GW emission isotropically, rather than from rotationally dominated dynamics of downflow.  

The most rapidly rotating models: \texttt{s40o1}, \texttt{s20o2}, and \texttt{s20o3}, and they behave similarly. 
 They begin with $\langle \cos^2 \theta \rangle_\mathrm{10\, ms} \sim 0$, due to the bounce and ringdown dynamics.  Over the timescale of milliseconds, they transition directly to GW emission predominantly along the pole; the rapid amount of angular momentum accretion is responsible.  As the PNS continuously spins up and receives perturbations from turbulent mass accretion, $\langle \cos^2 \theta \rangle_\mathrm{10\, ms}$ rapidly approaches a value of 1 for the duration of the simulation.  This behavior can be attributed to continued matter motion around the PNS, induced by rotation.  In particular, for model \texttt{s40o1} the onset of the low T/|W| instability---noted in \citet{pan:2018}---provides this condition.  While not formally forming a bar-mode, these perturbations, paired with rotation, provide the necessary conditions from Section \ref{ssec:toy_rotation}, mimicking approximate rotating ellipsoid evolution.  Figure \ref{fig:GW_costheta_evolution_mesa} does not show noteworthy deviation from an isotropic distribution---that is, the evolution of the preferred GW viewing angle remains nearly isotropic for the nonrotating \texttt{mesa} models.

The CCSN set from \citet{vartanyan:2023} begins with relatively stochastic values, with two exceptions.  During the prompt convective phase $\sim 50 \sim 100$ ms, models \texttt{v11} and \texttt{v23} show preferred GW emission along the equator. Continuing into the accretion phase, for all five models, PNS oscillations generate GW signals with $\langle \cos^2 \theta \rangle_\mathrm{10\, ms}$ oscillating about 0.5.  Beyond $\sim 350$ ms, the GW memory dominates the preferred direction of GW emission for models \texttt{v[9BW, 11, 12].} This direct offset signal settles around 1.75 sec pb for \texttt{v23}.  As different models settle to different $\langle \cos^2 \theta \rangle_\mathrm{10\, ms}$, the geometry of the matter motion dictates the final preferred viewing angle.  The only exception is model \texttt{v12.25}, which retains stochastic, nonsecular $\langle \cos^2 \theta \rangle_\mathrm{10\, ms}$ behavior, as it continues accreting and does not successfully explode.  As expected, highly asymmetric and energetic ejecta correspond to more extreme asymmetries in the GW emission.  The variety in $\langle \cos^2 \theta \rangle_\mathrm{10\, ms}$ values in the bottom panel of Figure \ref{fig:GW_costheta_evolution_vartanyan} shows yet another lens describing the highly nonspherical and almost chaotic differences that can manifest in CCSNe, due to differences in initial conditions.  We note the secular, low frequency drifts seen in the $\langle \cos^2 \theta \rangle_\mathrm{10\, ms}$ values provide directionality considerations for future space-based GW observatories, sensitive to frequencies of order $\lesssim10$ Hz.  When filtering out frequencies below 10 Hz, we see similar stochastic evolution in $\langle \cos^2 \theta \rangle_\mathrm{10\, ms}$ to the \texttt{mesa} suite of models in Figure \ref{fig:GW_costheta_evolution_mesa}.  This behavior is expected, since the models from \citet{vartanyan:2023} do not exhibit rotationally dominated dynamics, similar to the \texttt{mesa} suite in Figure \ref{fig:GW_costheta_evolution_mesa}.

\section{Quantifying the Evolution of Strain Surfaces}
\label{sec:surface_decomp}

\begin{figure*}
    \centering
\includegraphics{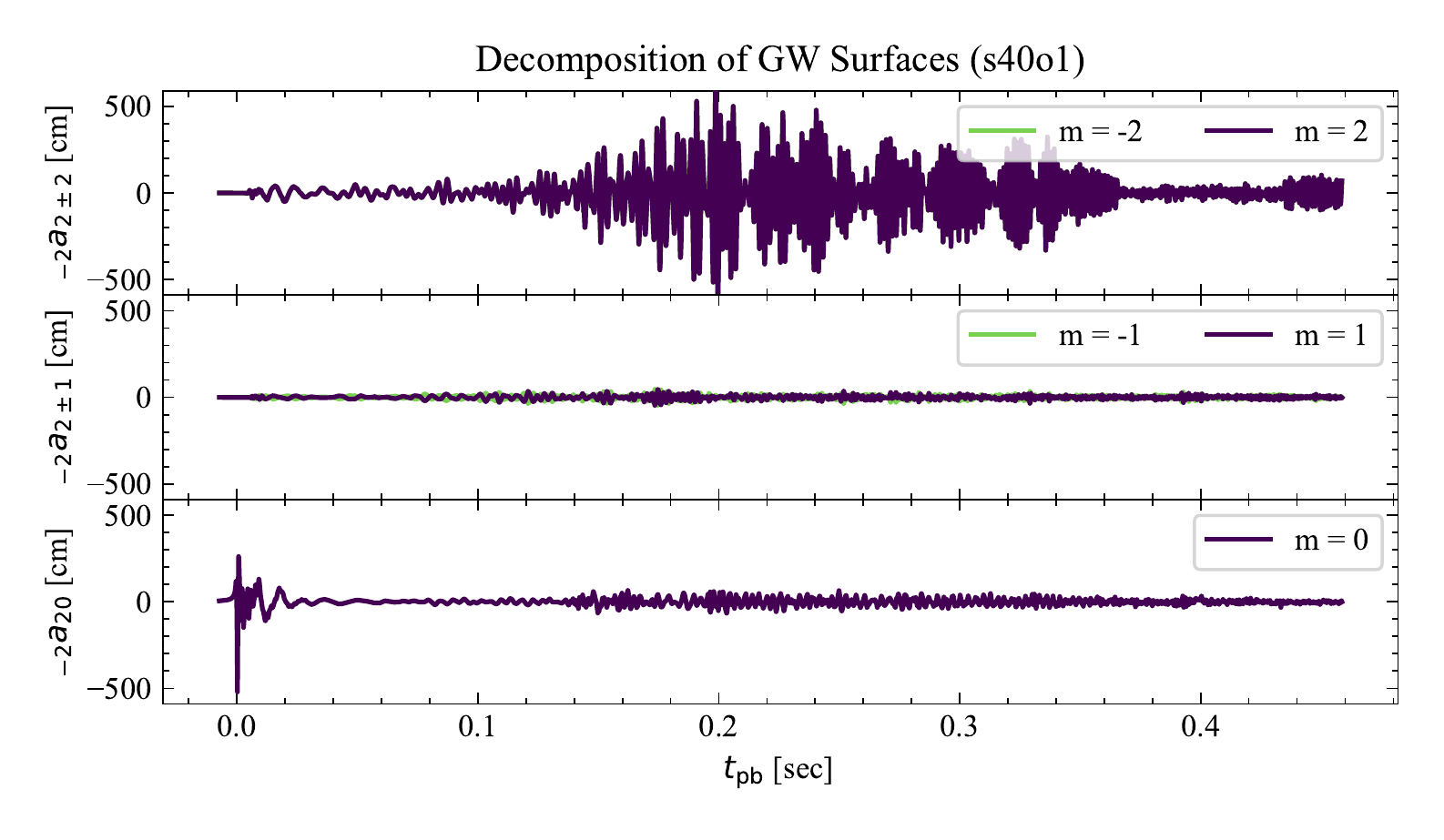}
    \caption{The real value of the decomposition of $h_+ - i h_\times$, scaled by the distance 10 kpc, into spin weighted spherical harmonics $_{-2}a_{2m}$.  Different rows correspond to different $m$ numbers.  Listed in Table \ref{table:spin_spher_harms} are the expressions for spin weighted spherical harmonics.  At early times, $_{-2}a_{20}$ shows dominant contributions to the GW surfaces.  Because of the $\sin^2\theta$ behavior, as noted in Table \ref{table:spin_spher_harms}, this behavior indicates dominant GW emission along the equator of the supernova---characteristic of the axisymmetric bounce signal.  Beyond 150 ms pb, the $m = \pm 2$ mode shows dominant contributions along the poles. }
    \label{fig:s40o1_decomp}
\end{figure*}

\begin{table*}[]
\begin{center}
\begin{tabular}{c|c|c}
 m & $_{-2}Y_{2m}$ & $_{-2}a_{2m}$\\
 \hline
 2 & $\frac{1}{8}\sqrt{\frac{5}{\pi}}(1 + \cos\theta)^2 e^{2i\phi}$ & $\sqrt{\frac{4 \pi}{5}}\frac{G}{c^4}\Big( \ddot{\mathcal{Q}}_{xx} - \ddot{\mathcal{Q}}_{yy} - 2i\ddot{\mathcal{Q}}_{xy}\Big)$ \\
 1 & $\frac{1}{4}\sqrt{\frac{5}{\pi}}\sin\theta(1 + \cos\theta) e^{i\phi}$ & $\sqrt{\frac{16 \pi}{5}}\frac{G}{c^4}\Big( -\ddot{\mathcal{Q}}_{xz} + i\ddot{\mathcal{Q}}_{yz}\Big)$ \\
 0  &  $\frac{3}{4}\sqrt{\frac{5}{6\pi}}\sin^2\theta$ & $\sqrt{\frac{32 \pi}{15}}\frac{G}{c^4}\Big( \ddot{\mathcal{Q}}_{zz} - \frac{1}{2}(\ddot{\mathcal{Q}}_{xx} + \ddot{\mathcal{Q}}_{yy})\Big)$\\
 -1  & $\frac{1}{4}\sqrt{\frac{5}{\pi}}\sin\theta(1 - \cos\theta) e^{-i\phi}$ & $\sqrt{\frac{16 \pi}{5}}\frac{G}{c^4}\Big(\ddot{\mathcal{Q}}_{xz} + i\ddot{\mathcal{Q}}_{yz}\Big)$ \\
 -2  & $\frac{1}{8}\sqrt{\frac{5}{\pi}}(1 - \cos\theta)^2 e^{-2i\phi}$ & $\sqrt{\frac{4 \pi}{5}}\frac{G}{c^4}\Big( \ddot{\mathcal{Q}}_{xx} - \ddot{\mathcal{Q}}_{yy} + 2i\ddot{\mathcal{Q}}_{xy}\Big)$\\
\end{tabular}
\end{center}
\caption{List of spin -2 weighted spherical harmonics, $_{-2}Y_{2m}$, and spin weighted spherical harmonic coefficients, $_{-2}a_{2m}$, used to construct the $_{-2}\tilde{P}_{2m}$ terms used in Figure \ref{fig:mini_power}, Figure \ref{fig:decomp_power_v23_combine}, and Figure \ref{fig:power_decomp}.   Expressions taken from \citet{ajith:2007}. }
\label{table:spin_spher_harms}
\end{table*}

\begin{figure*}
\centering
    \begin{subfigure}[t]{0.49\linewidth}
\includegraphics[width=\linewidth]{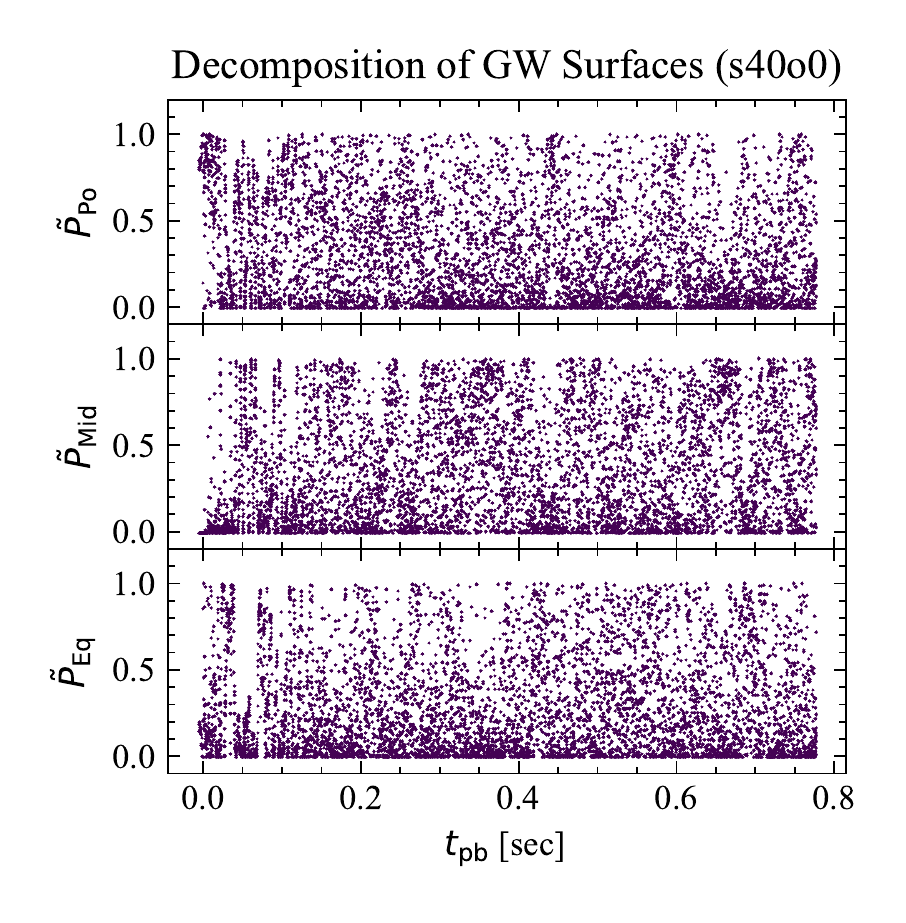} 
    \caption{By eye, model \texttt{s40o0} shows relatively equal contributions (i.e., similar density of points) from $\tilde{P}_\mathrm{Po}$ , $\tilde{P}_\mathrm{Mid}$, and $\tilde{P}_\mathrm{Eq}$; physically, contributions from all three modes correspond to more isotropic emission.}
\label{fig:decomp_power_s40o0}
    \end{subfigure}\hfill
    \begin{subfigure}[t]{0.49\linewidth}
\includegraphics[width=\linewidth]{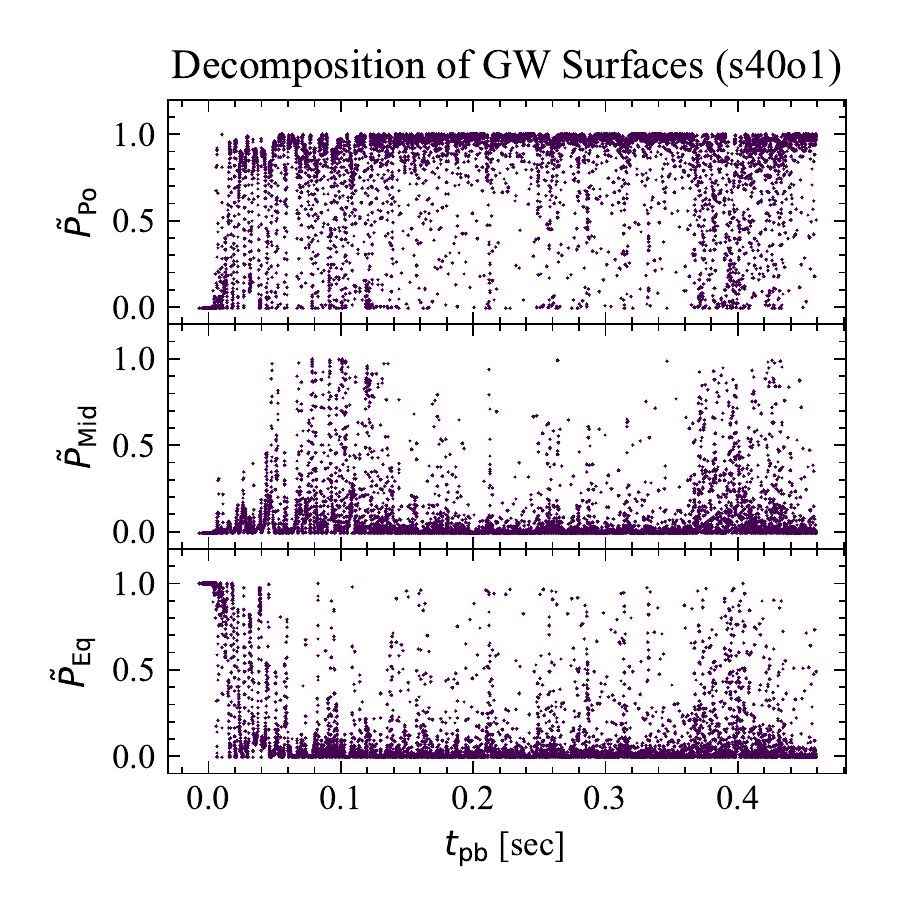}
    \caption{Compared to the top panel of Figure \ref{fig:decomp_power_s40o0}, $\tilde{P}_\mathrm{Po}$ for model \texttt{s40o1} shows a larger contribution to the overall GW signal emitted.  These increased $m = \pm2$ modes indicate increased GW amplitudes closer to the axis of rotation.}
\label{fig:decomp_power_s40o1}
    \end{subfigure}

\caption{The real value of the decomposition of $h_+ - i h_\times$ into spin weighted spherical harmonics $_{-2}a_{2m}$.  These coefficients are then squared and normalized by their sum, or $_{-2}\tilde{P}_{2|m|} = (_{-2}a_{2\pm m})^2 / \sum_i (_{-2}a_{2\pm m_i})^2$.  For convenience, we introduce, $\tilde{P}_\mathrm{Po} = _{-2}\tilde{P}_{2|2|}$, favoring emission along the poles, $\tilde{P}_\mathrm{Mid} =_{-2}\tilde{P}_{2|1|}$, favoring emission along the midlatitudes, and $\tilde{P}_\mathrm{Eq} = _{-2}\tilde{P}_{20}$, favoring emission along the equator.  Different rows correspond to different magnitudes of $m$.  Listed in Table \ref{table:spin_spher_harms} are the expressions for spin weighted spherical harmonics, $_{-2}Y_{2m}$.  Physically, the more points there are closer to a value of 1, for a given $m$, the more similar the GW distribution resembles $_{-2}Y_{2m}$.}

    \label{fig:mini_power}
    \end{figure*}


\begin{figure*}
\centering
    \begin{subfigure}[t]{0.49\linewidth}
    \centering
\includegraphics[width=\textwidth]{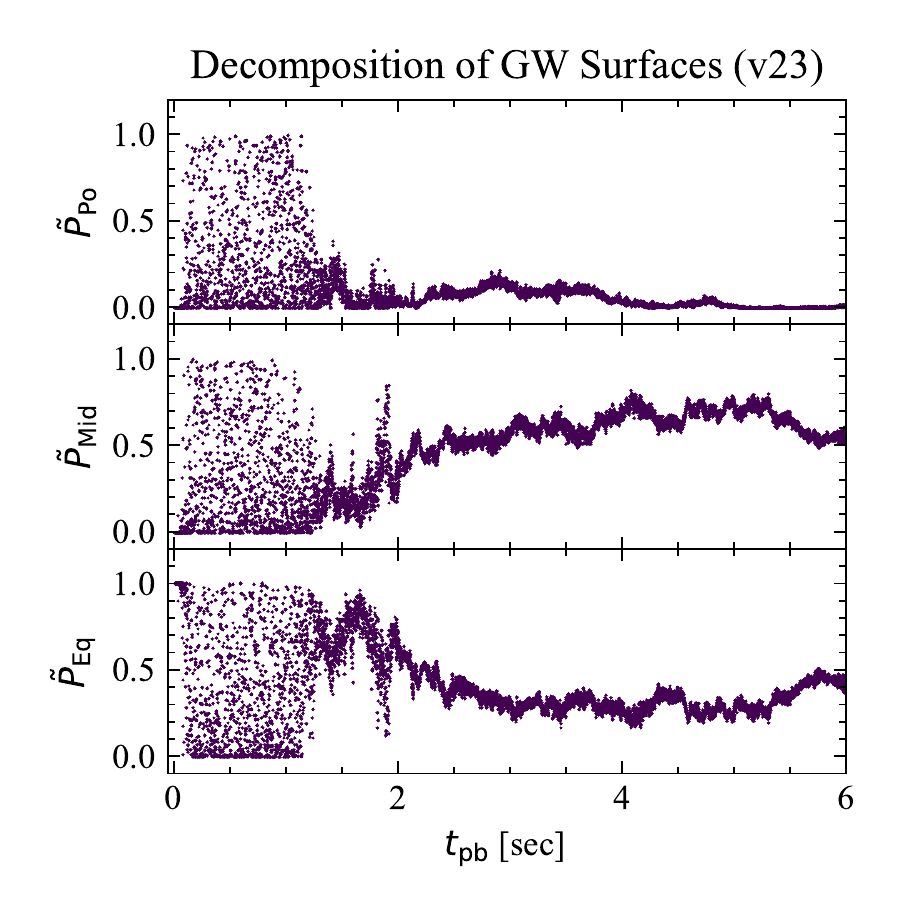}
    \caption{  While the PNS oscillations provide the stochasticity in the signal, after $t_{\mathrm{pb}}\sim 1.5$ seconds the asymmetry of the explosion ejecta dominates the GW signal, this secular drift of material provides slow changes in $\tilde{P}_\mathrm{Eq}$ and $\tilde{P}_\mathrm{Mid}$, compared to the models in Figure \ref{fig:mini_power}.  }
    \label{fig:mini_power_mem}
    \end{subfigure}\hfill
    \begin{subfigure}[t]{0.49\linewidth}
\includegraphics[width=\linewidth]{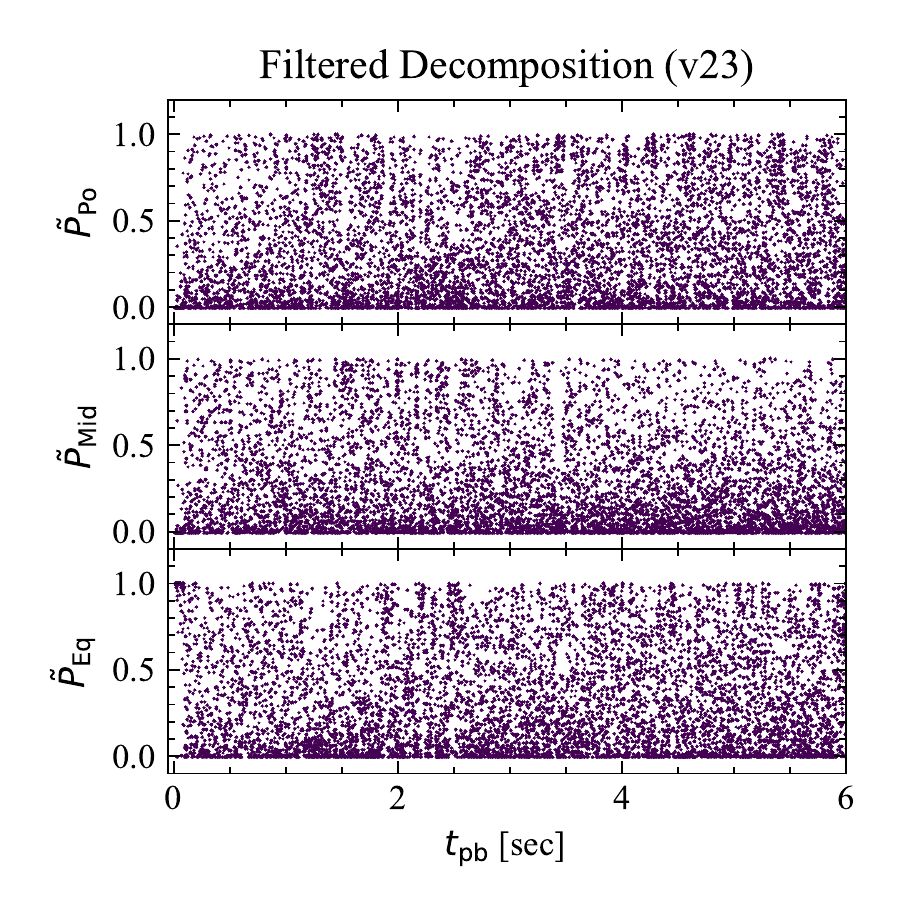}
    \caption{Similar to the left panel, however, after passing the signal through a 10 Hz high-pass filter.  As expected, the power is evenly distributed among $\tilde{P}_\mathrm{Eq}$, $\tilde{P}_\mathrm{Mid}$, and $\tilde{P}_\mathrm{Po}$ modes, similar to model \texttt{s40o0} in Figure \ref{fig:decomp_power_s40o0}.}
\label{fig:decomp_power_v23_filter}
    \end{subfigure}

\caption{Same as Figure \ref{fig:mini_power}, however for model $\texttt{v23}$.}

    \label{fig:decomp_power_v23_combine}
    \end{figure*}

\begin{figure*}
    \centering
\includegraphics[width=\textwidth]{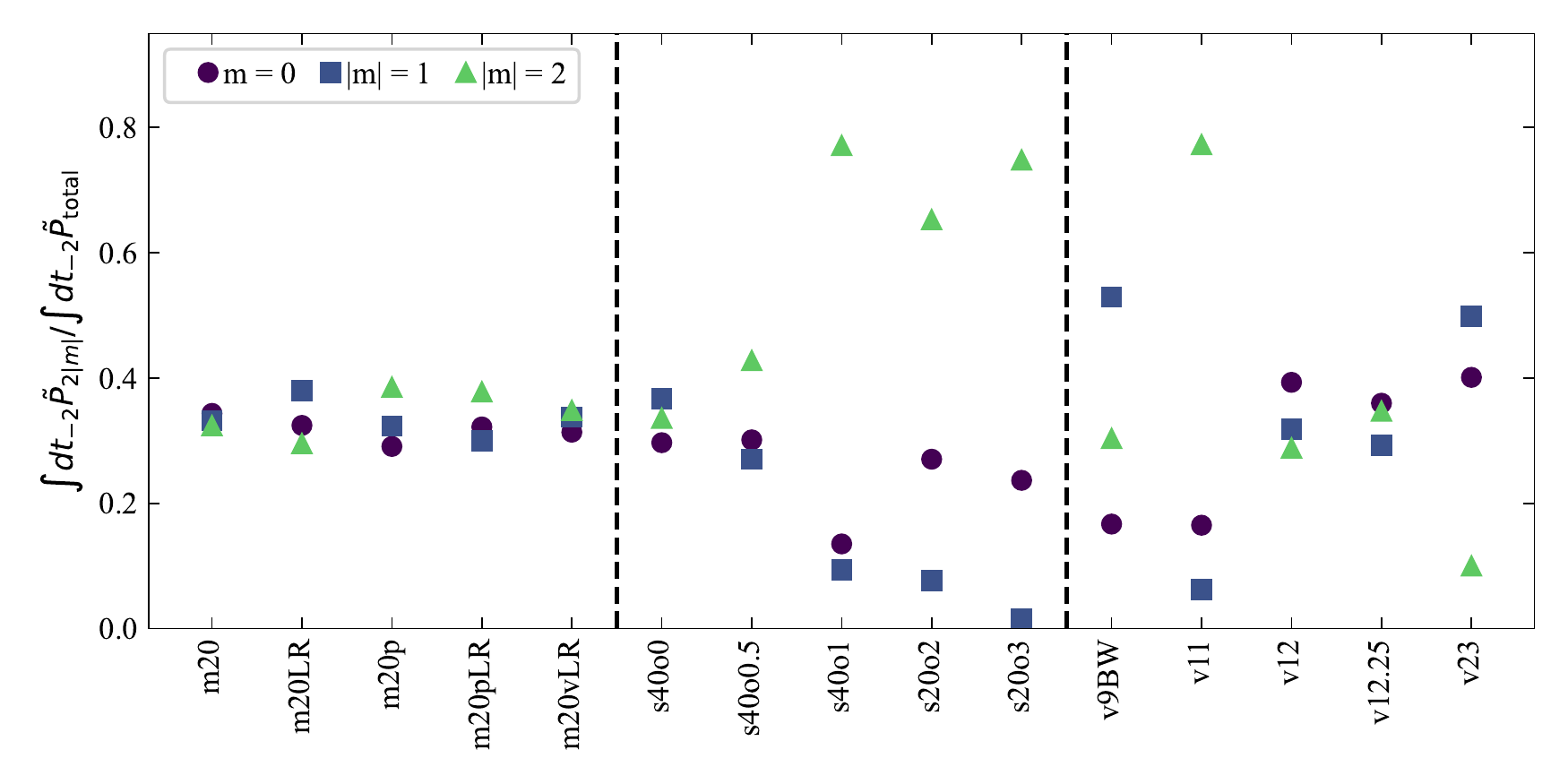}
\caption{Integrated values of $_{-2}\tilde{P}_{2|m|}$---the time series data seen in Figure \ref{fig:mini_power} and Figure \ref{fig:mini_power_mem} for all models in this study, normalized by the total integrated power over all $m$ values.  Physically, this value corresponds to what fraction of time the GW strain surface resembles a given spin weighted spherical harmonic.  Our datasets are split into thirds by the two vertical dashed lines. The left section indicates results from the \texttt{mesa} suite of models.  These nonrotating models act as control cases, displaying equal contributions among $m=0$, $m = \pm 1$, and $m = \pm 2$ modes.  Between both dashed lines are the rotating models, listed in order of initial rotation rate.  As rotation increases, there is a transition of GW amplitudes away from $|m| = 1$ modes (lower values of squares), towards $|m| = 2$ modes (increasing values of triangles).  The right section contains the models that show varying degrees of asymmetry of the supernova ejecta.  In particular, \texttt{v9BW} displays GW amplitudes resembling $_{-2}Y_{2{\pm1}}$, whereas \texttt{v11} GW distributions more strongly resemble $_{-2}Y_{2{\pm2}}$, due to unique explosion geometries.  While the remaining models show varying degrees of asymmetry, they show different contributions due to not only different morphologies, but also different relative orientations.  In summary, two main factors that influence the directionality of GW emission from CCSNe are the degree of rotation, along with explosion morphology and explosion orientation.  }
    \label{fig:power_decomp}
\end{figure*}


We now turn to the influence of polarization.  Thus far, $h_+$ strain surface plots have allowed for an intuitive description of the physical mechanisms driving the GW generation and the evolution of the preferred direction.  However, as predicted by general relativity, GW detectors will pick up a combination of $h_+$ and $h_\times$.  Thus, we take inspiration from the numerical relativity community by decomposing the GW emission into specific modes; for example, see \citet{boyle:2014}.  While the $S_3$ plots were useful for drawing connections between ellipsoid toy models, decomposing strain surface plots into spin weighted spherical harmonics grants an additional dimension of understanding.  This is because the $S_3$ plots were constructed with GW maxima, neglecting the remainder of the surface, that indeed have nonzero GW strain.  By observing $_s a_{lm}$ values, we gain a better understanding by \textit{quantifying 
 the GW strain of both polarizations over time}.

We take the surfaces for both modes by constructing $h = h_+ - i h_\times$, where $i$ is the imaginary number.  With a real and imaginary component to each GW strain surface, we now convolve these complex numbers with a complex conjugate of the spin weighted spherical harmonics, as introduced in Section \ref{ssec:spher_harm}.  A list of the spin -2 weighted spherical harmonics is found in the central column of Table \ref{table:spin_spher_harms}.  The right column of Table \ref{table:spin_spher_harms} lists the closed form expressions of $_{-2}a_{2m}$ in terms of the reduced mass quadrupole moment time derivatives, $\ddot{\mathcal{Q}}_{ij}$.
 The $_{-2}a_{20}$ mode is weighted by a factor of $\sin^2\theta$, indicating preferential GW amplitudes along the equator, with no azimuthal dependence.  For the $m = \pm 1$ modes, maxima occur for $\theta$ at $\pi / 3, 2\pi/3$ and for $\phi$ at $0, \pi/2$.  Physically, these correspond to maxima occurring between the equator and axis of rotation. 
 For $_{-2}a_{2\pm2}$, these distributions are weighted by $(1 \pm \cos\theta)^2e^{\pm 2 i \phi}$, which achieve maxima at $\theta$ set to integer multiples of $\pi$ and $\phi$ set to integer multiples of $\pi / 2$.  Physically, this behavior will track strong GW amplitudes along the z axis (or axis of rotation for rotating models).



Figure \ref{fig:s40o1_decomp} shows the spherical harmonic coefficients for model \texttt{s40o1}.  Near bounce, the $_{-2}a_{20}$ mode dominates the GW strain.  This is indicative of the strongest GW emission along the xy plane, when considering both polarizations, consistent with the visual provided in Figure \ref{fig:surface_sim_bounce}.  After a quiescent period of $\sim 120$ ms, moderate equatorial GW amplitudes arise.  By contrast, $_{-2}a_{2\pm2}$ dominate the GW signal beyond $\sim150$ ms pb.  This is due to rapid mass and angular momentum accretion, along with the presence of the low T/|W| instability, and is consistent with the columnated behavior in Figure \ref{fig:scatterC} and Figure \ref{fig:scatterD}.  The contributions between the pole and equator, from the $_{-2}a_{2\pm1}$ modes, remain subdominant.  In summary, the time scale that sets this transition is set by the transition of PNS dynamics from bounce and ringdown to excitations during the accretion phase: $\sim 150$ ms.  While comparing $_{-2}a_{2m}$s is illustrative for a single model, as different models emit differing amounts of GWs, a cross model comparison becomes difficult.  As a remedy, we appeal to the power emitted in each mode.

We define the power in each mode $m$ as the square of the real component of the spin weighted spherical harmonic coefficient, $_{-2}P_{l|m|} = (_{-2}a_{l\pm m})^2$.  Note, in this context, we use the term `power' \textit{not} to represent the time rate of change of GW energy, but the relative contributions of each $_{-2}a_{lm}$ to the strain surface.  In order to compare relative amplitudes between modes, we use a normalization factor $P_\mathrm{norm} = \sum_i (_{-2}a_{lm_i})^2$, which represents the power in all modes.  Normalizing each power yields $_{-2}\tilde{P}_{lm} = (_{-2}a_{lm})^2 / P_\mathrm{norm}$.  As we are concerned with directionality, we group the power by the magnitude of $m$ values $|m|$; explicitly, $_{-2}\tilde{P}_{l|m|} = _{-2}\tilde{P}_{lm} + _{-2}\tilde{P}_{l(-m)}$.  For convenience, we introduce shorthand symbols, $\tilde{P}_\mathrm{Po} = _{-2}\tilde{P}_{2|2|}$ which describes what fraction of the strain surface resembles $_{-2}Y_{2\pm 2}$, favoring emission along the poles.  $\tilde{P}_\mathrm{Mid} =_{-2}\tilde{P}_{2|1|}$ describes what fraction of the strain surface resembles $_{-2}Y_{2\pm 1}$, emission between the pole and equator---the midlatitudes.  $\tilde{P}_\mathrm{Eq} = _{-2}\tilde{P}_{20}$ describes what fraction of the strain surface resembles $_{-2}Y_{20}$, emission along the equator.  Armed with these quantities, we turn to Figure \ref{fig:mini_power}.  Each row corresponds to $_{-2}\tilde{P}_{l|m|}$ for a given $m$.  The purpose of these figures is to gain a quick intuition to identify in which modes the power resides.  For \texttt{s40o0}, Figure \ref{fig:decomp_power_s40o0} shows relatively equal contributions for $m = 0$, $|m| = 1$, and $|m| = 2$ modes.  By eye, there is a similar density of points near a value of 1 for all three rows.  By contrast, for model \texttt{s40o1} in Figure \ref{fig:decomp_power_s40o1}, more interesting features emerge.  In the bottom row, $\tilde{P}_\mathrm{Eq}$ shows a cluster of points near 1 for the first 20 ms.  This is indicative of the bounce signal and ringdown.  Moving attention to the middle row, $\tilde{P}_\mathrm{Mid}$ shows a distinct lack of power from $\sim 150$ ms to $\sim 350$ ms.  Simultaneously in the top panel, there is a surplus of power in the $\tilde{P}_\mathrm{Po}$ modes---rotation has facilitated a transition of power from the $|m| = 1$ modes to the $|m| = 2$ modes.  

Figure \ref{fig:mini_power_mem} shows the same quantities, for \texttt{v23}.  For the first $\sim 1.5$ seconds, similar to the nonrotating \texttt{s40o0}, the accretion phase exhibits relatively equal emission in the $m = 0$, $m = \pm1$, and $m = \pm2$ modes.  However, as the main source of GW emission transitions from PNS oscillations to explosion ejecta, the modes settle into a secular drift because the matter source of the GW amplitude is now dominated by expanding matter, rather than nearly chaotic PNS oscillations.  Noting the frequency sensitivity of current-age GW detectors to frequencies $\gtrsim 10$ Hz, in Figure \ref{fig:decomp_power_v23_filter}, we generate the same plot, after passing the $\ddot{\mathcal{Q}}_{ij}$ time series data through a 10 Hz high pass filter (specifically, a  tenth order \texttt{Butter} filter using \texttt{SciPy.signal} \citep{jones:2001}.  We note a similar distribution of points to the nonrotating model \texttt{s40o0} seen in Figure \ref{fig:decomp_power_s40o0}.  This behavior is expected as there is no coherent matter motion dominating the dynamics of the PNS, leaving a relatively stochastic signal.  The difference in the directionality of the GW signal between the filtered and unfiltered \texttt{v23} signal emphasizes that future space-based GW observatories, sensitive to signals of order $\lesssim 10$ Hz, will be subject to explosion geometry-dependent directional factors.

Lastly, we address how these transitions compound over the entire supernova evolution. 
 After calculating $_{-2}\tilde{P}_{2|m|}$ for all models over their entire duration, we integrate these quantities in time: $\int dt _{-2}\tilde{P}_{2|m|}$.  To obtain a normalization factor, we add all integrated powers together $\int dt _{-2}\tilde{P}_{\mathrm{total}} = \sum_i\int dt _{-2}\tilde{P}_{2|m_i|}$.  The quantity, $\int dt _{-2}\tilde{P}_{2|m|}/\int dt _{-2}\tilde{P}_{\mathrm{total}}$ shows the relative contributions of each $|m|$ mode over the entire CCSN evolution.  For all models, this quantity is plotted in Figure \ref{fig:power_decomp}.  The datasets are grouped into thirds, separated by two vertical dashed lines.  The leftmost section indicates the \texttt{mesa} models.  These models act as a control case to establish behavior for nonrotating, failed supernovae.  Among all models, the $m = 0$, $|m| = 1$, and $|m| = 2$ modes contain equal contributions $\sim 1/3$.

 Between both sets of dashed lines are models \texttt{s40o[0,0.5,1]} and \texttt{s20o[2,3]}. 
 The left-most model is nonrotating, and the rightmost model has the most rapid rotation. The nonrotating \texttt{s40o0} behaves similar to the \texttt{mesa} models, as expected.  With moderate rotation, \texttt{s40o0.5} shows a slight decrease in the $|m| = 1$ modes, and slight increase in the $|m| = 2$ modes.  Continuing rightward, the models with increasingly rapid initial rotation exhibit more power in the $|m| = 2$ mode and less in the $|m| = 1$ modes.  We note relative strength of the $m = 0$ remains around $\sim 1/3$, refining our physical picture.  While rotation helps create stronger amplitudes along the axis of rotation, the contributions of GWs along the equator still remain important.

To the right of both dashed lines contains models \texttt{v[9BW, 11, 12, 12.25, 23]}.  Models \texttt{v12}, \texttt{v12.25} show similar behavior to the \texttt{mesa} suite as well because they do not display drastic memory effects in the GW signal (see lower panel of Figure (\ref{fig:GW_costheta_evolution_vartanyan})).  However, models \texttt{v9BW}, \texttt{v11}, \texttt{v23} show a different picture.  Model \texttt{v9BW} exhibits most fractional power in the $|m| = 1$ modes and roughly half as much in the $m = 0$ and $|m| = 2$ modes.  This inversion is due to the effect of the direct offset GW signal.  In particular, it exhibits higher values of $\tilde{P}_\mathrm{Mid}$ for the evolution, particularly after $\sim 400$ ms pb.  The physical cause of the imprint is the explosion ejecta geometry favoring GW emission in a direction between the equator and pole.  By contrast, model \texttt{v11} shows the large majority of its power in the $|m| = 2$ modes, with slightly more relative power in the $|m| = 0$ mode, compared to $|m| = 1$.  Model \texttt{v23} shows yet another unique ordering, with most power being deposited in the $|m| = 1$ modes, then $m = 0$, and least in $|m| = 2$.  Due to the spread in $\int dt _{-2}\tilde{P}_{2|m|}/\int dt _{-2}\tilde{P}_{\mathrm{total}}$, this implies, in general, for each exploding supernova in Nature, these relative contributions will change, depending on the explosion outcome and geometry.  

For this section, we summarize: two important factors when considering the directionality of GWs---for both polarizations---are the degree of rotation and the explosion geometry.  Rotation has a tendency to transition stronger GW amplitudes along the pole.  The directionality of GW amplitudes for matter contributions to direct offset signals is dependent on the explosion morphology and orientation.

As a final note, we emphasize the potential for spin weighted spherical harmonics to analyze GW directionality data for future studies.  In the binary merger community, instead of multiple TDWFs at different viewing angles, $_{-2}a_{lm}$ coefficients are stored \citep{boyle:2019}.  Then, when a waveform at a specific viewing angle is needed, the coefficients can be used as weights for the $_{-2}Y_{lm}$ expressions, for example given in Table \ref{table:spin_spher_harms}; they are summed together, and a TDWF is generated.  In practice, for CCSN models that use the quadrupole formula for generating GW data, sharing the time derivative(s) of the mass quadrupole moments would provide the most precise angular GW information because Equation (\ref{eq:h+}) and Equation (\ref{eq:hx}) can be applied without information loss during the integral over angles for the $_{-2}a_{lm}$ calculation.  However, the value of tracking $_{-2}a_{lm}$s---or calculating them from quadrupole data---is that they provide a metric to quantify the directionality of the GW distribution, something the quadrupole data cannot concisely provide.  For those interested in performing their own decomposition of GW surfaces, we point them to the right column of Table \ref{table:spin_spher_harms}, which contains the complex valued, closed form $_{-2}a_{2m}$ expressions \citep{ajith:2007}. For this work, we are concerned with general angular distributions of GWs, that is, GW amplitudes along the equator, midlatitudes, or pole, for which $l=2$ modes are sufficient. 
 While an interesting question, investigating the contributions from $l > 2$ modes to the CCSN GW signal (in particular, for codes that directly evolve spacetime) is beyond the scope of this work and will be saved for future studies.


\section{Discussion and Summary}

\subsection{Relating GW Directionality to Specific Phases of Fluid Evolution}
\label{sec:physicality}


There are a variety of physical mechanisms responsible for driving the PNS to produce GWs.  \citet{vartanyan:2023} mention the presence of a prompt convective GW signal present in $\sim 50$ ms following bounce.  The authors establish the strain during this phase is only present in $h_+$.  After examining \texttt{v11}, we note the distribution of $h_+$ is concentrated in the equator, or strongly favors the $m = 0$ spin weighted spherical harmonic.  This behavior is indicative of nearly axisymmetric dynamics. 
 While the negative entropy gradient near the stalled shock, just after bounce, is responsible for generating this prompt convection, the GW distribution appears similar to the strain surface plot at bounce in Figure \ref{fig:surface_sim_bounce}.   In the bottom panel of Figure \ref{fig:GW_costheta_evolution_vartanyan}, between 0.05 and 0.1 sec, we note values for  $\langle \cos^2 \theta \rangle_\mathrm{10\, ms} < 0.5$ for most of the interim, corroborating these observations.  

\citet{pan:2021} also note several potential factors such as the SASI and the low T/|W| instability.  In particular, it has been shown that the spiral modes of the SASI can redistribute angular momentum, causing the PNS to spin up in an initially nonrotating progenitor \citep{guilet:2014,pan:2021}.  As the GW directionality for the \texttt{mesa} models with and without SASI were explored, with no clear preferred direction, we do not observe dominant GW viewing angle evolution associated with this hydrodynamic instability.  This can be expected as \citet{oconnor:2018b} note models exhibiting SASI, \texttt{m[20, 20LR, 20p, 20pLR]}, only yield slow rotating, $\sim1$ sec period PNSs from SASI spin-up.  For the initially nonrotating model \texttt{s40o0}, \citet{pan:2021} note a PNS forming with spin period $\sim 0.1$ sec.  We do not see preferred GW directionality with this model either (see bottom panel of Figure \ref{fig:GW_costheta_evolution} and Figure \ref{fig:power_decomp}).  We conclude it is not the mere presence of a rotating PNS, but the accretion of significant amounts of angular momentum, which assist in directing stronger GW amplitudes along the axis of rotation.  

In the fast rotating model, \citet{pan:2021} note the possible emergence of the low T/|W| instability, based on low frequency GW emission beginning $\sim 150$ ms pb.  As the fast rotating model exhibits a preferred direction of GW emission along the axis of rotation and coherent paths in the evolution of the preferred direction, it is possible this instability contributes.  However, the presence of the low T/|W| instability is not requisite for these two behaviors as the slow rotating model---which shows no sign of low T/|W|---also remains slightly columnated and exhibits clear paths in Figure \ref{fig:scatterB}.

In relation to the \texttt{mesa} models, we do not notice any coherent paths in Figure \ref{fig:GW_costheta_evolution_mesa} for preferred GW directions during instabilities like the SASI or Lepton number Self-sustained Asymmetry (LESA).

\begin{figure*}[t!]
    \centering
\includegraphics[width=\textwidth]{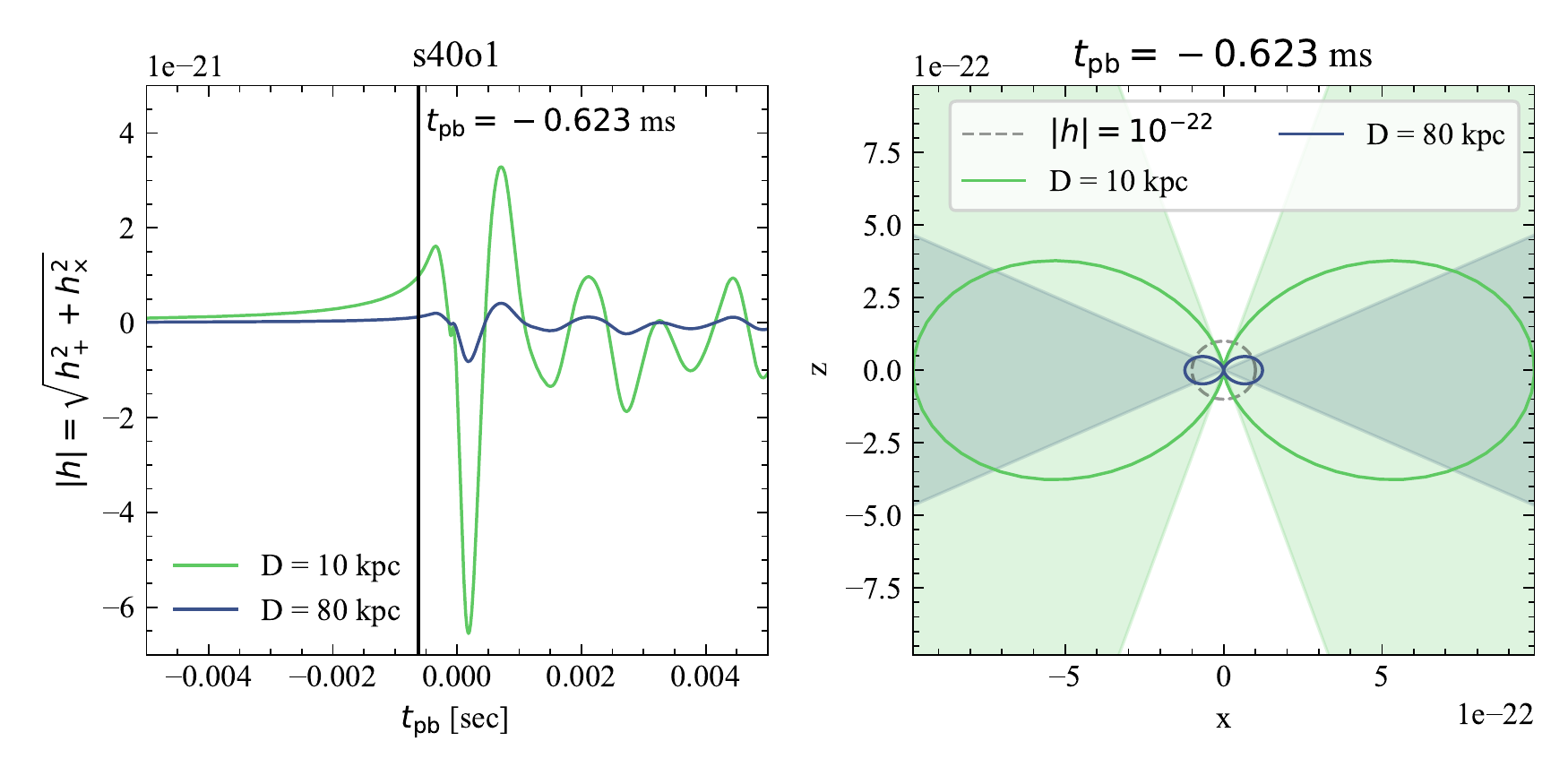}
    \caption{(Left) TDWF for model \texttt{s40o1 } when viewed along the equator of the CCSN.  Light green color is for a source distance of 10 kpc and dark blue is for a source distance of 80 kpc.  The vertical line marks $t_\mathrm{pb} = -0.623$ ms, when the strain for the 10 kpc source reaches $10^{-21}$.  (Right) Since the bounce creates an axisymmetric GW distribution, we show a slice through a strain surface plot---the x-z plane.  The distance from the origin to any point in this plane is the GW strain along that viewing angle.  The grey dashed circle shows a contour for $|h| = \sqrt{h_+^2 + h_\times^2} = 10^{-22}$.  The light green, solid, curved line is the strain surface at $t_\mathrm{pb} = -0.623$ ms for the 10 kpc source distance.  The dark blue line is for the 80 kpc source distance.  Shaded areas show the viewing angles where $|h| > 10^{-22}$ for the two source distances.  As the CCSN source distance increases, the opening angle for detectable GW strains decreases; the darker blue `detectable' region traces a smaller angle ($\sim 53^\circ$) than the lighter green `detectable' region ($\sim 143^\circ$).} 
    \label{fig:open_angle}
\end{figure*}

\subsection{Implications for Observability}
\label{sec:observability}

These directional dependencies help prepare for the first direct detection of GWs from CCSNe. As detector sensitivities improve and signal search algorithms become more advanced, identifying which components of the signal can be reconstructed is vital.  Importantly, the observability of a GW depends on not only the physical mechanism generating it but also the angle at which it is observed.  For a rotating supernova, the bounce signal is most visible when viewed along the supernova equator.  However, potentially stronger GW amplitudes along the axis of rotation may be missed during the accretion phase.  By contrast, if viewed along the axis of rotation, the broadband bounce signal will not be detected, but high frequency oscillations from the PNS may have stronger amplitudes.  For a randomly oriented supernova, the viewing angle will likely be off-axis, so GW detection algorithms should search for both components of the signal. 

During the accretion phase, the GW emission remains roughly isotropic for nonrotating or slowly rotating cases---i.e., the majority of expected supernovae.  For rarer, rapid rotators, GW amplitudes along the pole can be stronger, by nearly a factor of two, compared to the equator.  When considering rapid rotators, a balance must be considered.  For the bounce phase, a well templated GW strain can be detected, with relatively weaker GW amplitudes during the accretion phase when viewed along the equator.  Conversely, no well templated bounce signal would be detected when viewed along the pole but stronger accretion phase GW amplitudes would arise.  This sensitive interplay as to which direction gives the best chance for detection depends on the orientation, source distance, supernova dynamics, and detection algorithm \citep{szczepanczyk:2022}.

Lastly, for successful explosions, the explosion geometry will ultimately dictate along which direction a maximum GW amplitude is emitted.  Likewise, it will produce certain directions along which GW minima drop below detector sensitivity levels.  Since the explosion ejecta morphology relies upon the interaction between initial progenitor convective perturbations, magnetized fluid evolution, and neutrino transport, we expect this effect to vary as widely as the progenitors themselves.  While our work examined the influence of matter contributions to the GW signal, asymmetric neutrino production would also provide measurable direct offset signals, which also vary by viewing angle \citep{vartanyan:2023}.  In this context, an exciting possibility exists for multiwavelength astronomy with the future space-based GW observatories, such as the Laser Interferometer Space Antenna (LISA) \citep{amaro-seoane:2017}, to use high frequency GW observations with the LIGO-Virgo-KAGRA network \citep{abbott:2018}---either during the bounce or accretion phase---as a warning sign for LISA to expect a low frequency counterpart in the following $\sim$second.  

Contributions to the GW background are also affected by consequences of directionality.  For detectors attempting to quantify the spectrum of the GW background \citep{christensen:2019}, this work offers motivation to consider CCSN population studies.  While supernova GW signals produce weaker amplitude GWs compared to binary mergers \citep{finkel:2022}, theoretical works could construct populations of randomly oriented and distributed CCSNe, with varying rotation rates and explosion fates, to better quantify which components of the signal may be detectable for GWs from a superposition of sources.

\subsection{Applying Directionality Evolution to the Bounce Phase}
\label{sec:open_angle}

To provide a concrete example considering the observability of GW amplitudes, we consider the bounce phase of model $\texttt{s40o1}$.  The left panel of Figure \ref{fig:open_angle} shows the GW strain $|h| = \sqrt{h_+^2 + h_\times^2}$ during the bounce.  The light green line is for a CCSN 10 kpc from Earth, and the dark blue line is for a CCSN distance of 80 kpc.  To produce a strain surface plot, we select a time when the light green line produces $|h|\sim 10^{-21}$, a GW amplitude above which most of the bounce signal is present. 

In the right panel of Figure \ref{fig:open_angle}, we show a 2D strain surface plot at $t_\mathrm{pb} \sim -0.6$ ms.  The light green line is the strain surface for a source distance of 10 kpc, and the blue line is for 80 kpc.  Suppose a GW detector needed a strain of $|h|\sim 10^{-22}$ to robustly reconstruct a bounce signal.  Then, the colored lines greater than $10^{-22}$ away from the origin (outside the grey dashed circle) would represent viewing angles along which the GW amplitude is `detectable'.  For ease, we have marked the shaded regions as the viewing angles along which the signal would be detectable.  The opening angle of detectability is $\Delta \theta^\mathrm{10 kpc}_\mathrm{detect} \sim 142^\circ$ for the 10 kpc (light green) case and $\Delta \theta^\mathrm{80 kpc}_\mathrm{detect} \sim 53^\circ$ for the 80 kpc (dark blue) case.  An opening angle of $\Delta \theta^\mathrm{10 kpc}_\mathrm{detect} \sim 142^\circ$ corresponds to a solid angle of $\Delta\Omega^\mathrm{10 kpc}_\mathrm{detect} = 11.9$ steradians.  An opening angle of $\Delta \theta^\mathrm{80 kpc}_\mathrm{detect} \sim 53^\circ$ corresponds to a solid angle of $\Delta\Omega^\mathrm{80 kpc}_\mathrm{detect} \sim 5.6$ steradians.  Armed with these numbers, consider a CCSN event at 10 kpc and 80 kpc.  Assuming a random orientation for each, the likelihood the Earth falls within the GW detectability opening angle is 94\% ($\Delta\Omega^\mathrm{10 kpc}_\mathrm{detect} / 4\pi$) for the 10 kpc case.  When increasing the source distance out to 80 kpc, this chance drops to 42\% ($\Delta\Omega^\mathrm{80 kpc}_\mathrm{detect} / 4\pi$).

We emphasize these are not actual detection probabilities that consider signal reconstructions and frequency sensitivities.  This is a simplified, worked example to show how the directionality work here can be used to better understand how viewing angle affects the likelihood of detecting CCSN GWs.  We leave more robust population estimates that use multiple sources, consider different GW distributions beyond the bounce phase, and consider signal reconstruction algorithms for future work.

\subsection{A Note on Signal Injection}
\label{sec:injection}

Because of the aforementioned directional considerations, we encourage the observational community to continue using gravitational waveforms from 3D studies injected at a variety of random angles \citep[e.g.,][]{szczepanczyk:2023}, rather than injecting an angle averaged amplitude.  As we have shown, depending on the source characteristics and supernova phase, the expected GW amplitudes may not be isotropic over a given phase.  An angle averaged GW amplitude may be appropriate for slow/non rotating CCSNe, during the accretion phase, as the emission is roughly isotropic over timescales of hundreds of milliseconds.  The bounce amplitude, by contrast, varies with $\sin^2\theta_\mathrm{pole}$, where $\theta_\mathrm{pole}$ is the angle between the pole, or axis of rotation, and the viewing angle.  Consider the 80 kpc example from the previous Subsection \ref{sec:open_angle}.  Sufficiently strong GW amplitudes $\gtrsim 10^{-22}$ can be detected for $\sim 42\%$ of the viewing angles. 
 However, if an angle averaged GW amplitude, $h_\mathrm{avg}$, were used, $h_\mathrm{avg} \sim \int_0^\pi 1.25\times10^{-22} \sin^2\theta d\theta / \int_0^\pi d\theta = 6.25\times 10^{-23}$, this source would fall below the $10^{-22}$ threshold for all viewing angles, yielding a $0\%$ probability of detection.  Similar examples with closed form expressions for GW amplitudes, like rapidly rotating models during the accretion phase---Equation (\ref{eq:h+_rot}) and Equation (\ref{eq:hx_rot})---or late time exploding models (for next generation, low frequency detectors)---see \citet{richardson:2022}---emphasize the importance of using a distribution of GW signal injections, rather than injecting an angle averaged quantity.  In particular, they give more proper estimates of GW detectability for asymmetric explosions, like those seen in \citet{vartanyan:2023} and rapidly rotating models, like those seen in \citet{powell:2020} and \citet{powell:2023}.

\subsection{Summary}
\label{sec:summary}

This work has considered a new factor regarding the observability of GWs from CCSNe, particularly the impact of viewing angle.  In particular, this work serves as a bridge between the theory and observation communities: explaining how the source physics influences viewing angle effects on CCSN GWs and quantifying the time evolution of preferred directions of GW emission.  Here, we review the findings from our work: 

\begin{itemize}
    \item strain surface plots can characterize the nature of the mechanism generating GWs at a given point in time, Figure \ref{fig:GWsurfaces},
    \item viewing the collective distribution of preferred directions for GW emission shows CCSNe do not have a single `optimal' viewing angle, Figure \ref{fig:tri3Dscatter}, 
    \item in particular,  when considering all viewing angles, we quantitatively show nonrotating or slowly rotating cases exhibit roughly isotropic GW amplitudes throughout their accretion phase, whose prominent direction of emission transitions on the dynamical timescale of the PNS, Figure \ref{fig:GW_costheta_evolution_mesa},
    \item depending on the degree of rotation, when considering all viewing angles, we quantitatively show rotating CCSNe can exhibit strong GW amplitudes along the equator, then along the axis of rotation, with a transition timescale set by the time between ringdown and the accretion phase (order tens of ms), Figure \ref{fig:GW_costheta_evolution}, Figure \ref{fig:s40o1_decomp}, and Figure \ref{fig:mini_power},
    \item the dominant viewing angle of GW emission in rotating CCSNe may follow coherent paths that precess around the axis of rotation, Figure \ref{fig:scatterC},
    \item while PNSs are not solid-body ellipsoids, simplified analytic models, motivated by PNS dynamics, can describe each phase of GW generation (see \citet{richardson:2022} for memory effects); if the geometry of a simple PNS deformation---or other ellipsoid-like astrophysical scenarios---can be predicted, a similar toy model can provide an estimate of the resulting GW directionality, Appendix \ref{appendix:rotation_matrices},
    \item when considering spin weighted spherical harmonics (Table \ref{table:spin_spher_harms}), rotation facilitates a transition of GW strain surfaces (e.g., Figure \ref{fig:surface_sim_ringdown}) from $m = \pm 1$ to $m = \pm 2$ modes, or focusing stronger GW amplitudes closer to the axis of rotation, Figure \ref{fig:mini_power},
    \item two major factors influencing the directionality of $h_+$ and $h_\times$ from matter sources are the degree of rotation and geometry of the explosion ejecta, Figure \ref{fig:power_decomp},
    \item  these two factors can inform future population studies that include CCSN contributions to the GW background,
    \item we recommend future CCSN GW theory investigations to  report both $h_+$ and $h_\times$, as the distribution $\sqrt{h_+^2 + h_\times^2}$ will remain invariant for different detectors, whereas the individual distributions of $h_+$ and $h_\times$ will change for different sites.  For example, a GW bounce signal with mostly $h_+$ at one detector may have significant $h_\times$ components at a detector oriented differently,
    \item we note that the use of angle averaged GW strains can affect detectability estimates for rapidly rotating and asymmetric exploding models; injecting TDWFs at a variety of source angles is recommended.
\end{itemize}

By using high fidelity 3D simulations of CCSNe from multiple code bases, we have reviewed a new method towards visualizing, and consequently analyzing, GW emission from CCSNe.  While this study uses CCSNe, in general, any GW source can make use of strain surface plots.  For those who have $\ddot{\mathcal{Q}}_{ij}$ from their own simulations, and wish to decompose the GW strain surfaces into $_{-2}a_{2m}$ (to construct similar figures above), we recommend using the closed form expressions for spin weighted spherical harmonics specified in the right column of Table \ref{table:spin_spher_harms}.

The strength of this work lies in the diversity of computational grids, initial progenitors, EOSs, rotation rates, neutrino treatments, initial perturbations, and unique explosion morphologies.  Nevertheless, we acknowledge there is room for improvement.  One major component is the influence of magnetic fields.  These fields facilitate different fluid dynamics and provide another avenue for angular momentum transport away from the PNS (e.g., \citet{powell:2023}).  Furthermore, including fully dynamical spacetimes would allow us to follow the CCSN evolution closer to, and beyond, black hole formation.  While many of these models use sophisticated neutrino physics---M1---accounting for effects such as transitions between neutrino flavors---and associated flavor instabilities---remains an important factor not yet incorporated into 3D CCSN simulations. 

Acknowledging these caveats, this study offers another observable factor to consider when generating gravitational waveforms from CCSNe.  By considering both theoretical factors like CCSN source physics, alongside observational factors like the impact of the viewing angle and different detector sites, the GW community will be better prepared for the first direct detection of GWs from stellar explosions.

\acknowledgements

We thank Saul Teukolsky, Nils Vu, Mike Boyle, Michele Zanolin, Brian O'Shea, and Tony Piro for helpful discussions. 
 M.A.P. was supported in part by the Sherman Fairchild Foundation, NSF grant PHY-2309231, OAC-2209655 at Caltech, and a Michigan State University Distinguished Fellowship.  KCP is supported under the NSTC grant NSTC 111-2112-M-007-037. The FLASH-IDSA simulations are performed by the {\tt Taiwania} supercomputer in the National Center for High-Performance Computing (NCHC) in Taiwan. 
 N.D. was supported in part by the Sherman Fairchild Foundation and by NSF Grants PHY-2011961, PHY-2011968, and OAC-2209655 at Caltech.  D.V. acknowledges support from the NASA Hubble Fellowship Program grant HST-HF2-51520.
S.M.C. is supported by the U.S. Department of Energy, Office of Science, Office of Nuclear Physics,
under award Nos. DE-SC0015904 and DE-SC0017955.
This research was supported by the Exascale Computing Project (17-SC-20-SC), a collaborative effort of two U.S. Department of Energy organizations (Office of Science and the National Nuclear Security Administration) that are responsible for the planning and preparation of a capable exascale ecosystem, including software, applications, hardware, advanced system engineering, and early testbed platforms, in support of the nation's exascale computing imperative.
The software used in this
work was in part developed by the DOE NNSA-ASC OASCR Flash Center at
the University of Chicago.  

    \software{FLASH (see footnote 7) \citep{fryxell:2000,fryxell:2010}, Matplotlib\footnote[8]{\url{https://matplotlib.org/}} \citep{hunter:2007},
    NuLib\footnote[9]{\url{http://www.nulib.org}}
    \citep{oconnor:2015},
    NumPy\footnote[10]{\url{http://www.numpy.org/}} \citep{vanderwalt:2011}, SciPy\footnote[11]{\url{https://www.scipy.org/}} \citep{jones:2001}}.


\bibliography{ms}

\appendix

\section{General Expression for GW Strain from Tilted, Rotating, Oscillating Ellipsoid}
\label{appendix:rotation_matrices}

Begin by observing the mass quadrupole moment for a uniform density ellipsoid with radii of $a$, $b$, and $c$, along the respective x, y, and z axes \citep{creighton:2011,ferrari:2019},
\begin{equation}
    I_{ij} = \int x^i x^j \rho d^3x,
\end{equation}
\begin{equation}
    I_{ij} = 
\begin{pmatrix}
I_{xx} & 0 & 0 \\
0 & I_{yy} & 0 \\
0 & 0 & I_{zz}
\end{pmatrix}
= \frac{1}{5}M 
\begin{pmatrix}
a^2 & 0 & 0 \\
0 & b^2 & 0 \\
0 & 0 & c^2
\end{pmatrix}.
\label{eqn:ellipsoid_quad_mmnt}
\end{equation}
Since Equation (\ref{eqn:ellipsoid_quad_mmnt}) has a closed form, the time derivatives yield simple expressions,
\begin{equation}
    \dot{I}_{ij} 
= \frac{1}{5}\dot{M}
\begin{pmatrix}
a^2 & 0 & 0 \\
0 & b^2 & 0 \\
0 & 0 & c^2
\end{pmatrix} + 
\frac{2}{5}M
\begin{pmatrix}
a\dot{a} & 0 & 0 \\
0 & b\dot{b} & 0 \\
0 & 0 & c\dot{c}
\end{pmatrix}.
\end{equation}
A second time derivative yields
\begin{equation}
    \Ddot{I}_{ij} 
= 
\frac{1}{5}\ddot{M} 
\begin{pmatrix}
a^2 & 0 & 0 \\
0 & b^2 & 0 \\
0 & 0 & c^2
\end{pmatrix} +
\frac{4}{5}\dot{M}
\begin{pmatrix}
a\dot{a} & 0 & 0 \\
0 & b\dot{b} & 0 \\
0 & 0 & c\dot{c}
\end{pmatrix} + 
\frac{2}{5}M
\begin{pmatrix}
\dot{a}^2 + a\Ddot{a} & 0 & 0 \\
0 & \dot{b}^2 + b\Ddot{b} & 0 \\
0 & 0 & \dot{c}^2 + c\Ddot{c}
\end{pmatrix}.
\label{eqn:ddotI_toy}
\end{equation}

In partial generality, we describe a toy system in which a tilted ellipsoid changes size and `wobbles' around the axis of rotation---in our case, we select the z axis.  That is, in general, the axis of rotation does not align with any of the three axes of symmetry, and the ellipsoid radii are not necessarily constant in time.  We also neglect mass changes, ignoring the $\ddot{M}$ and $\dot{M}$ terms from above.  To account for a tilted ellipsoid, similar to \citet{creighton:2011}, we apply a rotation matrix $\textbf{R}$ to the quadrupole moment matrix $\textbf{I}$ about the x axis, by a tilt angle $\tau$
\begin{align}
    \textbf{I}_\mathrm{tilt} &= \textbf{R}_\mathrm{tilt} \textbf{I} \textbf{R}^\mathrm{T}_\mathrm{tilt}\\
    &= \begin{pmatrix}
        1 & 0 & 0 \\
        0 & \cos \tau & \sin \tau \\
        0 & -\sin \tau & \cos \tau
    \end{pmatrix}
    \begin{pmatrix}
        I_{xx} & 0 & 0 \\
        0 & I_{yy} & 0 \\
        0 & 0 & I_{zz}
    \end{pmatrix}
    \begin{pmatrix}
        1 & 0 & 0 \\
        0 & \cos \tau & -\sin \tau \\
        0 & \sin \tau & \cos \tau
    \end{pmatrix}\\
    &=    \begin{pmatrix}
        I_{xx} & 0 & 0 \\
        0 & I_{yy}\cos^2\tau + I_{zz}\sin^2\tau & \frac{1}{2}(-I_{yy} + I_{zz})\sin 2\tau \\
        0 & \frac{1}{2}(-I_{yy} + I_{zz})\sin 2\tau & I_{yy}\sin^2\tau + I_{zz}\cos^2\tau
    \end{pmatrix}.
\label{eq:tilted_quad_matrix}
\end{align}

Next, to account for rotation, we apply a time-dependent rotation matrix around the z axis with a constant angular velocity $\Omega$, so $\phi = \Omega t$, to form a wobbling system,
\begin{align}
    \textbf{I}_\mathrm{wob} &= \textbf{R}_\mathrm{spin} \textbf{I}_\mathrm{tilt} \textbf{R}^\mathrm{T}_\mathrm{spin}\\
    &= \begin{pmatrix}
        \cos \Omega t & \sin \Omega t & 0\\
        -\sin \Omega t & \cos \Omega t & 0 \\
        0 & 0 & 1
    \end{pmatrix}
    \begin{pmatrix}
        I_{\mathrm{tilt},xx} & 0 & 0 \\
        0 & I_{\mathrm{tilt},yy} & I_{\mathrm{tilt},yz} \\
        0 & I_{\mathrm{tilt},zy} & I_{\mathrm{tilt},zz}
    \end{pmatrix}
    \begin{pmatrix}
        \cos \Omega t & -\sin \Omega t & 0\\
        \sin \Omega t & \cos \Omega t & 0 \\
        0 & 0 & 1
    \end{pmatrix}\\
    &=     \begin{pmatrix}
        I_{\mathrm{tilt},xx}\cos^2\phi + I_{\mathrm{tilt},yy}\sin^2\phi & \frac{1}{2} (-I_{\mathrm{tilt},xx} + I_{\mathrm{tilt},yy})\sin 2\phi & I_{\mathrm{tilt},yz}\sin\phi \\
        \frac{1}{2} (-I_{\mathrm{tilt},xx} + I_{\mathrm{tilt},yy})\sin 2\phi & I_{\mathrm{tilt},xx}\sin^2\phi + I_{\mathrm{tilt},yy}\cos^2\phi & I_{\mathrm{tilt},yz}\cos\phi \\
        I_{\mathrm{tilt},yz}\sin\phi & I_{\mathrm{tilt},yz}\cos\phi & I_{\mathrm{tilt},zz}
    \end{pmatrix}.
\end{align}
$\textbf{I}_\mathrm{wob}$ contains the relevant time varying information; one time derivative yields
\begin{align}
\dot{I}_{\mathrm{wob},xx} &= (\dot{I}_{\mathrm{tilt},xx}\cos^2\phi - \Omega I_{\mathrm{tilt},xx}\sin 2\phi)+( \dot{I}_{\mathrm{tilt},yy}\sin^2\phi + \Omega I_{\mathrm{tilt},yy}\sin2\phi)\\
\dot{I}_{\mathrm{wob},xy} &= \frac{1}{2}\Big[ (-\dot{I}_{\mathrm{tilt},xx} + \dot{I}_{\mathrm{tilt},yy})\sin 2\phi + 2\Omega(-I_{\mathrm{tilt},xx} + I_{\mathrm{tilt},yy})\cos 2\phi\Big]\\
\dot{I}_{\mathrm{wob},xz} &= \dot{I}_{\mathrm{tilt},yz}\sin\phi +\Omega I_{\mathrm{tilt},yz}\cos\phi \\
\dot{I}_{\mathrm{wob},yy} &= (\dot{I}_{\mathrm{tilt},xx}\sin^2\phi + \Omega I_{\mathrm{tilt},xx}\sin2\phi) + (\dot{I}_{\mathrm{tilt},yy}\cos^2\phi - \Omega I_{\mathrm{tilt},yy}\sin 2\phi)\\
\dot{I}_{\mathrm{wob},yz} &= \dot{I}_{\mathrm{tilt},yz}\cos\phi - \Omega I_{\mathrm{tilt},yz}\sin\phi \\
\dot{I}_{\mathrm{wob},zz} &= \dot{I}_{\mathrm{tilt},zz},\\
\end{align}
and a second time derivative ($\dot{\Omega} = 0$ for constant rotation) yields
\begin{align}
\ddot{I}_{\mathrm{wob},xx} &= (\ddot{I}_{\mathrm{tilt},xx}\cos^2\phi - 2\Omega \dot{I}_{\mathrm{tilt},xx}\sin 2\phi - 2\Omega^2 I_{\mathrm{tilt},xx}\cos 2\phi)\nonumber\\ &+( \ddot{I}_{\mathrm{tilt},yy}\sin^2\phi + 2\Omega \dot{I}_{\mathrm{tilt},yy}\sin2\phi + 2\Omega^2 I_{\mathrm{tilt},yy}\cos 2\phi)\\
\ddot{I}_{\mathrm{wob},xy} &= \frac{1}{2}\Big[ (-\ddot{I}_{\mathrm{tilt},xx} + \ddot{I}_{\mathrm{tilt},yy})\sin 2\phi + 4\Omega(-\dot{I}_{\mathrm{tilt},xx} + \dot{I}_{\mathrm{tilt},yy})\cos 2\phi - 4\Omega^2(-I_{\mathrm{tilt},xx} + I_{\mathrm{tilt},yy})\sin 2\phi\Big]\\
\ddot{I}_{\mathrm{wob},xz} &= \ddot{I}_{\mathrm{tilt},yz}\sin\phi +2\Omega \dot{I}_{\mathrm{tilt},yz}\cos\phi - \Omega^2 {I}_{\mathrm{tilt},yz}\sin\phi\\
\ddot{I}_{\mathrm{wob},yy} &= (\ddot{I}_{\mathrm{tilt},xx}\sin^2\phi + 2\Omega \dot{I}_{\mathrm{tilt},xx}\sin2\phi + 2\Omega^2 I_{\mathrm{tilt},xx}\cos2\phi) \nonumber\\ &+ (\ddot{I}_{\mathrm{tilt},yy}\cos^2\phi - 2\Omega \dot{I}_{\mathrm{tilt},yy}\sin 2\phi - 2\Omega^2 I_{\mathrm{tilt},yy}\cos 2\phi) \\
\ddot{I}_{\mathrm{wob},yz} &= \ddot{I}_{\mathrm{tilt},yz}\cos\phi - 2\Omega \dot{I}_{\mathrm{tilt},yz}\sin\phi - \Omega^2 I_{\mathrm{tilt},yz}\cos\phi\\
\ddot{I}_{\mathrm{wob},zz} &= \ddot{I}_{\mathrm{tilt},zz}.
\end{align}
As we assume no change in tilt $\tau$, the first time derivative of $\textbf{I}_\mathrm{tilt}$ is 
\begin{align}
      \dot{\textbf{I}}_\mathrm{tilt}  &=    \begin{pmatrix}
        \dot{I}_{xx} & 0 & 0 \\
        0 & \dot{I}_{yy}\cos^2\tau + \dot{I}_{zz}\sin^2\tau & \frac{1}{2}(-\dot{I}_{yy} + \dot{I}_{zz})\sin 2\tau \\
        0 & \frac{1}{2}(-\dot{I}_{yy} + \dot{I}_{zz})\sin 2\tau & \dot{I}_{yy}\sin^2\tau + \dot{I}_{zz}\cos^2\tau
    \end{pmatrix},
\end{align}
with a similar form for the second time derivative, by changing $\dot{I}_{ij}$ with $\ddot{I}_{ij}$.  The time derivatives for $I_{ij}$ are acquired from Equation (\ref{eqn:ddotI_toy}) with simulation informed quantities describing the ellipsoid.

For an observer at a general altitudinal viewing angle $\theta$ with respect to the z axis, we apply a final rotation matrix about the x axis
\begin{align}
    \ddot{\textbf{I}}' &= \textbf{R}_\mathrm{view} \ddot{\textbf{I}}_{\mathrm{wob}} \textbf{R}^\mathrm{T}_\mathrm{view}\\
    &= \begin{pmatrix}
        1 & 0 & 0 \\
        0 & \cos \theta & \sin \theta \\
        0 & -\sin \theta & \cos \theta
    \end{pmatrix}
    \begin{pmatrix}
        \ddot{I}_{\mathrm{wob},xx} & \ddot{I}_{\mathrm{wob},xy} & \ddot{I}_{\mathrm{wob},xz} \\
        \ddot{I}_{\mathrm{wob},yx} & \ddot{I}_{\mathrm{wob},yy} & \ddot{I}_{\mathrm{wob},yz} \\
        \ddot{I}_{\mathrm{wob},zx} & \ddot{I}_{\mathrm{wob},zy} & \ddot{I}_{\mathrm{wob},zz}
    \end{pmatrix}
    \begin{pmatrix}
        1 & 0 & 0 \\
        0 & \cos \theta & -\sin \theta \\
        0 & \sin \theta & \cos \theta
    \end{pmatrix}.
\end{align}
Next, take the transverse projection, by removing the last row and column from $\ddot{\textbf{I}}'$.  For formatting considerations, we explicitly list the nonzero components $\ddot{I}'_{ij}$, rather than $\ddot{\textbf{I}}'$ in its entirety,
\begin{align}
    \ddot{I}_{xx}^{\prime\mathrm{T}} &= \ddot{I}_{\mathrm{wob},xx}\\
    \ddot{I}_{xy}^{\prime\mathrm{T}}    &= \ddot{I}_{\mathrm{wob},xy}\cos\theta + \ddot{I}_{\mathrm{wob},xz}\sin\theta \\
    \ddot{I}_{yy}^{\prime\mathrm{T}} &= \ddot{I}_{\mathrm{wob},yy}\cos^2\theta + \ddot{I}_{\mathrm{wob},yz}\sin 2\theta + \ddot{I}_{\mathrm{wob},zz}\sin^2\theta. 
\end{align}
To make the new transverse matrix $\textbf{I}^{\prime\mathrm{T}}$ traceless, subtract $1/2 \mathrm{Tr}(\textbf{I}^{\prime\mathrm{T}})$ from the nonzero diagonal elements of $\textbf{I}^{\prime\mathrm{T}}$ yielding,
\begin{align}
    \ddot{I}_{xx}^{\prime\mathrm{TT}} &= \frac{1}{2}\Big( \ddot{I}_{\mathrm{wob},xx} - \ddot{I}_{\mathrm{wob},yy}\cos^2\theta - \ddot{I}_{\mathrm{wob},yz}\sin2\theta - \ddot{I}_{\mathrm{wob},zz}\sin^2\theta \Big)\\
    \ddot{I}_{xy}^{\prime\mathrm{TT}} &= \ddot{I}_{yx}^{\prime\mathrm{TT}} = \ddot{I}_{xy}^{\prime\mathrm{T}} \\
    \ddot{I}_{yy}^{\prime\mathrm{TT}} &= -\ddot{I}_{xx}^{\prime\mathrm{TT}}.
\end{align}
With a known form for $\Ddot{\textbf{I}}'^{\mathrm{TT}}$, the two polarizations for the GW amplitude can is expressed as
\begin{equation}
    h_+ = \frac{2G}{c^4 D}\Ddot{I}_{xx}^{\prime\mathrm{TT}}
    \label{eqn:h+_general}
\end{equation}
and
\begin{equation}
    h_\times = \frac{2G}{c^4 D}\Ddot{I}_{xy}^{\prime\mathrm{TT}}.
    \label{eqn:hx_general}
\end{equation}
From Equations (\ref{eqn:h+_general}-\ref{eqn:hx_general}) the strain surface plots for the wobbling, oscillating ellipsoid are constructed.

\section{Angular Velocities of Numerical Simulations}
\label{app:rotational_velocity}

\begin{figure*}[h]
\centering
    \begin{subfigure}[t]{0.495\linewidth}
\includegraphics[width=\linewidth]{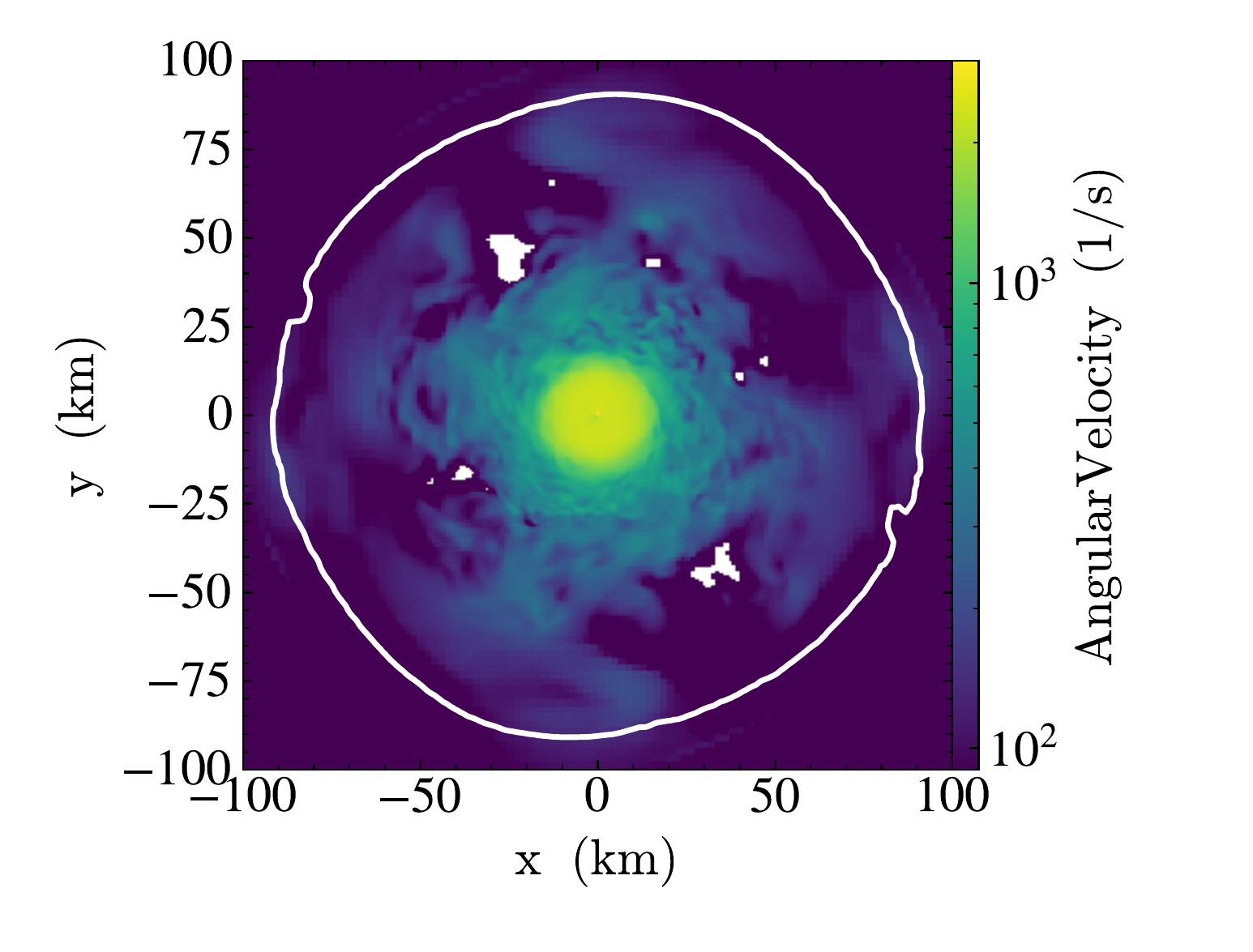}
    \end{subfigure}\hfill
    \begin{subfigure}[t]{0.495\linewidth}
\includegraphics[width=\linewidth]{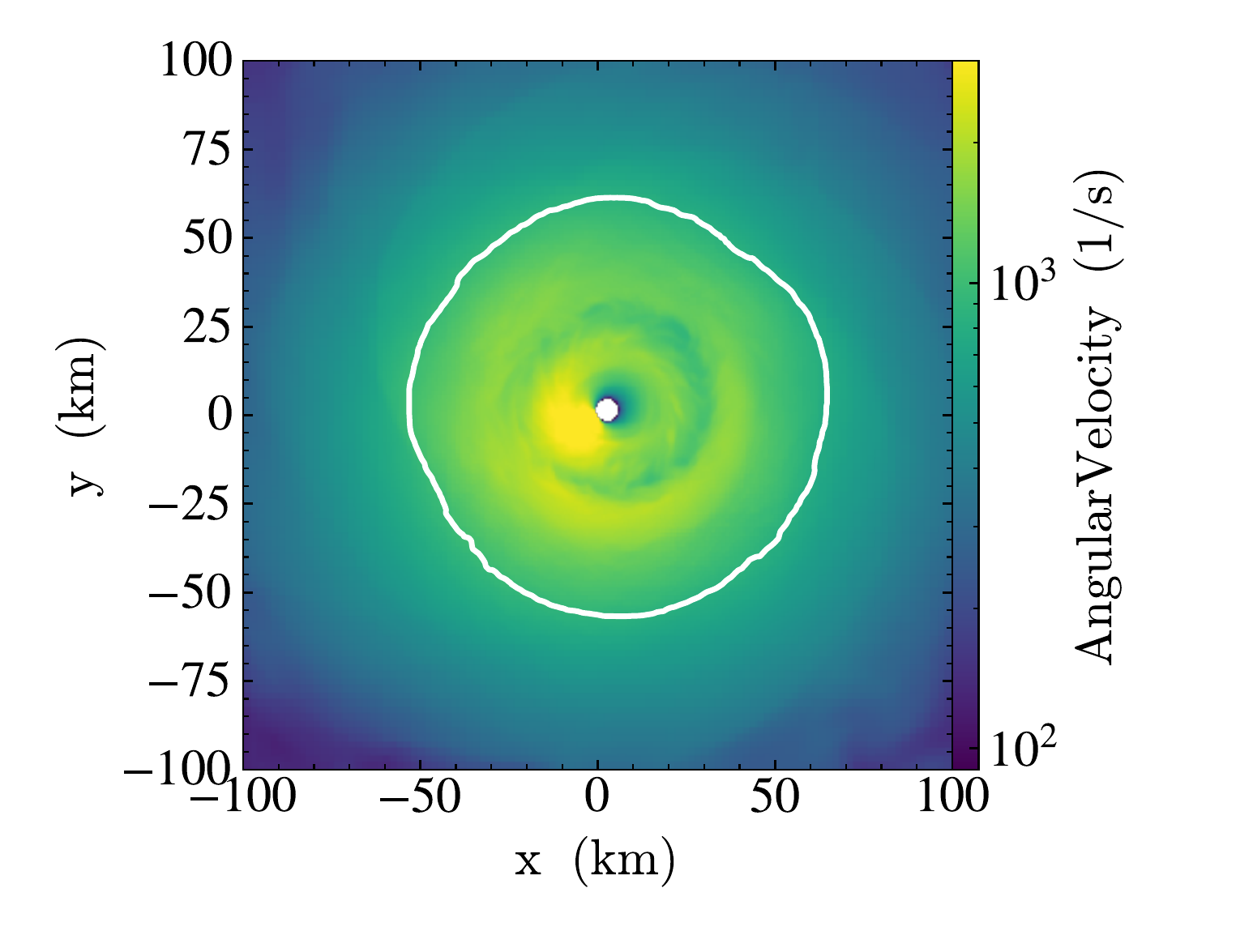}
    \end{subfigure}
\caption{ Nonuniformities in the angular velocity profiles for model \texttt{s40o1} during the ringdown phase (left) and accretion phase (right).  Slice plots show the equatorial plane.  White lines indicate the rest mass density contour for $10^{11}$ g cm$^{-3}$.}
    \label{fig:omega_eq_slices}
    \end{figure*}

In Section \ref{ssec:toy_ringdown} and Section \ref{ssec:toy_rotation}, we note a mismatch between the amplitudes of the strain surface plots for the simulations and toy model.  We attribute this difference to the solid-body assumption of the toy model.  In Figure \ref{fig:omega_eq_slices} we provide angular velocities for model \texttt{s40o1} during the ringdown phase ($\sim 13$ ms pb) on the left and the accretion phase ($\sim 285$ ms pb) on the right.  We denote the $10^{11}$ g cm$^{-3}$ density contour in white.  The small white patches within each slice plot are due to negative angular velocities (clockwise rotation) not rendering on the log scale.

To justify our reasoning regarding differences in the GW amplitudes between the strain surfaces from the simulations and toy models, we provide slices of the angular velocities. 
 In the left panel, we note a rapidly rotating inner 10 km, with nonuniformities in the angular velocity in the outer 90 km.  For the accretion phase case on the right, we notice differences that vary radially outward.  Neither case exhibits a uniform, solid-body angular velocity profile.  Thus, we justify our difference in amplitudes for the strain surface plots in Figure \ref{fig:GWsurfaces_ringdown} and Figure \ref{fig:GWsurface_solid_rot}.

\end{document}